\documentclass{aa}
\usepackage{txfonts}
\usepackage{natbib}
\usepackage{graphicx,amssymb,subfigure}
\usepackage[normalem]{ulem}
\usepackage{color}
\pdfoutput=1

\newcommand{\wave}[1]{$\lambda#1\,\mathrm{cm}$}  
\newcommand{\wwwave}[3]{$\lambda\lambda#1,#2,#3\,\mathrm{cm}$}  

\newcommand{\HI}{\mathrm{H\,\scriptstyle I}}
\newcommand{\HII}{\mathrm{H\,\scriptstyle II}}

\bibpunct{(}{)}{;}{a}{}{,} 
\def\ccm{\,{\rm cm^{-3}}}

\def\kms{\,{\rm km\,s^{-1}}}

\def\degr{\hbox{$^\circ$}}
\def\farcm{\hbox{$.\mkern-4mu^\prime$}}
\def\farcs{\hbox{$.\!\!^{\prime\prime}$}}

\def\muG{\,\mu{\rm G}}

\def\pc{\,{\rm pc,}}

\def\radm{\,\mathrm{rad\,m^{-2}}}

\begin{document}

\title{Magnetic fields in the nearby spiral galaxy IC~342:\\
A multi-frequency radio polarization study
\thanks{Based on observations with the VLA of the NRAO at Socorro and the
100-m telescope of the Max-Planck-Institut f\"ur Radioastronomie at Effelsberg.}
}

\titlerunning{Magnetic fields in IC~342}
\authorrunning{R.~Beck}

\author{Rainer~Beck \inst{}}

\institute{Max-Planck-Institut f\"ur Radioastronomie, Auf dem H\"ugel 69,
53121 Bonn, Germany, rbeck@mpifr-bonn.mpg.de}


\abstract{Magnetic fields play an important role in the formation and stabilization of spiral structures
in galaxies, but the interaction between interstellar gas and magnetic fields has not yet been understood.
In particular, the phenomenon of ``magnetic arms'' located between material arms is a mystery.}
{The strength and structure of interstellar magnetic fields and their relation to spiral arms in gas and
dust are investigated in the nearby and almost face-on spiral galaxy IC~342.}
{The total and polarized radio continuum emission of IC~342 was observed with high spatial resolution
in four wavelength bands with the Effelsberg and VLA telescopes. At \wave{6.2} the data from both
telescopes were combined. I separated thermal and nonthermal (synchrotron) emission components with the help
of the spectral index distribution and derived maps of the magnetic field strength, degree of magnetic field order,
magnetic pitch angle, Faraday rotation measure, and Faraday depolarization.}
{IC~342 hosts a diffuse radio disk with an intensity that decreases exponentially with increasing radius.
The frequency dependence of the scalelength of synchrotron emission indicates energy-dependent propagation
of the cosmic-ray electrons, probably via the streaming instability. The equipartition strength of the
total field in the main spiral arms is typically $15\muG$, that of the ordered field about $5\muG$.
The total radio emission, observed with the VLA's high resolution, closely follows the dust emission in
the infrared at 8\,$\mu$m (SPITZER telescope) and 22\,$\mu$m (WISE telescope).
The polarized emission is not diffuse, but concentrated in spiral arms of various types:
(1) a narrow arm of about 300\,pc width, displaced inwards with respect to the eastern arm by about
200\,pc, indicating magnetic fields compressed by a density wave;
(2) a broad arm of 300--500\,pc width around the northern arm with systematic variations in polarized
emission, polarization angles, and Faraday rotation measures on a scale of about 2\,kpc, indicative of a
helically twisted flux tube generated by the Parker instability;
(3) a rudimentary magnetic arm in an interarm region in the north-west;
(4) several broad spiral arms in the outer galaxy, related to spiral arms in the total neutral gas;
(5) short features in the outer south-western galaxy, probably distorted by tidal interaction.
Faraday rotation of the polarization angles reveals an underlying regular field of only $\simeq0.5\muG$ strength
with a large-scale axisymmetric spiral pattern, probably a signature of a mean-field $\alpha-\Omega$ dynamo,
and an about $10\times$ stronger field that fluctuates on scales of a few 100\,pc. The magnetic field around the bar in the central region of IC~342 resembles that of large barred galaxies;
it has a regular spiral pattern with a large pitch angle, is directed outwards, and is opposite to the large-scale
regular field in the disk. Polarized emission at \wave{20.1} is strongly affected by Faraday depolarization in the western and northern
parts of the galaxy. Helical fields extending from disk to halo may account for this asymmetry.}
{Interstellar magnetic fields interact with the gas and gas flows. Density-wave compression
generates polarized radio emission at the inner edge of some spiral arms. Fast MHD density waves can generate
coincident spiral arms in gas and magnetic fields in the outer parts of IC~342. Magnetic armsar e offset
from the spiral pattern in gas and dust; their generation and development by mean-field dynamo action probably
need a spiral pattern that is stable over a few galactic rotation periods, which is probably the case for
the galaxy NGC~6946. The mean-field dynamo in IC~342 is slow and weak, probably disturbed by the bar,
tidal interaction, or a transient spiral pattern.
}


\maketitle


\section{Introduction}
\label{sect:intro}

The role of magnetic fields in the interstellar medium (ISM) of galaxies is far from being understood.
Radio polarization observations have revealed magnetic fields of considerable strength in all galaxies
containing a significant level of star formation \citep{beck+wielebinski13,beck15b}.
The magnetic energy density, measured with the help of radio continuum emission, was found to be comparable to that
of turbulent gas motions and cosmic rays \citep{beck07,taba08}. However, the processes cannot be identified
in detail by observations with the resolution of current radio telescopes.

Numerical simulations of density-wave perturbations of the ISM in galaxies, including magnetic fields, are scarce and
inconclusive so far. According to \citet{gomez02,gomez04} a moderately strong magnetic field modifies the gas response
to a spiral density perturbation, causing radial and vertical gas flows outwards and upwards in front of the
spiral arm, and inwards and downwards behind the arm. The magnetic field is carried along with the flow,
and the magnetic pitch angle changes sign when moving across the spiral arm.
\citet{dobbs08} find that gaseous spiral arms are smoother and better defined when strong magnetic fields
are included, because instabilities are suppressed and that the magnetic field follows the gaseous spiral arm.
In the MHD simulation of an evolving galaxy \citep{pakmor13}, spiral arms develop dynamically
in gas and magnetic fields.
All simulations so far agree that the magnetic field is mostly ordered in the regions of the highest gas density
in the spiral arms, while radio polarization observations show that the strongest ordered fields are
located mostly {\em \emph{between}}\ the spiral arms.


IC~342 is an ideal object for investigating magnetic fields and their relation to gas and dust with high spatial
resolution. It is an outstanding late-type spiral galaxy because of its large angular size (30\arcmin\ in optical
light and in the infrared, more than 90\arcmin\ in the $\HI$ line of atomic neutral gas) and its favourably
low inclination.
Beyond M~31 and M~33, IC~342 is the nearest spiral galaxy.\footnote{Assuming a distance of $D\simeq3.5$\,Mpc \citep{saha02,karachentsev03,tikhonov10}, 1\arcmin\ corresponds to $\simeq1.0$\,kpc and the optical radius
$r_{25}$ to 11\,kpc.}
Optical observations of IC~342 are hampered by the location behind the Galactic disk ($b=10.6\degr$) and
hence high extinction along the line of sight.
A moderately high overall star-formation rate is inferred from the far--infrared luminosity of
$9 \cdot 10^9$\,L$_{\odot}$ \citep{young96}, scaled to the distance of 3.5\,Mpc assumed in this paper.

According to kinematic $\HI$ data, the galaxy plane is inclined
by 31\degr\ to the plane of the sky and has a position angle of 37\degr\ \citep{crosthwaite00}.
The spiral structure is multi-armed and partly broken into segments in the optical range
\citep{white03}, in the infrared \citep{jarrett13}, in molecular gas \citep{crosthwaite01} and in neutral
hydrogen ($\HI$) gas \citep{newton80b,crosthwaite00}. Assuming trailing spiral arms,
the north-western side of IC~342 is nearest to us, and the galaxy is rotating counter-clockwise. The spiral pattern
is two-armed in the inner disk and four-armed in the outer disk, possibly with two different
pattern speeds \citep{meidt09}.
Velocity residuals indicate a large-scale bipolar pattern, which indicates a bar potential, so that IC~342 may
be in the transition phase from a barred to a two-armed spiral galaxy \citep{crosthwaite00}.
The lack of velocity perturbations $>5$\,km/s indicates that density waves are weak \citep{newton80b}.

The first radio continuum study of IC~342 by \citet{baker77} revealed a diffuse disk with a steep spectrum
of synchrotron origin and emission ridges mostly following the optical arms.
The average intensity of the radio disk is much higher than that of the nearest spiral galaxy, M~31, though
not as high as that of NGC~6946. The high star-formation rate per surface area of about
0.016\,M$_{\odot}$\,yr$^{-1}$\,kpc$^{-2}$ \citep{calzetti10} also yields a high surface brightness of the
disk of IC~342 in the infrared bands \citep{jarrett13}.
IC~342 was one of the first four galaxies for which a tight correlation between radio and infrared intensities
{\em \emph{within}}\ the disks was found \citep{beck+golla88}, indicating that the energy densities of cosmic rays
and magnetic fields are related to star-forming activity \citep{niklas97}.

\begin{table*}           
\caption{Radio continuum observations of IC~342. The Effelsberg observations were performed in scanning mode and
hence have no specific pointings. ``Resolution'' means diameter of the telescope beam at half power.}
\centering
\begin{tabular}{lcc}
\hline
\  & VLA & Effelsberg \\
\hline
Frequency (GHz)       & 8.435 \& 8.485  & 10.55 \\
Wavelength (cm)       & 3.55  \& 3.53   & 2.84 \\
Bandwidth (MHz)       & 100             & 500 \\
Array configuration      & D            & --  \\
Pointings                & 3            & -- \\
Observing dates \hfill   & 1991 Mar 23, 2003 Feb 24  & 1994 June -- Oct \\
Project IDs              & AB591, AB953 & 122-93 \\
On-source observing time (h)   & 13     & 79 \\
Resolution of final maps & 12\arcsec\ \& 15\arcsec & 68\arcsec\ \& 90\arcsec\\
Rms noise in $I$; $Q$ and $U$ ($\mu{\rm Jy/beam\,area}$)  & 10; 5 \& 13; 7  & 650; 250 \& 500; 150 \\
\hline

Frequency (GHz)       & 4.835 \& 4.885  & 4.85 \\
Wavelength (cm)       & 6.20  \& 6.14   & 6.18 \\
Bandwidth (MHz)       & 100             & 500 \\
Array configuration      & D            & --  \\
Pointings                & 5            & -- \\
Observing dates \hfill   & 1988 Sep 23 + 2003 Feb 28 + 2007 May 20 & 2000 Apr -- Aug \\
Project ID               & AK201, AB1066, AB1228 &  18-00 \\
On-source observing time (h)   & 24        & 29 \\
Resolution of final maps & 15\arcsec\ (NW pointing) \& 25\arcsec       & 147\arcsec\ \& 180\arcsec \\
Rms noise in $I$; $Q$ and $U$ ($\mu{\rm Jy/beam\,area}$) & 20; 10 \& 20; 12 & 600; 80 \& 500; 70\\
Reference                & \citet{krause93} (NW pointing) \\
\hline

Frequency (GHz)       &                 & 2.675 \\
Wavelength (cm)       &                 & 11.2  \\
Bandwidth (MHz)       &                 & 80 \\
Observing dates \hfill   &              & 2000 Apr to Aug \\
Project ID            &                 & 19-00 \\
On-source observing time (h)   &        & 24 \\
Resolution of final maps &              & 260\arcsec\ \& 300\arcsec \\
Rms noise in $I$; $Q$ and $U$ ($\mu{\rm Jy/beam\,area}$)  &       & 1000; 650 \& 800; 450 \\
\hline

Frequency (GHz)       & 1.490              & 1.400 \\
Wavelength (cm)       & 20.1               & 21.4 \\
Bandwidth (MHz)       & 100                & 40 \\
Array configuration      & D + C           & -- \\
Pointings                & 3               & -- \\
Observing date  \hfill   & 1984 Jul 29 \& 1986 Dec 20 + 1987 Mar 16 &  2000 Nov -- Dec \\
Project ID               &  AK107, AH248   &  19-00 \\
Net observing time (h)   &  16             &  7 \\
Resolution of final maps & 15\arcsec\ \& 25\arcsec\ \& 51\arcsec\ & 560\arcsec\ \& 600\arcsec \\
Rms noise in $I$; $Q$ and $U$ ($\mu{\rm Jy/beam\,area}$) & 30; 20 \& 45; 25 \& 100; 40 & 2500; 2000 \& 2000; 1500 \\
Reference                & \citet{krause89a} (D array, central pointing) \\
\hline

\label{tab:obs}
\end{tabular}
\end{table*}

\begin{figure*}[htbp]
\includegraphics[width=0.44\textwidth]{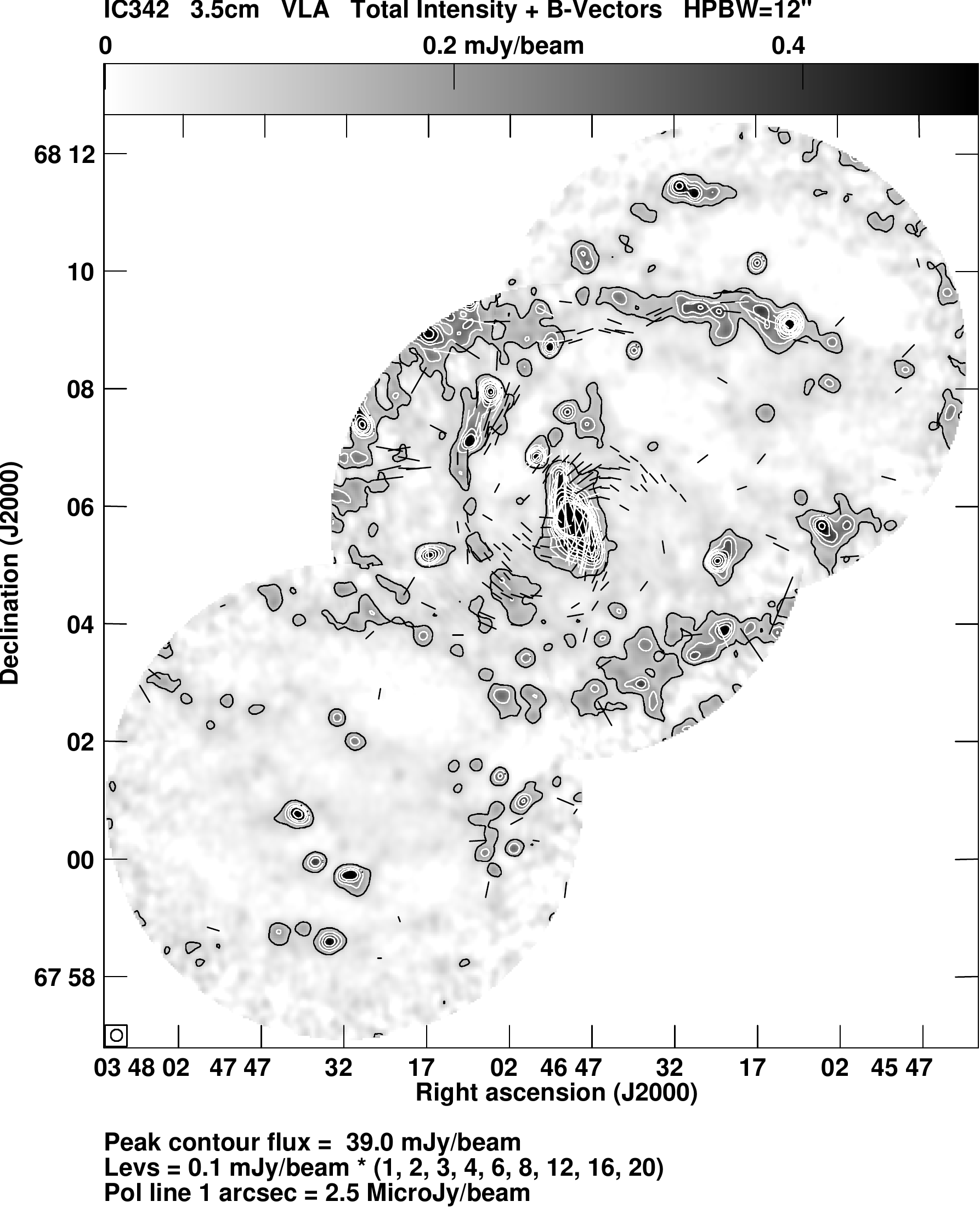}
\hfill
\includegraphics[width=0.48\textwidth]{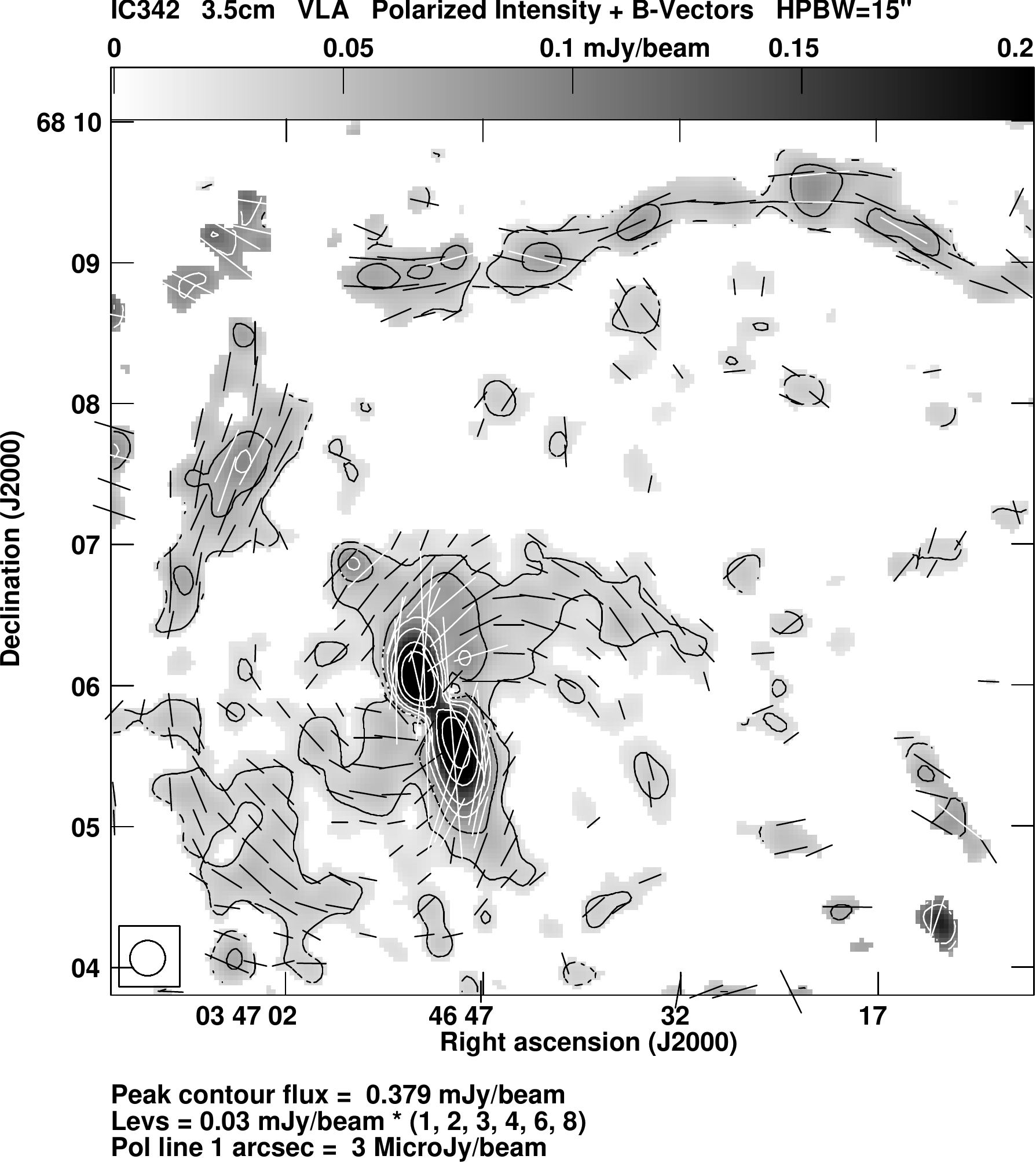}
\caption{ {\it Left:\/} Total intensity (contours and greyscale) and observed
$B$ vectors ($E$+90\degr) of IC~342 at \wave{3.5}
and at 12\arcsec\ resolution, combined from three VLA pointings (D array).
Here and in the following maps, the beam size is shown in the bottom
left corner of each panel.
{\it Right:\/} Linearly polarized intensity and observed
$B$ vectors ($E$+90\degr) in the central and northern regions
at \wave{3.5} and at 15\arcsec\ resolution.
The galaxy centre is located at
RA, DEC (J2000) = $03^\mathrm{h}\ 46^\mathrm{m}\ 48\fs 1$, +68\degr\ 05\arcmin\ 47\arcsec.
}
\label{cm3}
\end{figure*}

\begin{figure*}[htbp]
\includegraphics[width=0.46\textwidth]{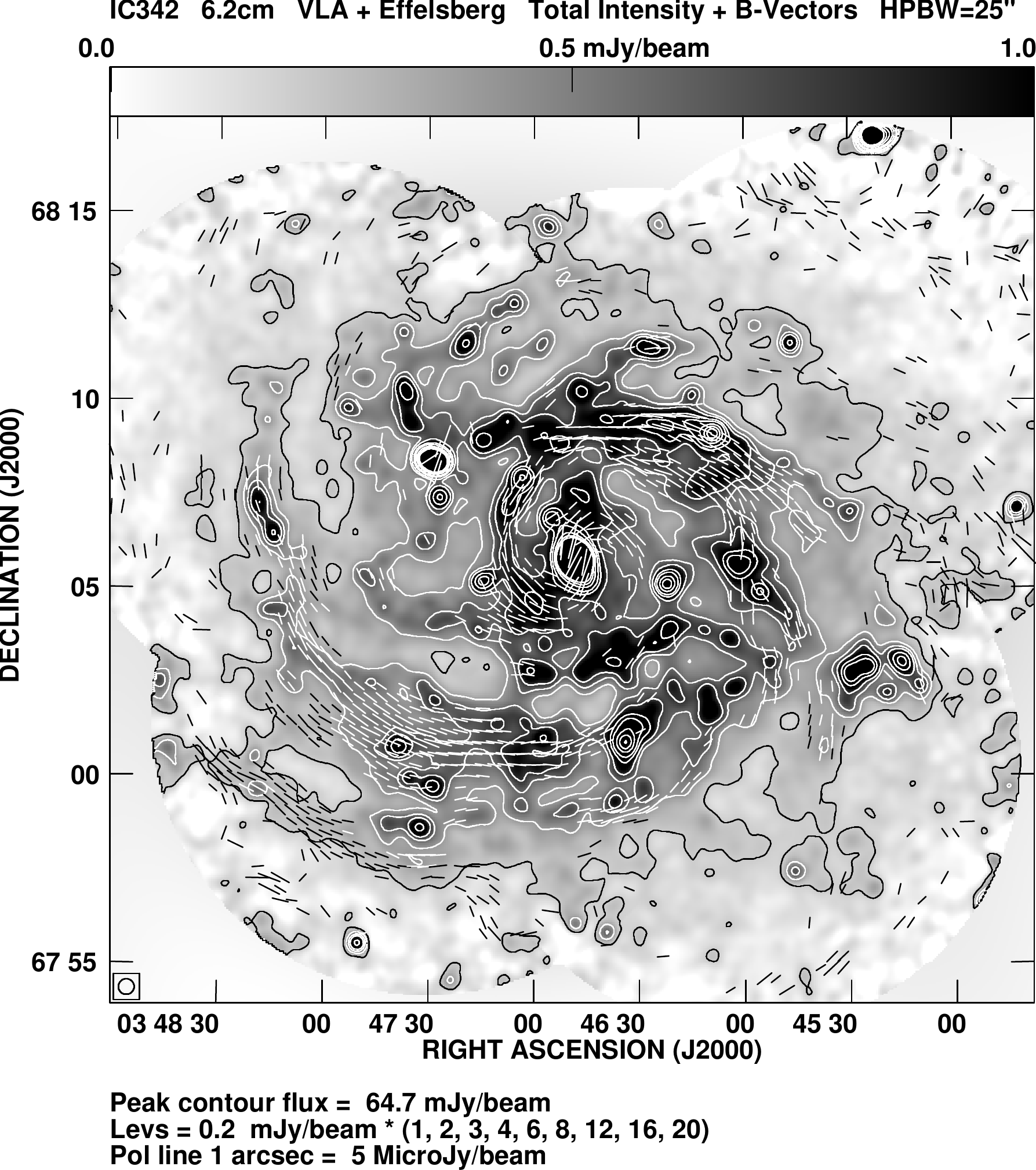}
\hfill
\includegraphics[width=0.445\textwidth]{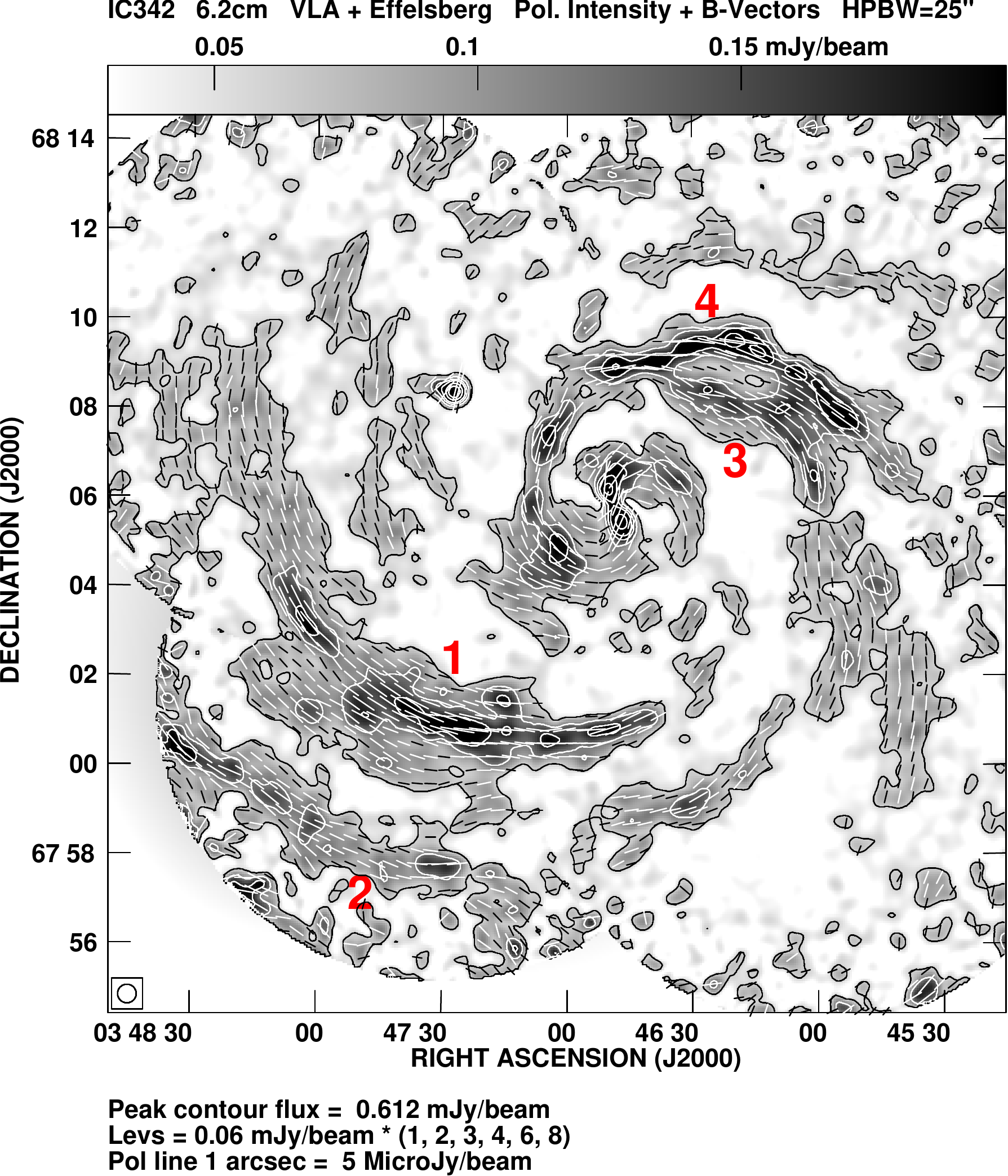}
\caption{ {\it Left:\/} Total intensity (contours and greyscale) and observed
$B$ vectors ($E$+90\degr) of IC~342 at \wave{6.2}
and at 25\arcsec\ resolution, combined from five VLA pointings (D array) and the Effelsberg data.
{\it Right:\/} Linearly polarized intensity and observed
$B$ vectors ($E$+90\degr) at \wave{6.2} and at 25\arcsec\ resolution.
Numbers refer to the spiral arms listed in Table~\ref{tab:pitch}.
}
\label{cm6}
\end{figure*}

\begin{figure*}[htbp]
\includegraphics[width=0.45\textwidth]{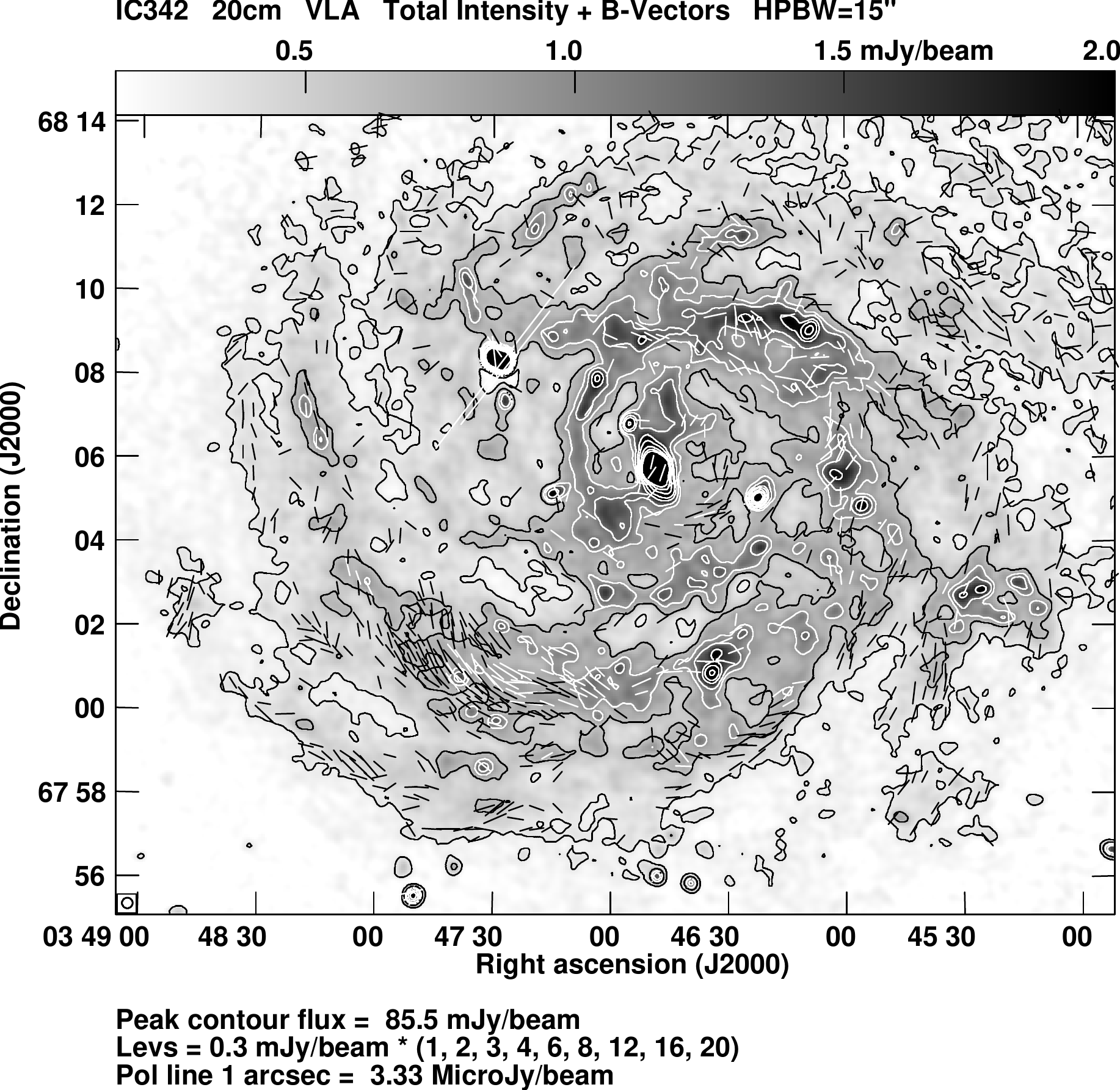}
\hfill
\includegraphics[width=0.45\textwidth]{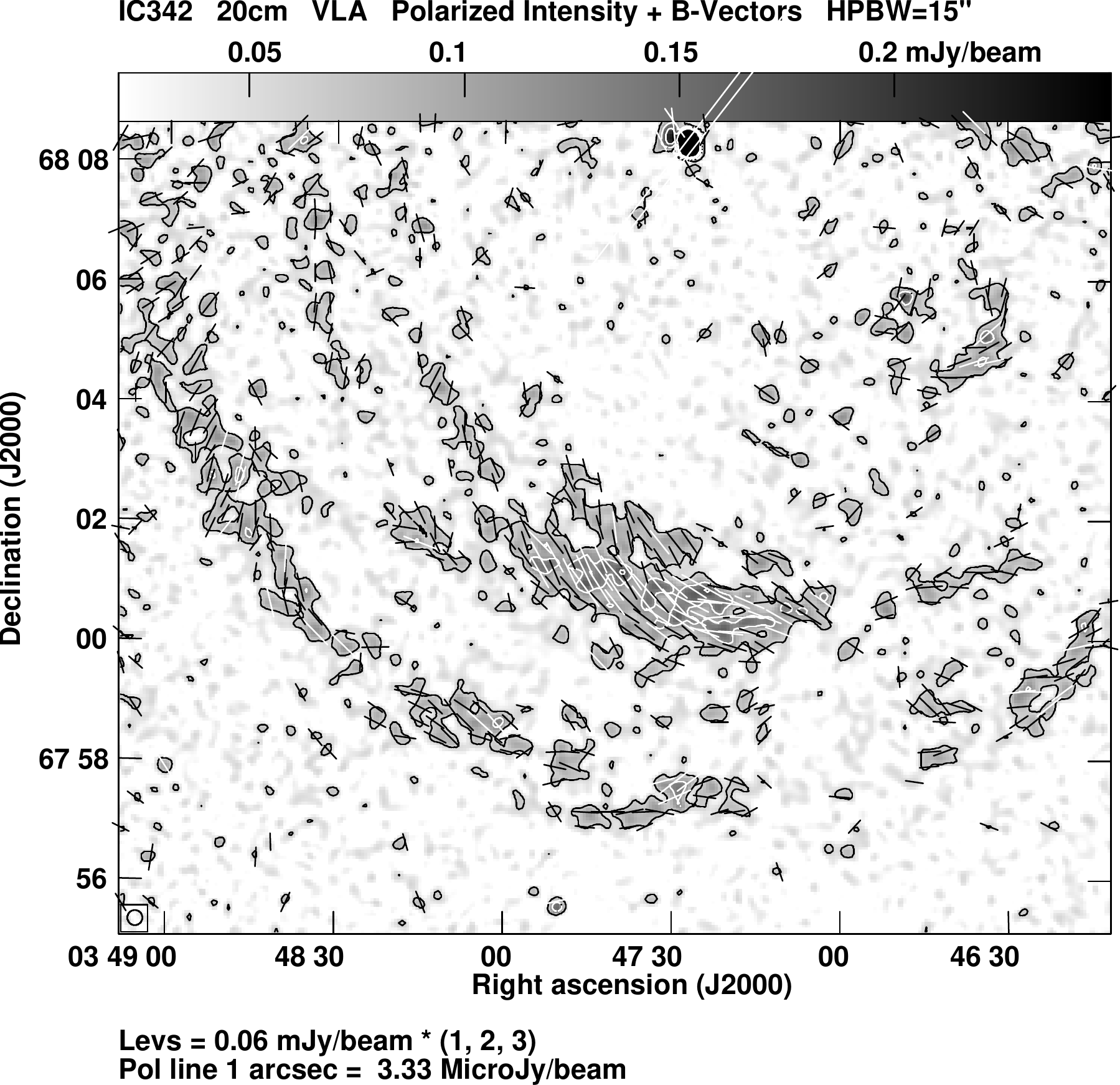}
\caption{ {\it Left:\/} Total intensity (contours and greyscale) and observed
$B$ vectors ($E$+90\degr) of IC~342 at \wave{20.1} and at 15\arcsec\ resolution,
combined from two VLA data sets at the same pointing position (C array and D array).
{\it Right:\/} Linearly polarized intensity and observed $B$ vectors ($E$+90\degr)
in the south-eastern region at \wave{20.1} and at 15\arcsec\ resolution.
}
\label{cm20a}
\end{figure*}

\begin{figure*}[htbp]
\includegraphics[width=0.44\textwidth]{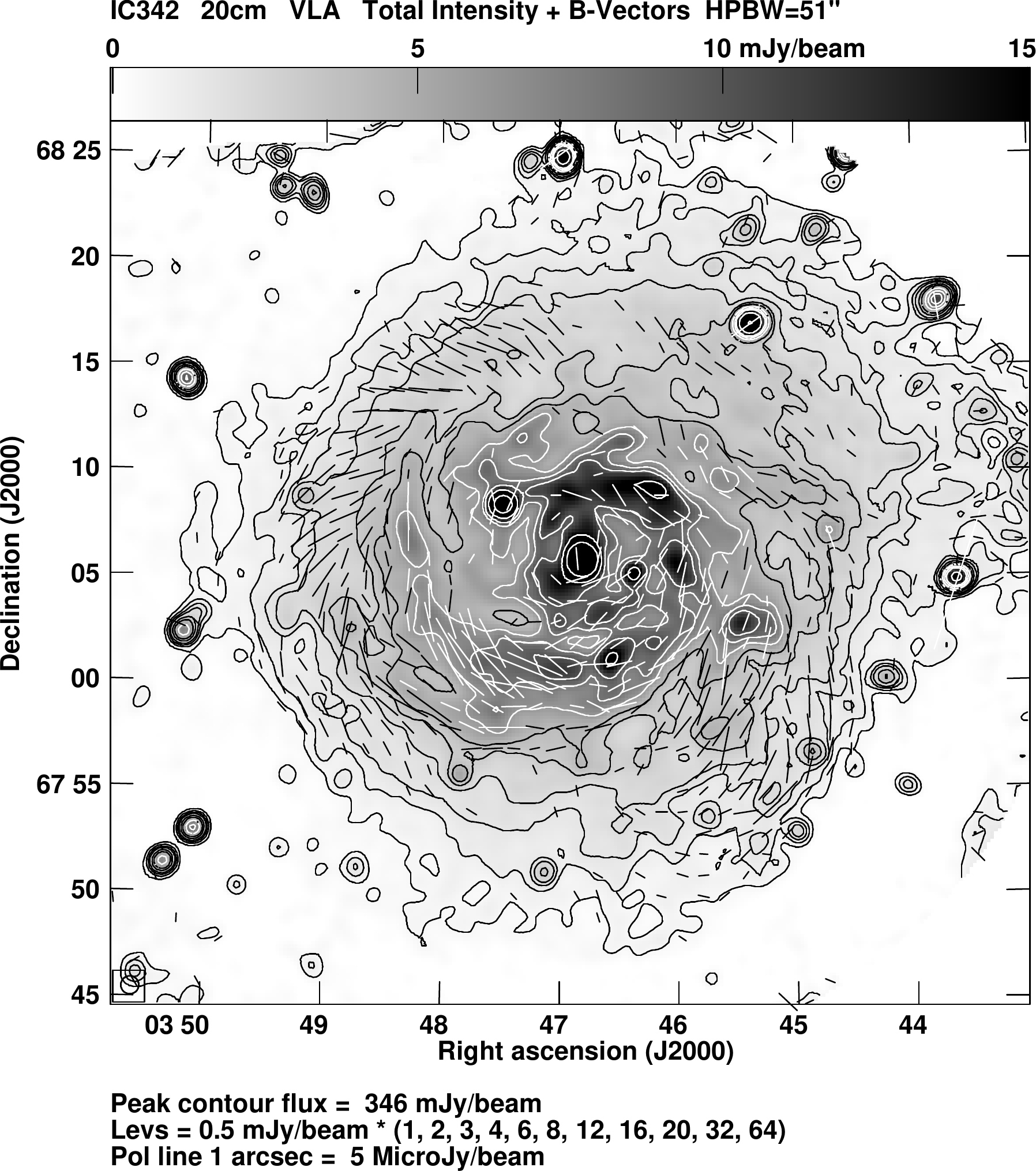}
\hfill
\includegraphics[width=0.47\textwidth]{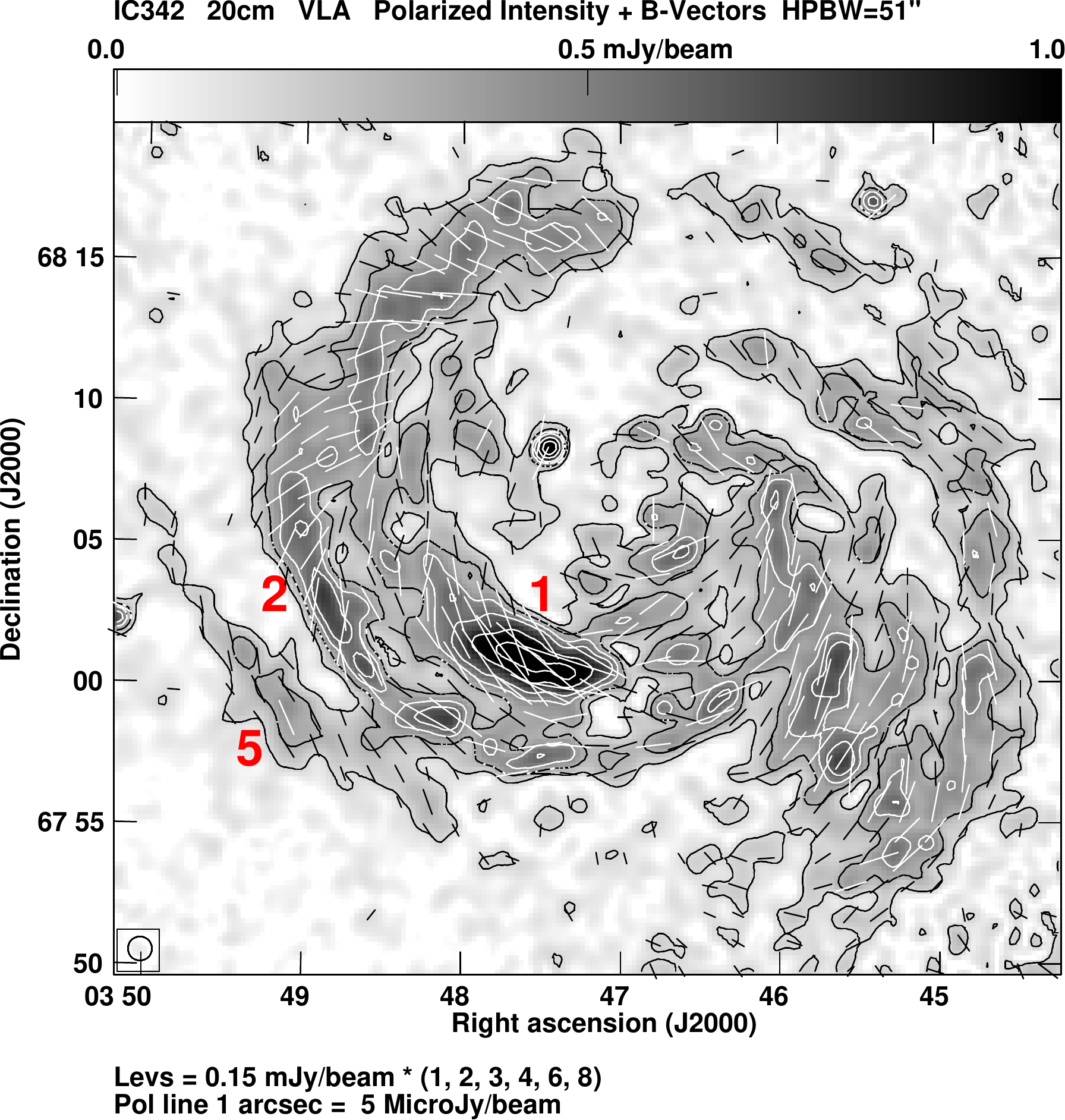}
\caption{ {\it Left:\/} Total intensity (contours and greyscale) and observed
$B$ vectors ($E$+90\degr) of IC~342 at \wave{20.1}
and at 51\arcsec\ resolution, combined from data of two VLA pointings (D array).
{\it Right:\/} Linearly polarized intensity and observed
$B$ vectors ($E$+90\degr) at \wave{20.1} and at 51\arcsec\ resolution.
Numbers refer to the spiral arms listed in Table~\ref{tab:pitch}.
}
\label{cm20b}
\end{figure*}

\begin{figure*}[htbp]
\includegraphics[width=0.45\textwidth]{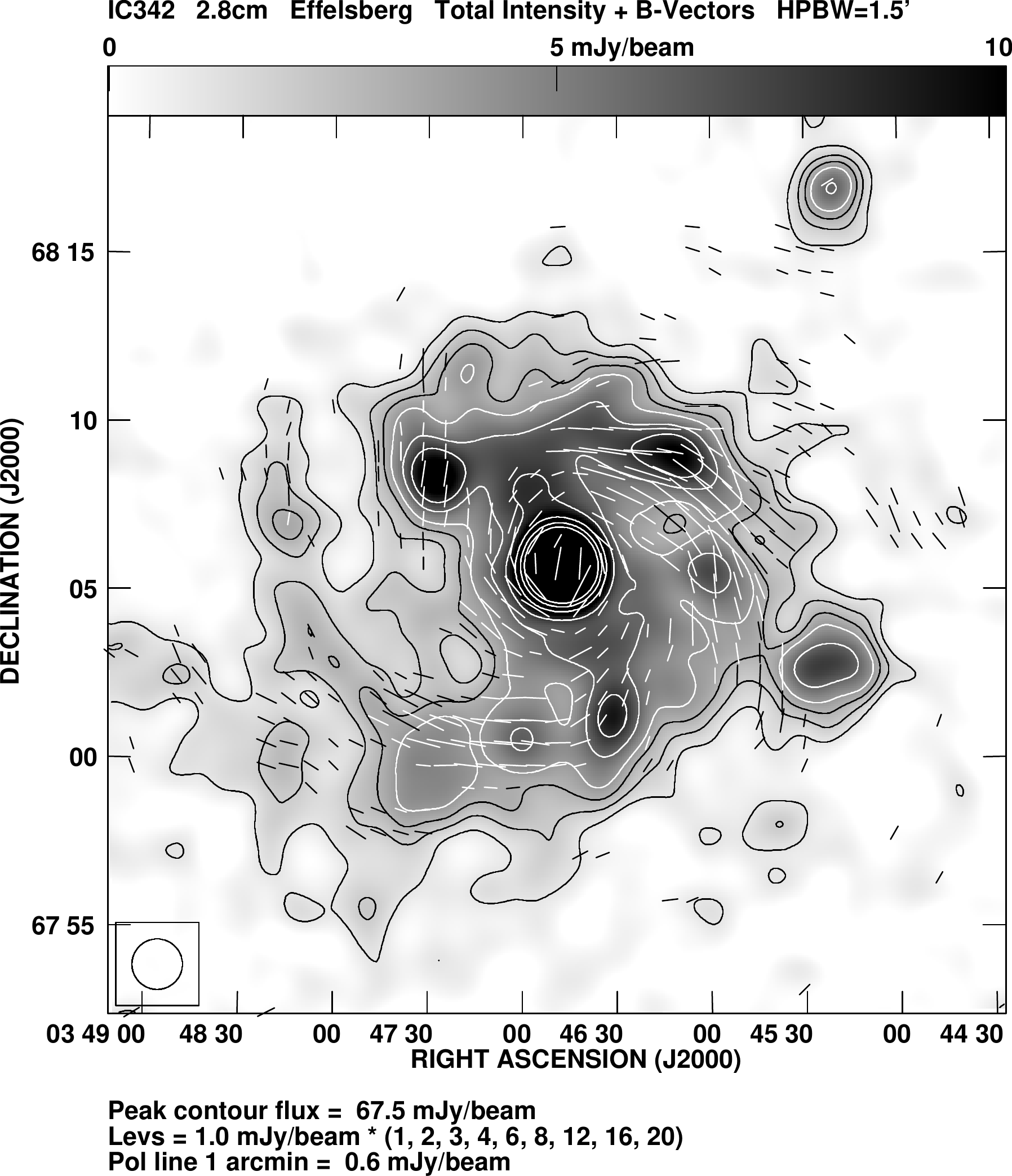}
\hfill
\includegraphics[width=0.45\textwidth]{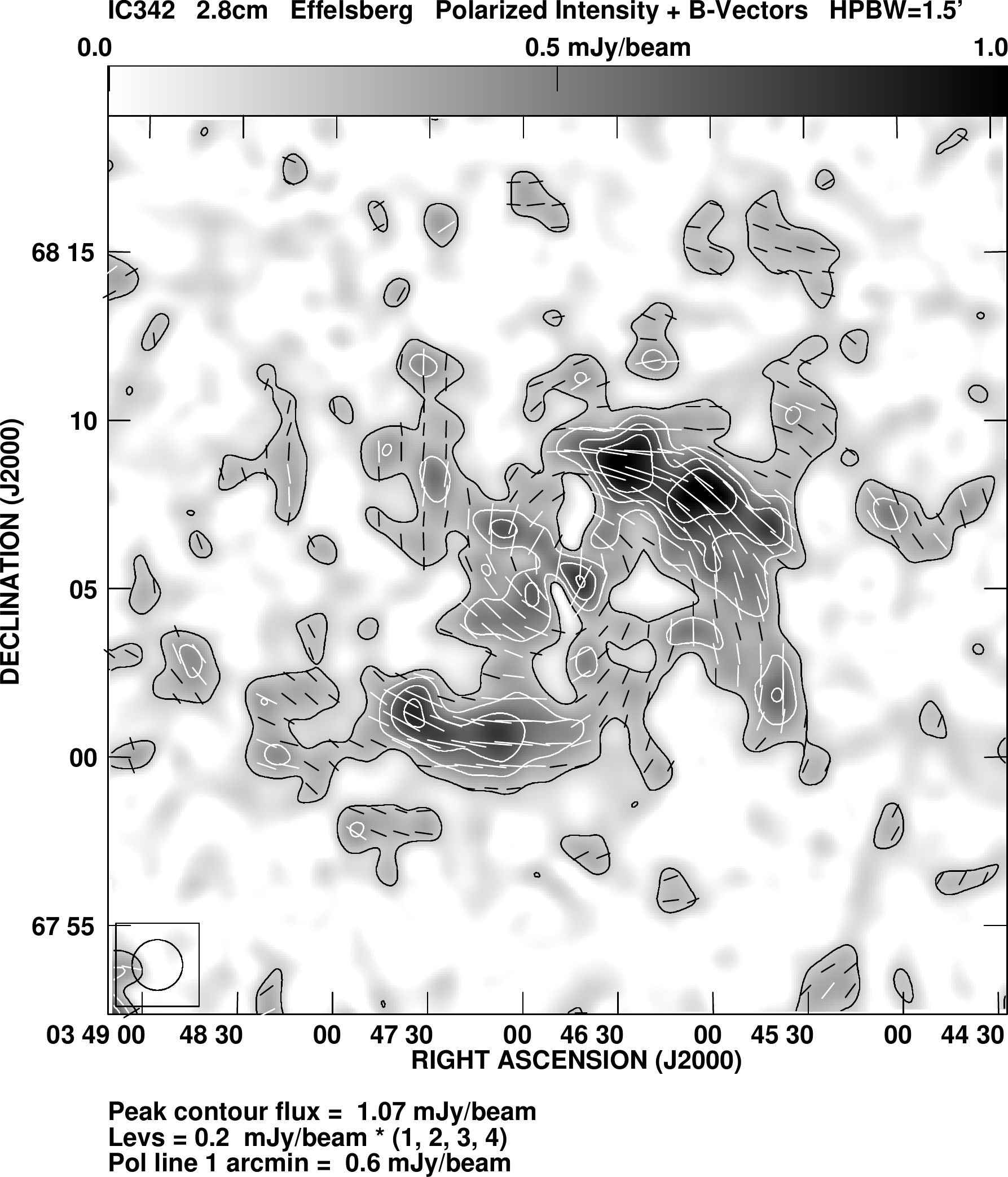}
\caption{ {\it Left:\/} Total intensity (contours and greyscale) and observed
$B$ vectors ($E$+90\degr) of IC~342 at
\wave{2.84} and at 90\arcsec\ resolution with the Effelsberg telescope.
{\it Right:\/} Linearly polarized intensity and observed
$B$ vectors ($E$+90\degr) at \wave{2.84} and at 90\arcsec\ resolution.
}
\label{cm28eff}
\end{figure*}

\begin{figure*}[htbp]
\includegraphics[width=0.45\textwidth]{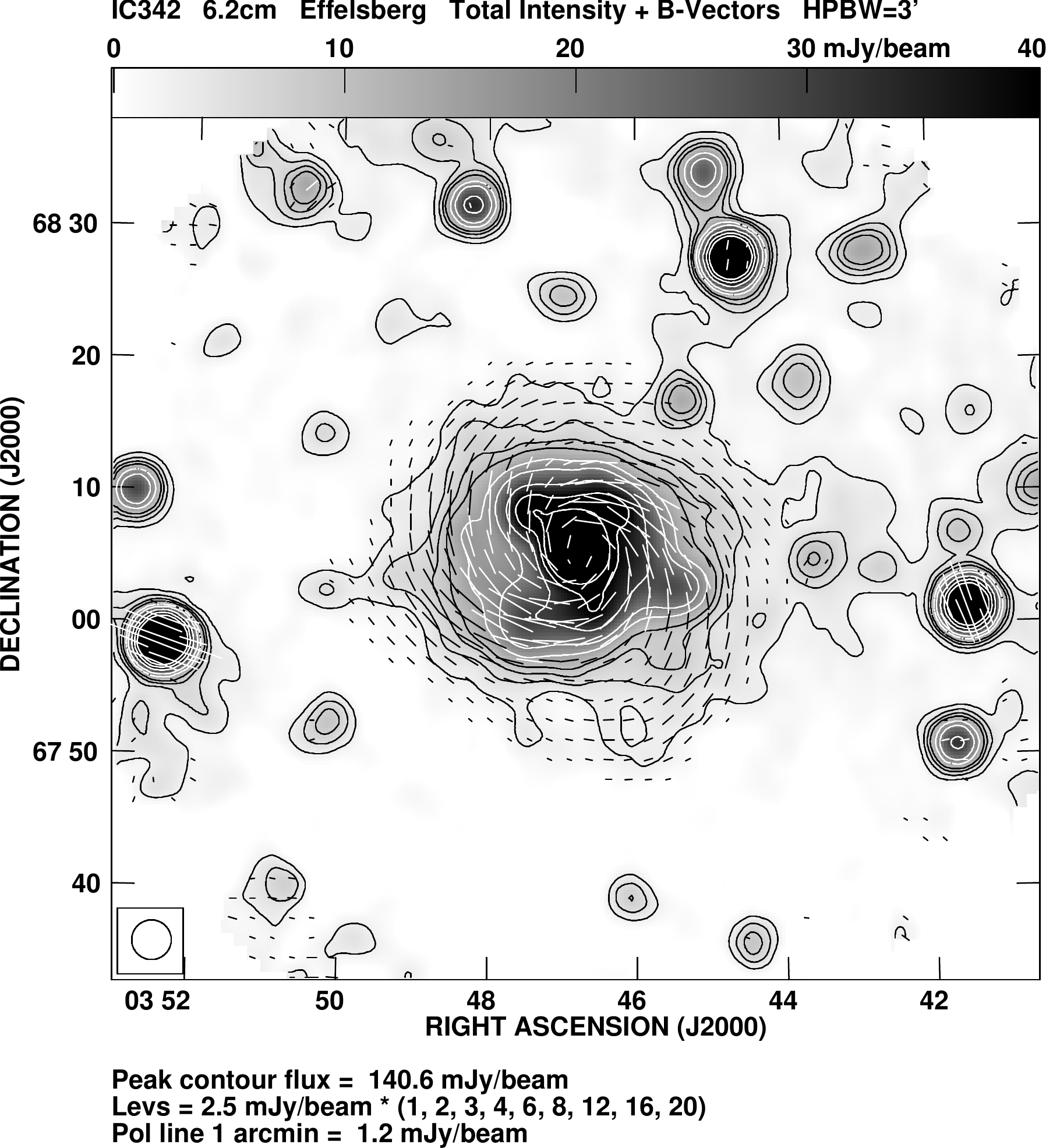}
\hfill
\includegraphics[width=0.45\textwidth]{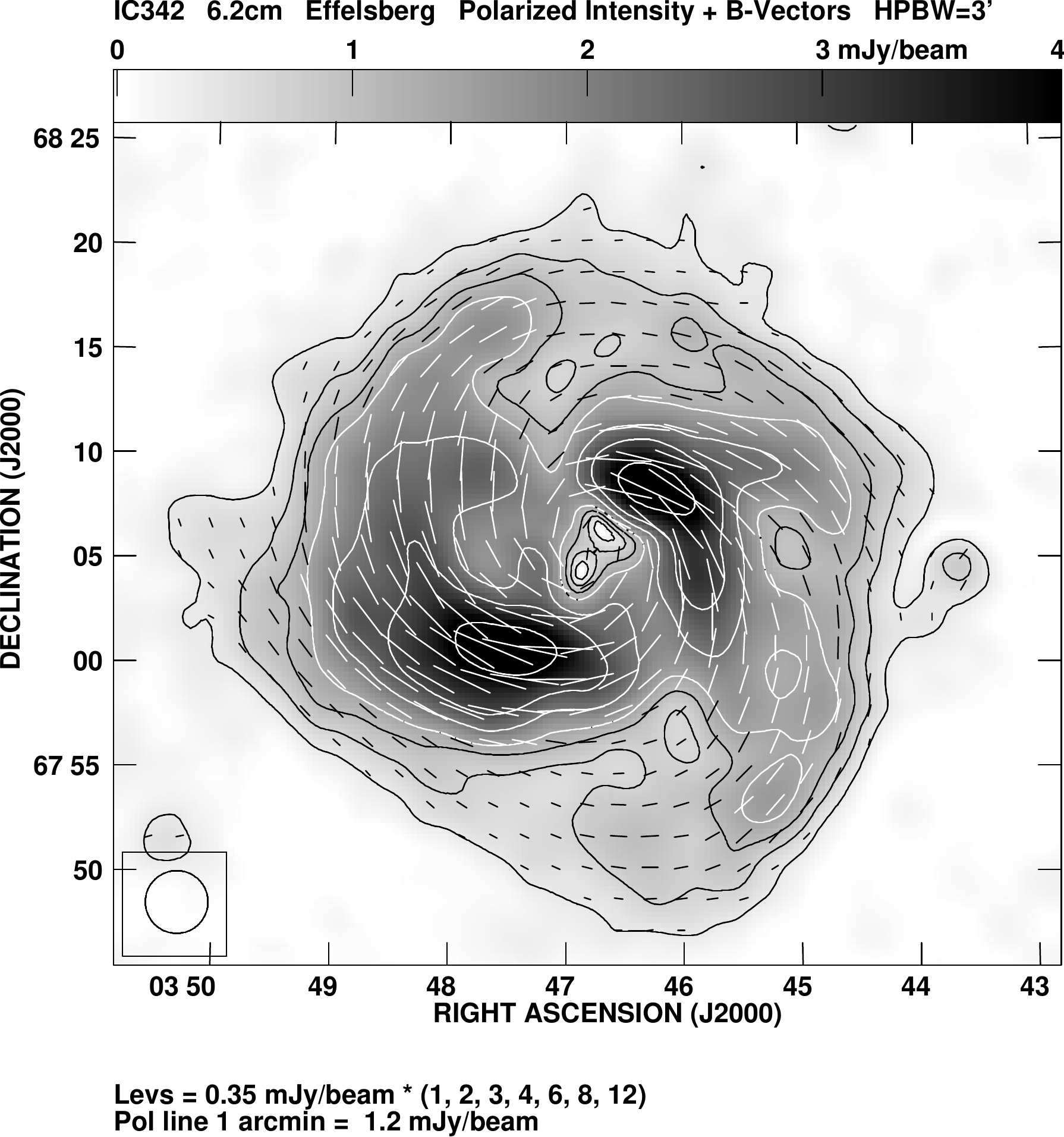}
\caption{ {\it Left:\/} Total intensity (contours and greyscale) and observed
$B$ vectors ($E$+90\degr) of IC~342 at
\wave{6.2} and at 3\arcmin\ resolution with the Effelsberg telescope.
{\it Right:\/} Linearly polarized intensity and observed
$B$ vectors ($E$+90\degr) of the inner region at \wave{6.2} and at 3\arcmin\ resolution.
}
\label{cm6eff}
\end{figure*}

\begin{figure*}[htbp]
\includegraphics[width=0.45\textwidth]{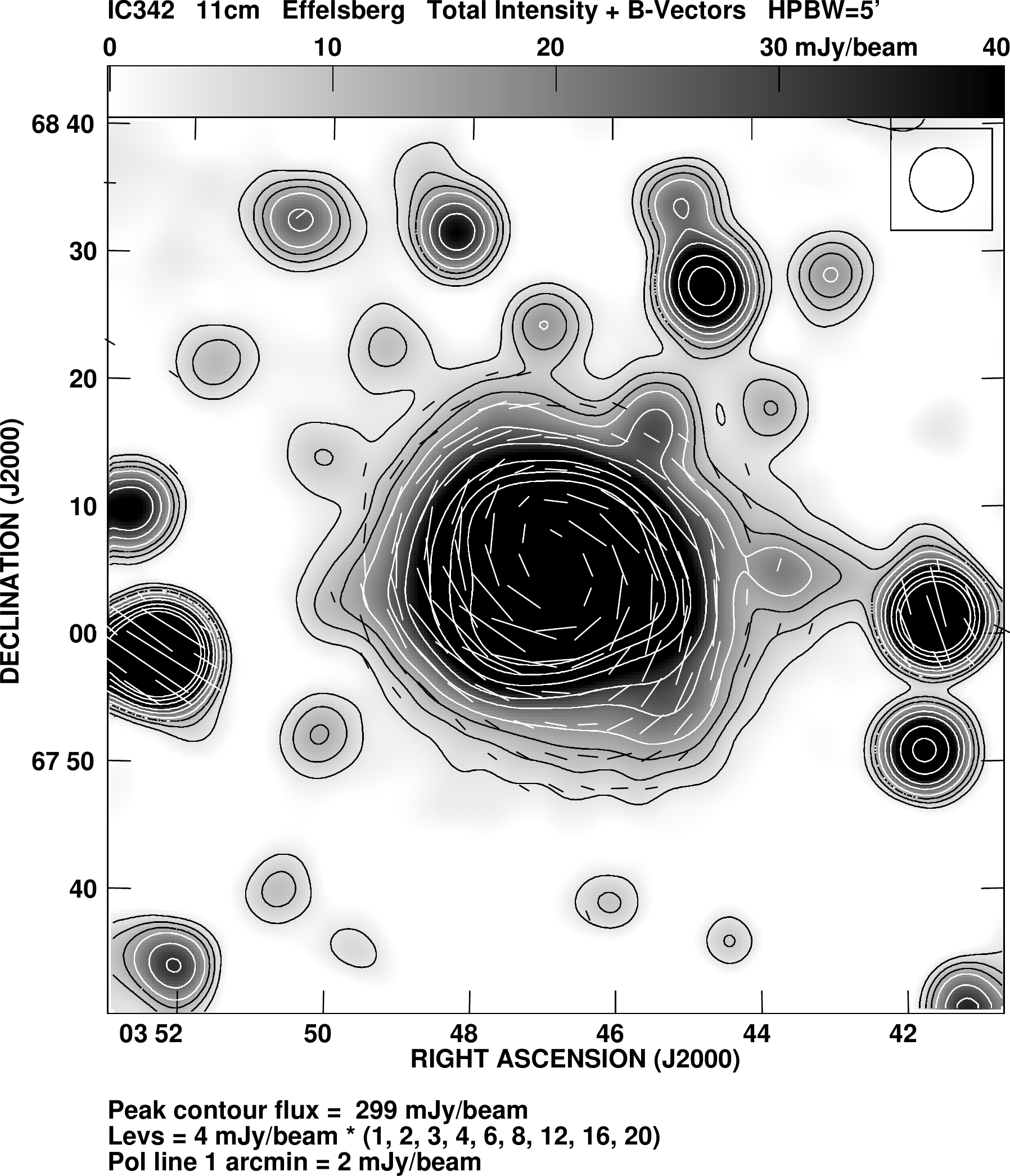}
\hfill
\includegraphics[width=0.45\textwidth]{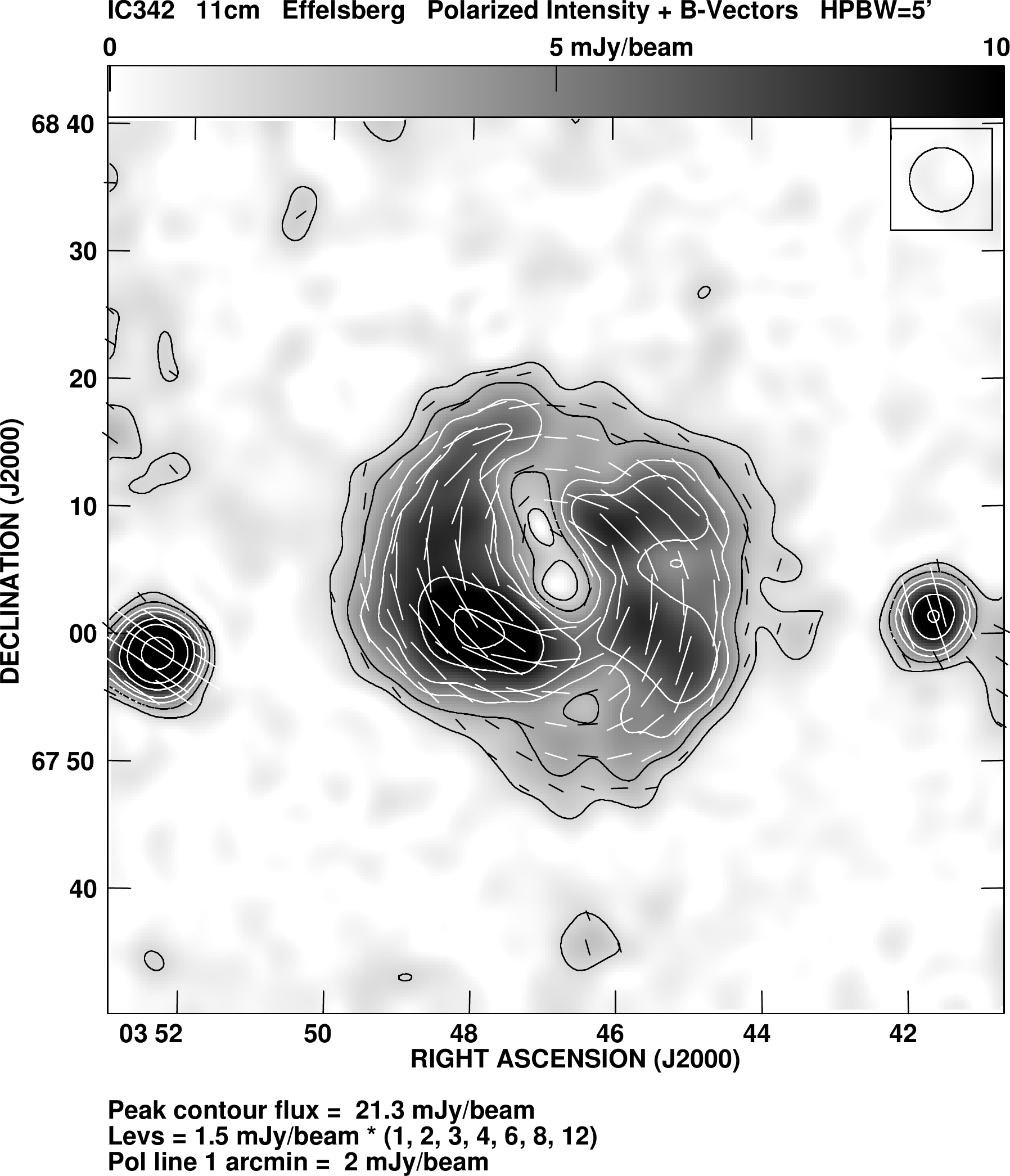}
\caption{ {\it Left:\/} Total intensity (contours and greyscale) and observed
$B$ vectors ($E$+90\degr) of IC~342 at
\wave{11.2} and at 5\arcmin\ resolution with the Effelsberg telescope.
{\it Right:\/} Linearly polarized intensity and observed
$B$ vectors ($E$+90\degr) at \wave{11.2} and at 5\arcmin\ resolution.
}
\label{cm11eff}
\end{figure*}

\begin{figure*}[htbp]
\includegraphics[width=0.45\textwidth]{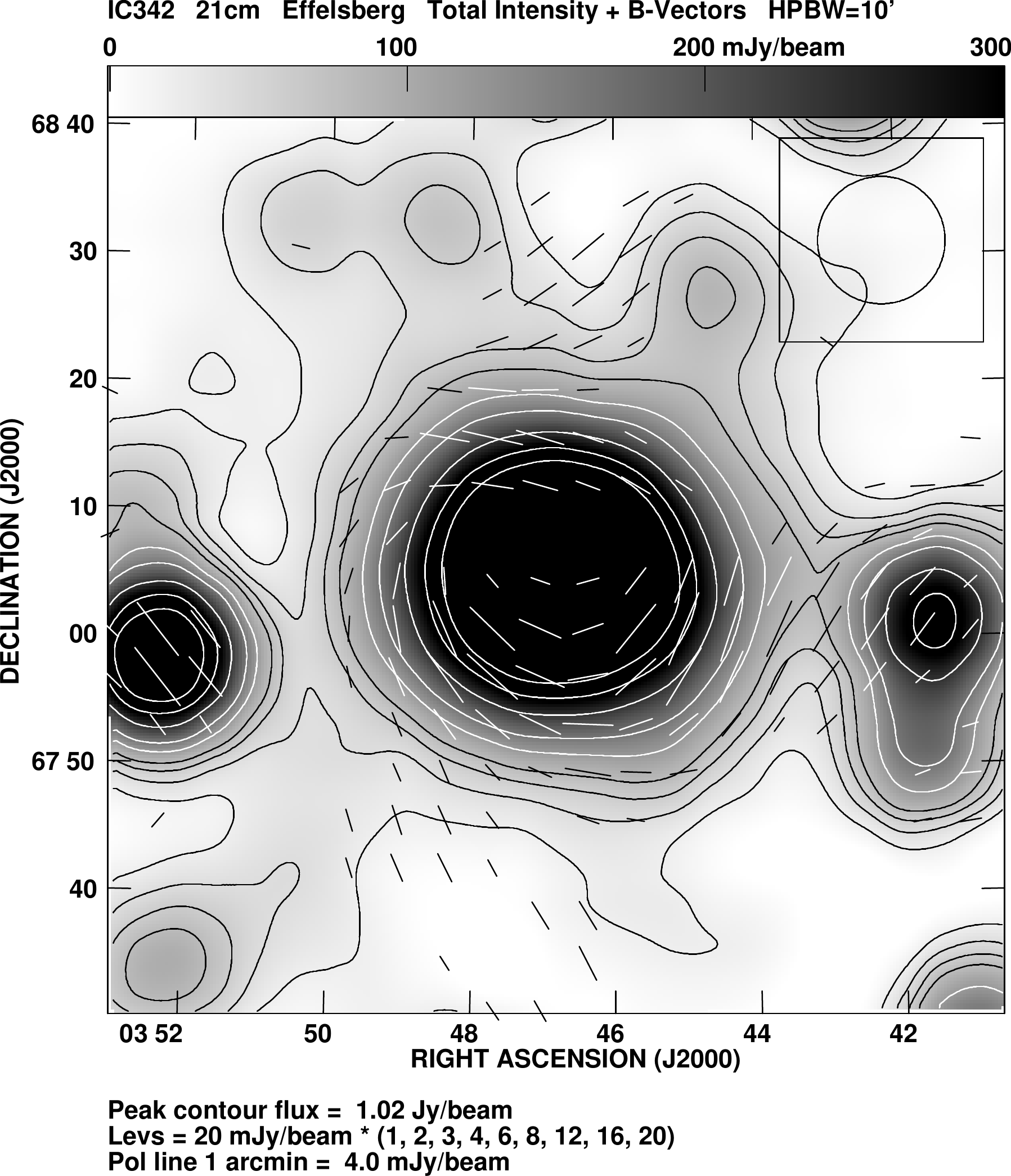}
\hfill
\includegraphics[width=0.45\textwidth]{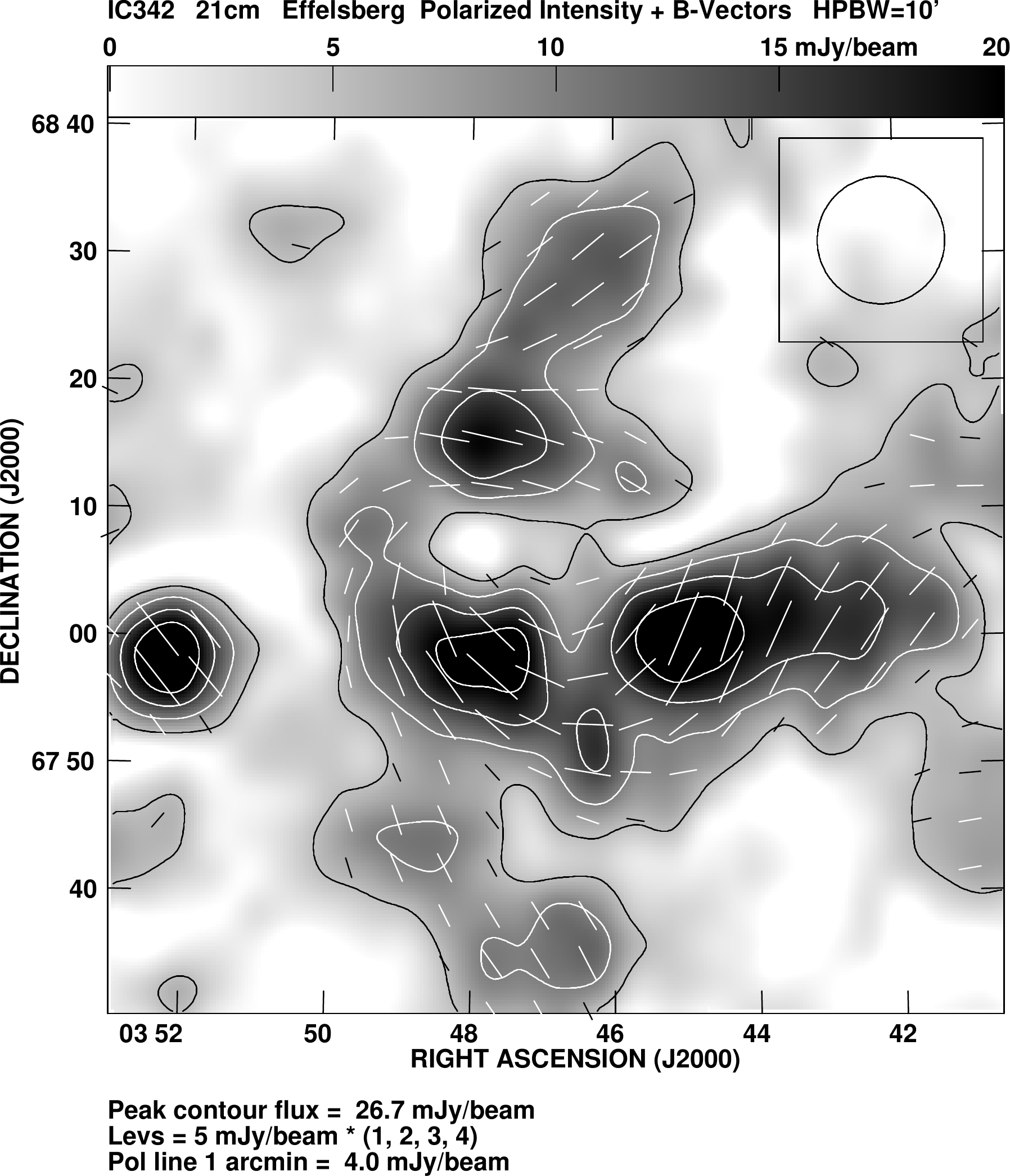}
\caption{ {\it Left:\/} Total intensity (contours and greyscale) and observed
$B$ vectors ($E$+90\degr) of IC~342 at
\wave{21.4} and at 10\arcmin\ resolution with the Effelsberg telescope.
{\it Right:\/} Linearly polarized intensity and observed
$B$ vectors ($E$+90\degr) at \wave{21.4} and at 10\arcmin\ resolution.
}
\label{cm21eff}
\end{figure*}

\begin{figure*}[htbp]
\includegraphics[width=0.45\textwidth]{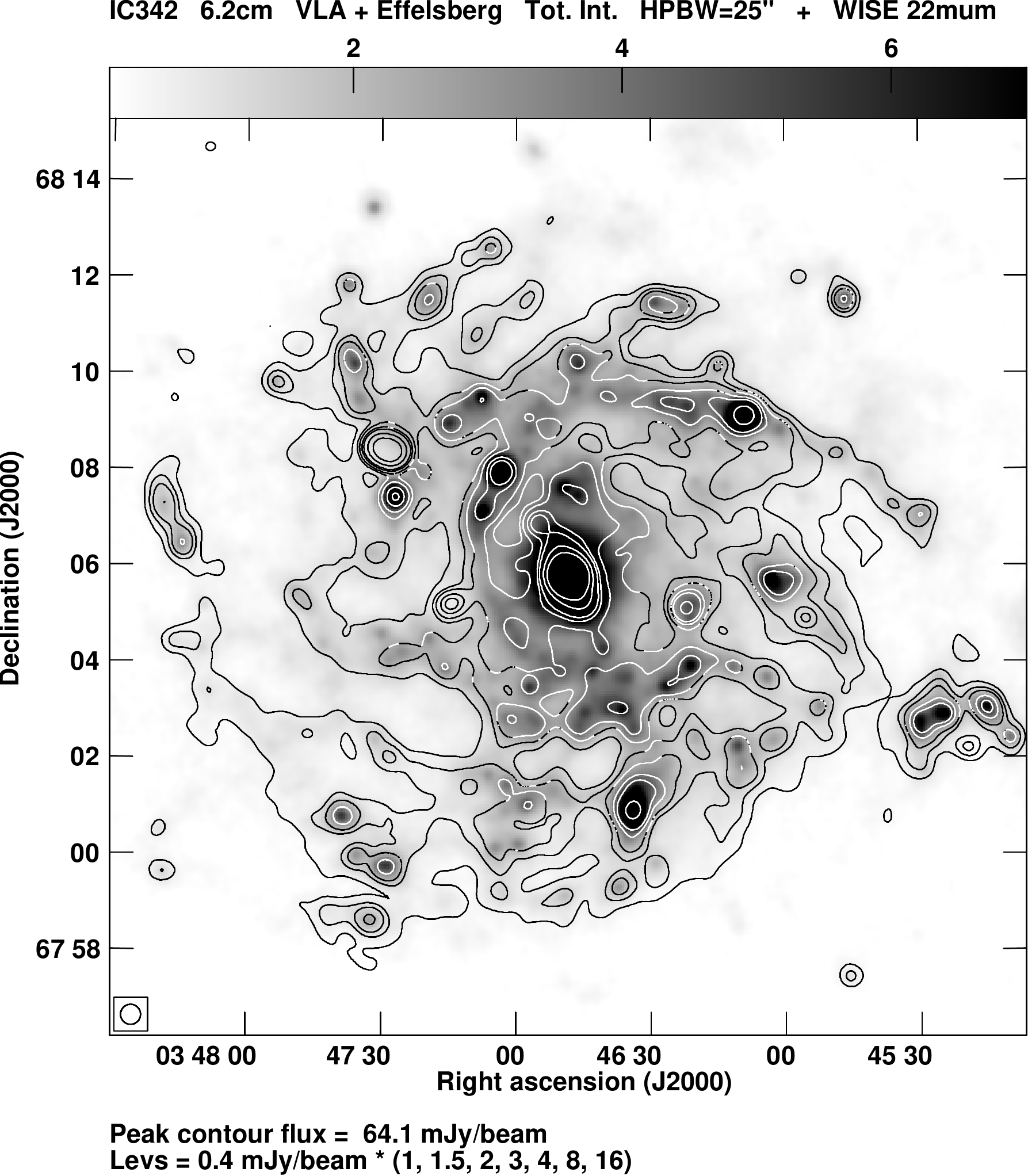}
\hfill
\includegraphics[width=0.45\textwidth]{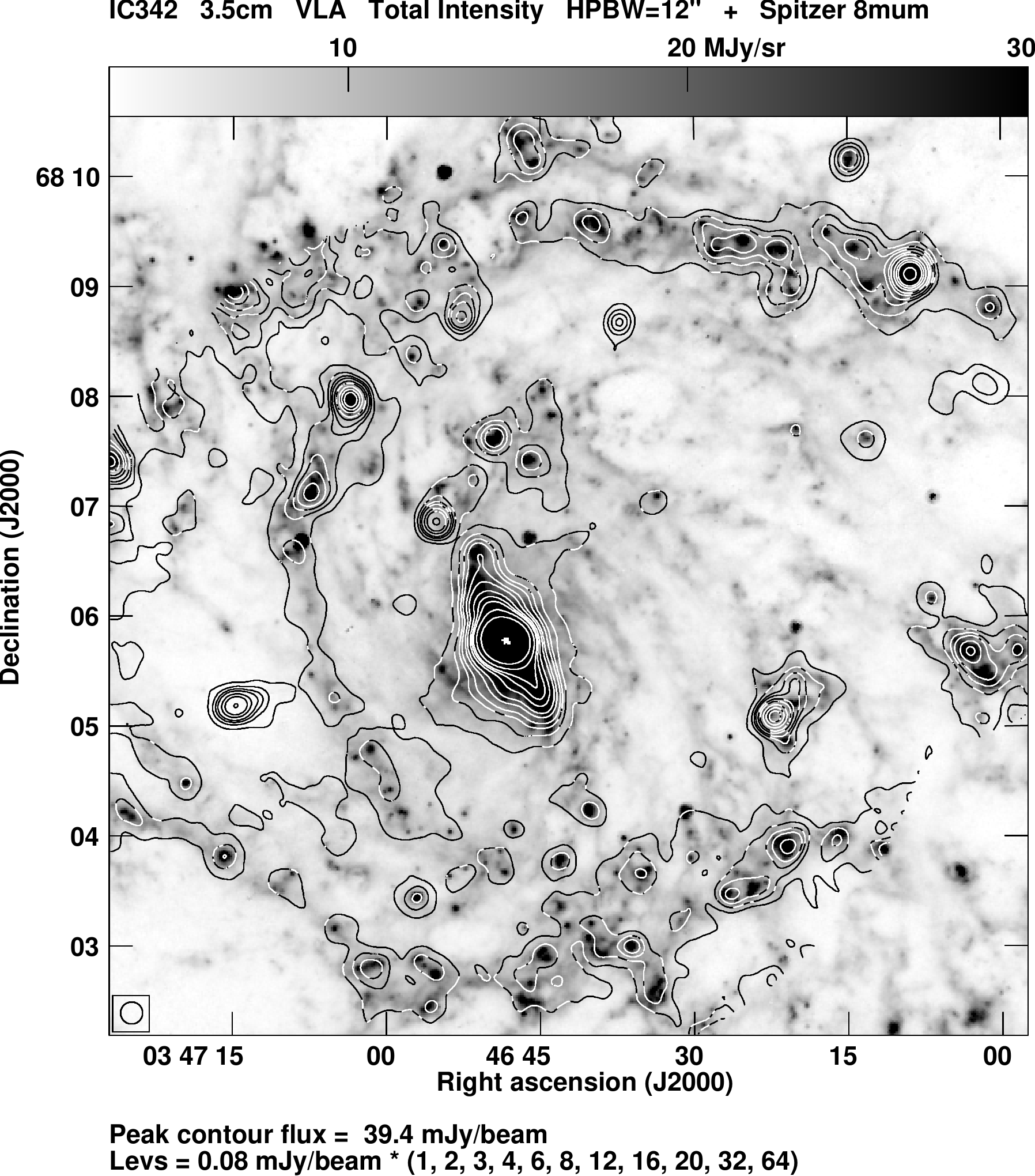}
\caption{ {\it Left:\/} Total emission (contours) at \wave{6.2} (VLA+Effelsberg) at 25\arcsec\ resolution,
overlaid onto a greyscale presentation (in arbitrary units) of the infrared map at $22\,\mu$m from the
WISE telescope \citep{wright10} at 17\arcsec\ resolution, kindly provided by Seppo Laine.
{\it Right:\/} Total emission (contours) at \wave{3.5} (VLA only) at 12\arcsec\ resolution,
overlaid onto a greyscale presentation (in arbitrary units) of the infrared map at $8\,\mu$m from the
IRAC camera of the SPITZER telescope at 2\arcsec\ resolution \citep{fazio04}. The radio map is incomplete
in the north-eastern and south-western corners.}
\label{wise}
\end{figure*}

Linearly polarized emission from IC~342 was first detected with the VLA at \wave{20} by \citet{krause87} and with
the Effelsberg telescope at \wave{6.2} and \wave{11} by \citet{graeve88}, who
found an ordered magnetic field with a smooth spiral pattern.
According to the detection of a large-scale pattern of Faraday rotation measures between \wave{6.2} and \wave{20}
and between \wave{6.2} and \wave{11} \citep{krause87, graeve88, krause89a, sokoloff92};
some part of the ordered field is {\em \emph{regular}} \footnote{The orientation of
polarization ``vectors'' is ambiguous by multiples of 180\degr. As a consequence,
``ordered'' magnetic fields as traced by linearly polarized emission can be either ``regular''
(``large-scale'') fields, preserving their direction over large scales, ``anisotropic turbulent'', or
``anisotropic tangled'' fields with multiple field reversals within the telescope beam. To distinguish
between these fundamentally different types of magnetic field observationally, additional Faraday rotation
data is needed.}
on spatial scales larger than a few kpc. These azimuthal variation of rotation measures signified for the first
time that the large-scale field pattern in a spiral galaxy can be described by an ``axisymmetric spiral''
(ASS), probably generated by a mean-field $\alpha-\Omega$ dynamo \footnote{A mean-field dynamo can
also be excited by the $\alpha^2$ effect, which is weaker in galactic disks than the $\alpha-\Omega$ effect
\citep{elstner92}. In the following, ``mean-field dynamo'' means ``mean-field $\alpha-\Omega$ dynamo''.}.
The variation in rotation measures between \wave{6.2} and \wave{20} has a lower amplitude than between
\wave{6.2} and \wave{11}, indicating significant Faraday depolarization at \wave{20}
\citep{krause89a,sokoloff92}.

Until today, IC~342, NGC~6946 \citep{ehle93} and M~31 \citep{fletcher04}
are candidates of an ASS-type field pattern, whereas many other spiral galaxies
show more complicated field structures \citep{fletcher10,beck+wielebinski13}. The ASS field is the basic mode
predominantly excited by the mean-field galactic dynamo \citep{ruzmaikin88,beck96}, but it takes several
billion years before full coherence is reached \citep{arshakian09}. Fields of opposite polarity need to
merge (see e.g. the simulations by \citet{hanasz09} and \citet{moss12}. A major distortion by the passage of
companion galaxy may delay or interrupt the field evolution \citep{moss14}. IC~342 lacks a companion, so that
the mean-field dynamo can operate without major disturbances. Still, the previous results on IC~342 were based on
low-resolution observations and affected by Faraday and beam depolarization. The detection of higher dynamo
modes and field patterns that are generated by other processes, such as non-axisymmetric gas flows or instabilities,
needs higher resolution.

The \wave{20} observations by \citet{krause89a} revealed at least two long, polarized spiral arms in
the outer south-eastern and eastern regions of IC~342. These features have smaller pitch angles than the
$\HI$ spiral arms delineated by \citet{newton80b} and seem to cross them. This behaviour is
different from the ``magnetic arms'' in NGC~6946 that are located between the gaseous arms and aligned
parallel to the arms \citep{beck+hoernes96,beck07}.
The western and north-western regions showed very little polarized
emission at \wave{20} owing to strong Faraday depolarization. At \wave{6.2}, where depolarization is lower,
\citet{krause93} discovered three narrow filaments in the inner galaxy, one of them located between gaseous
arms, resembling the magnetic arms of NGC~6946 \citep{beck07}.

Studying the phenomenon of magnetic arms in many galaxies may help to develop and
constrain future models of the evolution
of the magnetic ISM in galaxies. The nearest spiral galaxy M~31 is not suited owing to its high inclination.
M~33 has ill-defined material arms and only a radio-weak magnetic arm in the northern region
\citep{taba08}. IC~342 is a nearby, radio-bright, almost face-on galaxy, and it allows observations with high
spatial resolution.

This paper presents observations from the VLA and Effelsberg radio telescopes in four wavelength bands with
higher resolution and higher sensitivity than with the previous data. The highest resolution of 12\arcsec\
corresponds to a spatial resolution of about 200\,pc at the assumed distance of 3.5\,Mpc. Similar spatial resolutions
in radio continuum have been achieved at \wave{20} in M~31 \citep{beck98} and in M~33 \citep{taba07a}, but
magnetic field investigations were hampered by Faraday depolarization that is strong in that wavelength band.

The radio maps in total and polarized intensity are presented in Section~\ref{sect:total},
maps of spectral index, thermal and nonthermal emission in Sect.~\ref{sect:thermal},
polarized emission in Sect.~\ref{sect:pol}, magnetic field strength in Sect.~\ref{sect:mf},
Faraday rotation measures in Sect.~\ref{sect:rm}, and Faraday depolarization in Sect.~\ref{sect:dp}.
Polarized background sources are discussed in Sect.~\ref{sect:sources} and the unusual central region
in Sect.~\ref{sect:central}.
The propagation of cosmic-ray electrons (CREs) is discussed in Sect.~\ref{sect:prop},
energy densities in Sect.~\ref{sect:energy},
the extent of magnetic fields in Sect.~\ref{sect:eq},
dynamo action in Sect.~\ref{sect:dynamo},
the detection of a large-scale helical field in Sect.~\ref{sect:parker},
and the origin of magnetic arms in Sect.~\ref{sect:ma}.

\section{Observations and data reduction}
\label{sect:obs}

The main parameters of the radio continuum observations with the
Very Large Array (VLA), operated by the NRAO \footnote{The NRAO is a
facility of the National Science Foundation operated under
cooperative agreement by Associated Universities, Inc.}, and the
Effelsberg telescope \footnote{The Effelsberg 100-m telescope is
operated by the Max-Planck-Institut f\"ur Radioastronomie in Bonn on
behalf of the Max-Planck-Gesellschaft (MPG).}
are given in Table~\ref{tab:obs}.
Data was processed with the standard routines of the software packages AIPS and NOD2, respectively.

The final Effelsberg map at \wave{2.8} is a combination of 45 coverages scanned in azimuthal direction
with the four-horn secondary-focus system, using software beam-switching \citep{emerson79}.
The average baselevels were subtracted by fitting linear baselines.
All coverages were transformed into the RA, DEC coordinate system.
The final Effelsberg map at \wave{6.2} is a combination of 20 coverages scanned in
azimuthal direction with the two-horn secondary-focus system, using software beam-switching,
baselevel subtraction, and transformation into the RA, DEC coordinate system.
The final Effelsberg maps at \wave{11.2} and \wave{21.4} were obtained from 21 and 7 coverages,
respectively, scanned alternating in RA and DEC with the one-horn secondary-focus systems
and combined using the spatial-frequency weighting method by \citet{emerson88}.
The radio sources 3C138 and 3C286 were used for the calibration of flux density and polarization angle.
During the VLA observations, the secondary calibrator 0410+769 was observed once every 20--30 minutes
and used for calibrating telescope gains and phases and for correcting instrumental
polarization.

The accuracy of polarization angle calibration is better than
1\degr\ for both telescopes. The level of instrumental polarization
of the VLA is less than 1\% at the centre of the primary beam, but
increases to a few percentage points at the half-power radius \citep{condon98}.
Because the observations presented here were obtained
from data observed over a wide range of parallactic angles,
the instrumental polarization is smoothed out. The instrumental
polarization of the Effelsberg telescope emerges from the polarized
sidelobes with 0.3--0.5\% of the peak total intensity at the
frequencies of the observations presented in this paper.

At \wave{3.5} the VLA D array maps in Stokes $I$, $Q,$ and $U$ from three pointings
with centres separated by 5\arcmin\ were combined in the image plane and corrected
for primary beam attenuation. Using natural weighting (Brigg's robust=5) of the $uv$ data, we obtained
maps with an angular resolution of 12\arcsec\ (about 200\,pc) that were further smoothed
to 15\arcsec\ (about 250\,pc) in order to increase the signal-to-noise ratio (Fig.~\ref{cm3}).
Merging with the Effelsberg data at \wave{2.8} was not possible because the region covered
by the VLA observations is much smaller than the Effelsberg map.

At \wave{6.2} four different pointings were observed with the VLA D array, centred on the
south-east, south-west, north-east, and far north-west of the centre and separated by about 8\arcmin.
Two pointings in the north-west and south-east were already observed by \citet{krause93}.
The data from the north-western pointing were included in this work, while the south-eastern field
was re-observed to improve the signal-to-noise ratio. Robust weighting (Brigg's robust=0)
of the $uv$ data was applied only
to the north-western pointing (with the strongest intensity), which gave an angular resolution
of about 14\arcsec; the resulting maps were smoothed to a resolution of 15\arcsec.
Natural weighting of the $uv$ data from the five pointings gave maps with angular resolutions
between 19\arcsec\ and 23\arcsec, which were smoothed to a common resolution of 25\arcsec.
The VLA maps in Stokes $I$, $Q,$ and $U$ from all pointings were corrected for primary beam
attenuation and combined in the image plane.

The Effelsberg maps at \wave{6.2} and \wave{11.2} (Fig.~\ref{cm6eff} and \ref{cm11eff})
reached rms noise levels that are several times lower than for the previous maps by \citet{graeve88}.
At \wave{6.2} the maps in Stokes $I$, $Q,$ and $U$ from both telescopes were combined (Fig.~\ref{cm6})
with an overlap in the $uv$ data for antenna baselines of 300--400$\lambda$.
At this wavelength the minimum baseline of the VLA is about 300$\lambda$, while the maximum
baseline of the Effelsberg telescope is about 1600$\lambda$. The VLA map alone contains about 38\%
of the total flux density and 75\% of the polarized flux density within 7\arcmin\ radius.

At \wave{20.1} the previous VLA D array data from \citet{krause89a} centred on the nucleus
(at RA, DEC (J2000) = $03^\mathrm{h}\ 46^\mathrm{m}\ 48\fs 1$, +68\degr\ 05\arcmin\ 47\arcsec)
were combined with the C array data in the $uv$ plane.
Maps were generated at 15\arcsec\ resolution (uniform weighting) and
corrected for primary beam attenuation (Fig.~\ref{cm20a}). Effelsberg data were
observed at a similar wavelength (Fig.~\ref{cm21eff}). Integration of the Effelsberg map
gave a lower total flux density than for the VLA map, indicating that some extended
emission is missing in the Effelsberg map owing to the baselevel subtraction. There is no indication
of any missing large-scale emission in the VLA maps, so that no combination was performed.

To search for emission in the outer disk east of IC~342, another pointing aimed at
RA, DEC (J2000) = $03^\mathrm{h}\ 48^\mathrm{m}\ 16\fs 9$, +68\degr\ 05\arcmin\ 48\arcsec\
was observed at \wave{20.1} with the VLA D array and combined with the central pointing,
to obtain maps with 51\arcsec\ resolution (Fig.~\ref{cm20b}).
The maps in Stokes $Q$ and $U$ at each wavelength and at each resolution were combined with
maps of linearly polarized intensity $PI$ including the correction for positive bias
due to noise. In the $PI$ maps the noise distribution is
non-Gaussian, and computing a Gaussian standard deviation would underestimate the
noise. As a result, only the rms noise values for the maps in Stokes $I$, $Q,$
and $U$ are given in Table~\ref{tab:obs}.

\begin{figure}[htbp]
\centerline{\includegraphics[width=0.45\textwidth]{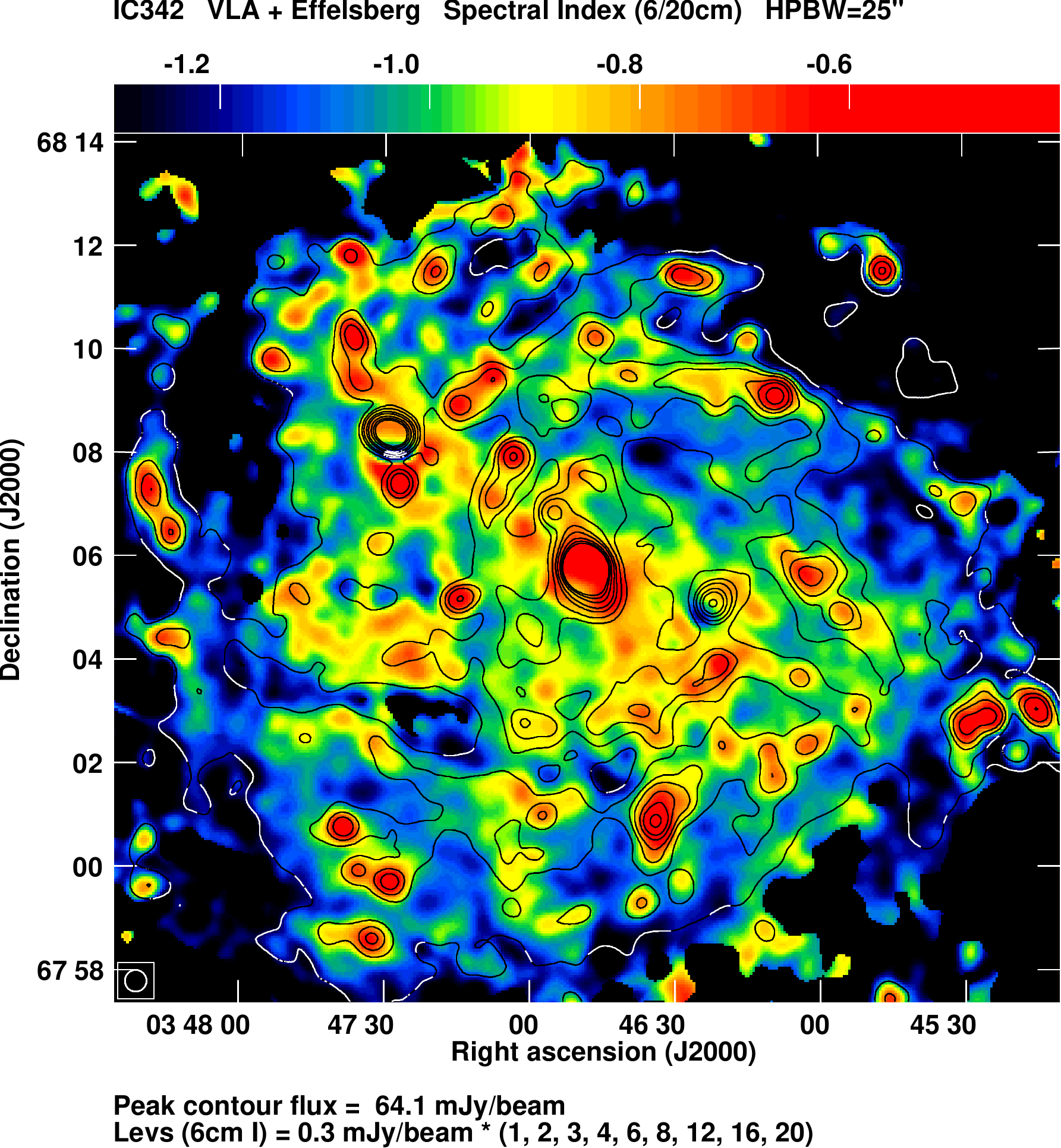}}
\caption{Spectral index distribution (greyscale) between \wave{6.2} and
\wave{20.1} at 25\arcsec\ resolution, determined for points where the
intensities are larger than 10\,times the rms noise at both wavelengths, to restrict
the maximum error due to noise to $\pm0.08$. Contours show the total intensity at \wave{6.2}.}
\label{spec}
\end{figure}

\begin{figure*}[htbp]
\includegraphics[width=0.45\textwidth]{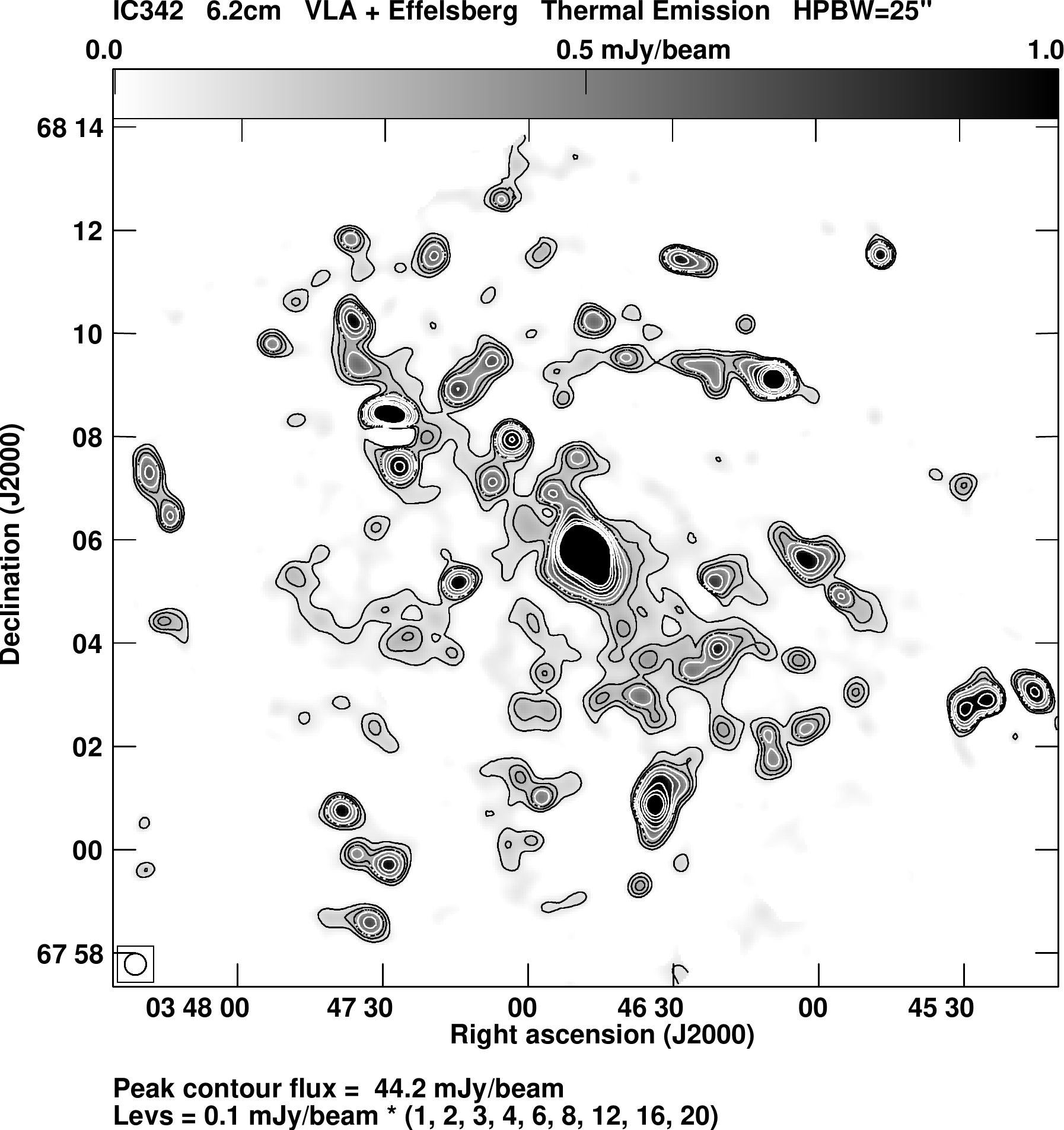}
\hfill
\includegraphics[width=0.45\textwidth]{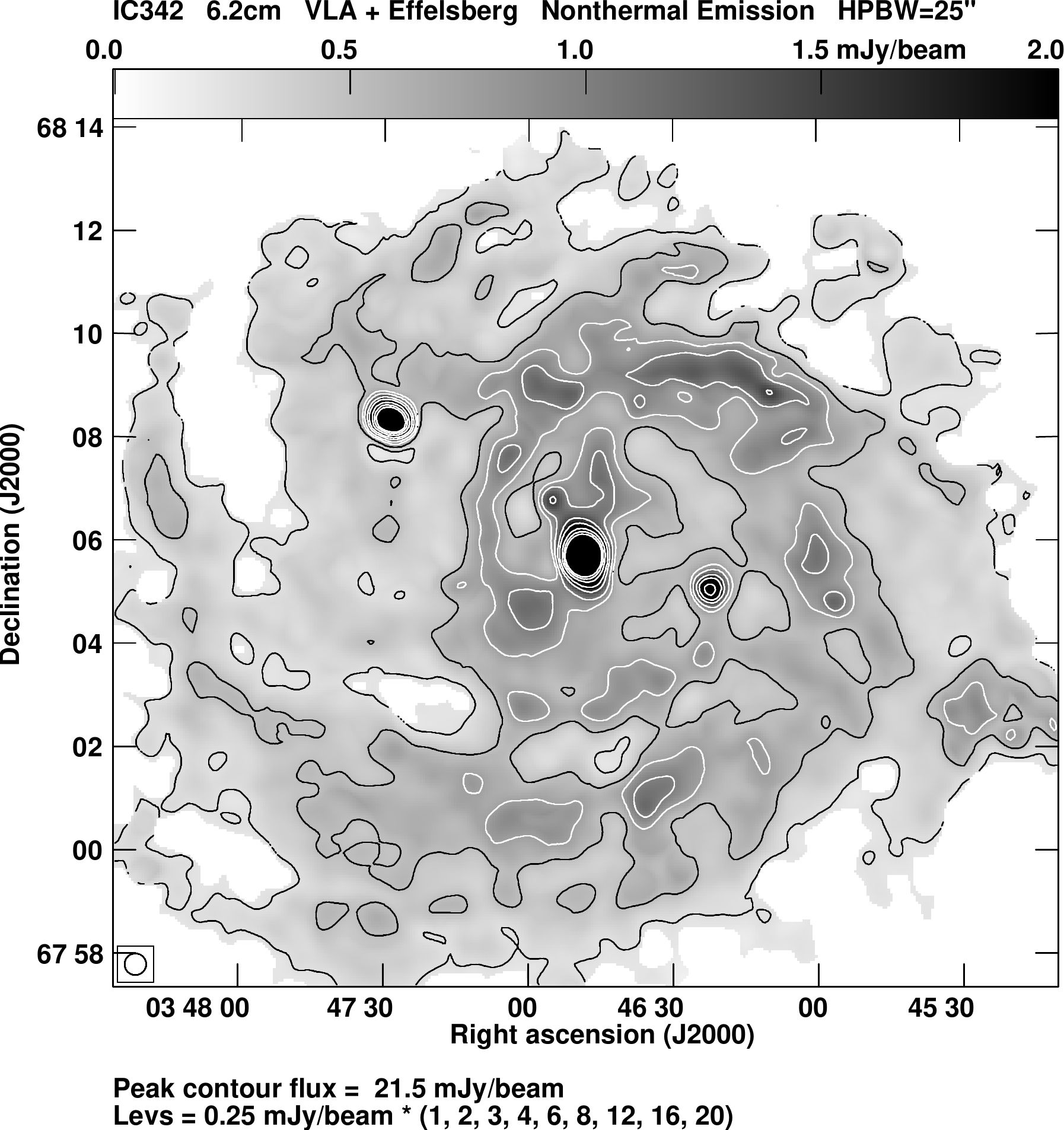}
\caption{Thermal intensity (left) and nonthermal intensity (right)
of IC~342 (contours and greyscale) at \wave{6.2} at 25\arcsec\ resolution.
}
\label{th}
\end{figure*}

\section{Results}
\label{sect:results}

\subsection{Total intensity}
\label{sect:total}

Figures~\ref{cm3}--\ref{cm21eff} show the maps of total (left panels) and polarized (right panels) radio
intensities. Figure~\ref{cm3} is based on VLA data alone, Fig.~\ref{cm6} on combined VLA+Effelsberg data, and
Figs.~\ref{cm28eff}--\ref{cm21eff} on Effelsberg data alone.

The total intensity in the maps at \wwwave{3.5}{6.2}{20.1} is strongest in the nuclear bar and in the inner
eastern and northern spiral arms, the regions that are also brightest in the optical and
infrared spectral ranges. The northern -- north-western arm, centred on
RA, DEC (J2000) = $03^\mathrm{h}\ 46^\mathrm{m}\ 27^\mathrm{s}$, +68\degr\ 09\arcmin\ 20\arcsec,
is the most pronounced and the narrowest one, with a width of only $\simeq 280$\,pc measured at \wave{3.5}
(Fig.~\ref{cm3} left and Fig.~\ref{wise} right), corrected for beam smearing.
The narrow inner arm just east of the central region, centred on
RA, DEC (J2000) = $03^\mathrm{h}\ 47^\mathrm{m}\ 08^\mathrm{s}$, +68\degr\ 06\arcmin\ 40\arcsec,
has a similar width of $\simeq 350$\,pc measured at \wave{3.5}.
The main southern -- south-eastern spiral arm, centred on
RA, DEC (J2000) = $03^\mathrm{h}\ 47^\mathrm{m}\ 45^\mathrm{s}$, +68\degr\ 01\arcmin\ 30\arcsec\
(Figs.~\ref{cm6} left and \ref{cm20a} left, named no.~1 in Fig.~\ref{cm6} right)
is much broader, with a width of $\simeq3$\,kpc measured at \wave{20.1}, but may consist of several arms.
Another broad but faint spiral arm is visible in the far south -- south-east (Figs.~\ref{cm6} left and
\ref{cm20a} left; no.~2 in Fig.~\ref{cm6} right).
Only the peaks of these broad southern arms are visible at \wave{3.5}
(Fig.~\ref{cm3} left), delineating the major star-forming regions (Fig.~\ref{wise} right).
The two outer arms are also visible (though unresolved) in the Effelsberg map at \wave{2.8} (Fig.~\ref{cm28eff}).

The total radio intensity is locally correlated with the infrared emission
(Fig.~\ref{wise}), an observational result found in many spiral galaxies \citep[e.g.][]{taba13b}.
This will be discussed in detail in a forthcoming paper.

\subsection{Thermal and nonthermal emission}
\label{sect:thermal}

The maps of total intensities $I_{\nu}$ at \wave{6.2} and \wave{20.1} at
25\arcsec\ resolution are used to determine the radio spectral
index $\alpha$, defined as $I_{\nu} \propto \nu^{\alpha}$ (Fig.~\ref{spec}).
The spectrum is generally flatter (i.e. a lower absolute value of $\alpha$)
in the spiral arms than outside the arms, indicating a greater
fraction of thermal emission in the arms and/or a flatter spectrum of the
nonthermal (synchrotron) emission.

The thermal and nonthermal components are separated in the ``classical'' way by assuming a constant
spectral index of $-0.1$ for the thermal and $-1.0$ for the diffuse nonthermal emission,
taken from the average spectral index of the interarm regions, resulting in the maps shown in Fig.~\ref{th}.
The typical thermal fraction at \wave{6.2} is about 50\% in the central region, 20\% to 30\% in the
spiral arms and 10\% or less in the interarm regions.

The separation method based on the spectral index means that all spectral variations
are interpreted as variations in the thermal fraction, while it is known that the
synchrotron spectrum depends on the age of cosmic-ray electrons (CREs) propagating from the spiral arms, so is expected to be flatter in the spiral arms than in the interarm regions.
As a result, the thermal fraction is overestimated in the spiral arms.

The H$\alpha$ map by \citet{hernandez05} is similar to the map of thermal intensity (Fig.~\ref{th} left),
but it shows more extended diffuse emission than the radio map. Spectral steepening due to energy losses
of cosmic-ray electrons in the interarm regions leads to a synchrotron spectrum that can be steeper than
the assumed slope of $-1.0$, so that the thermal fraction is underestimated in such regions.

For the purpose of the present study that determines magnetic field strengths (Sect.~\ref{sect:mf}),
the classical approach to separate thermal and nonthermal components is sufficient.
The advanced method introduced by \citet{taba07b}, which was first applied to the galaxy M~33, uses an extinction-corrected
H$\alpha$ map as a template for thermal emission and will be applied to IC~342 in a future paper.
Low-frequency observation obtained with the LOFAR Highband Array (115--175\,MHz) will reveal an almost purely
synchrotron map of IC~342 and allow an improved investigation of the variation in synchrotron spectral index,
as recently performed for M~51 \citep{mulcahy14}.

\subsection{Radial distributions of radio intensity}
\label{sect:radial}

\begin{figure*}[htbp]
\vspace{0.7cm}
\includegraphics[width=0.49\textwidth]{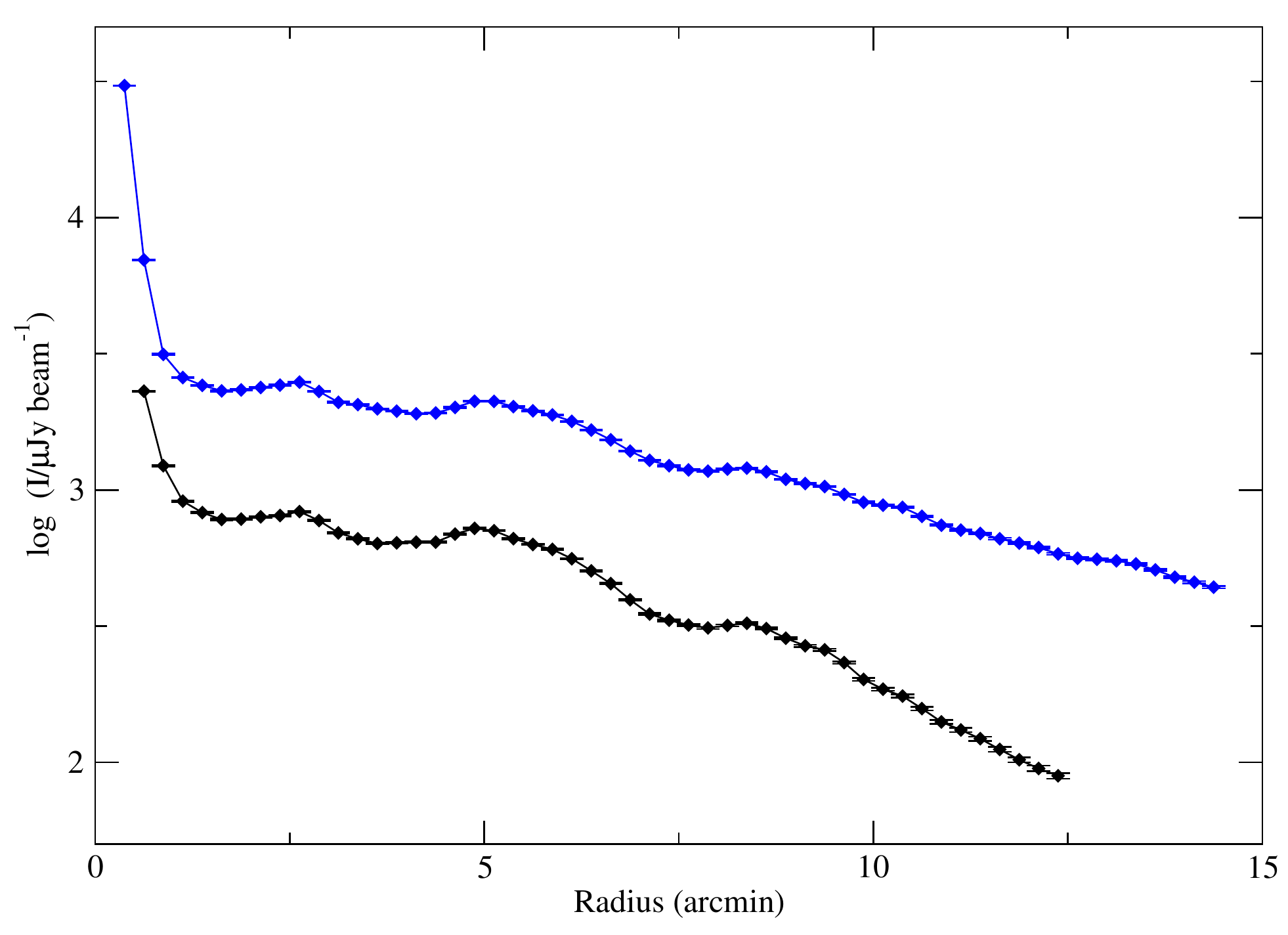}
\hfill
\includegraphics[width=0.49\textwidth]{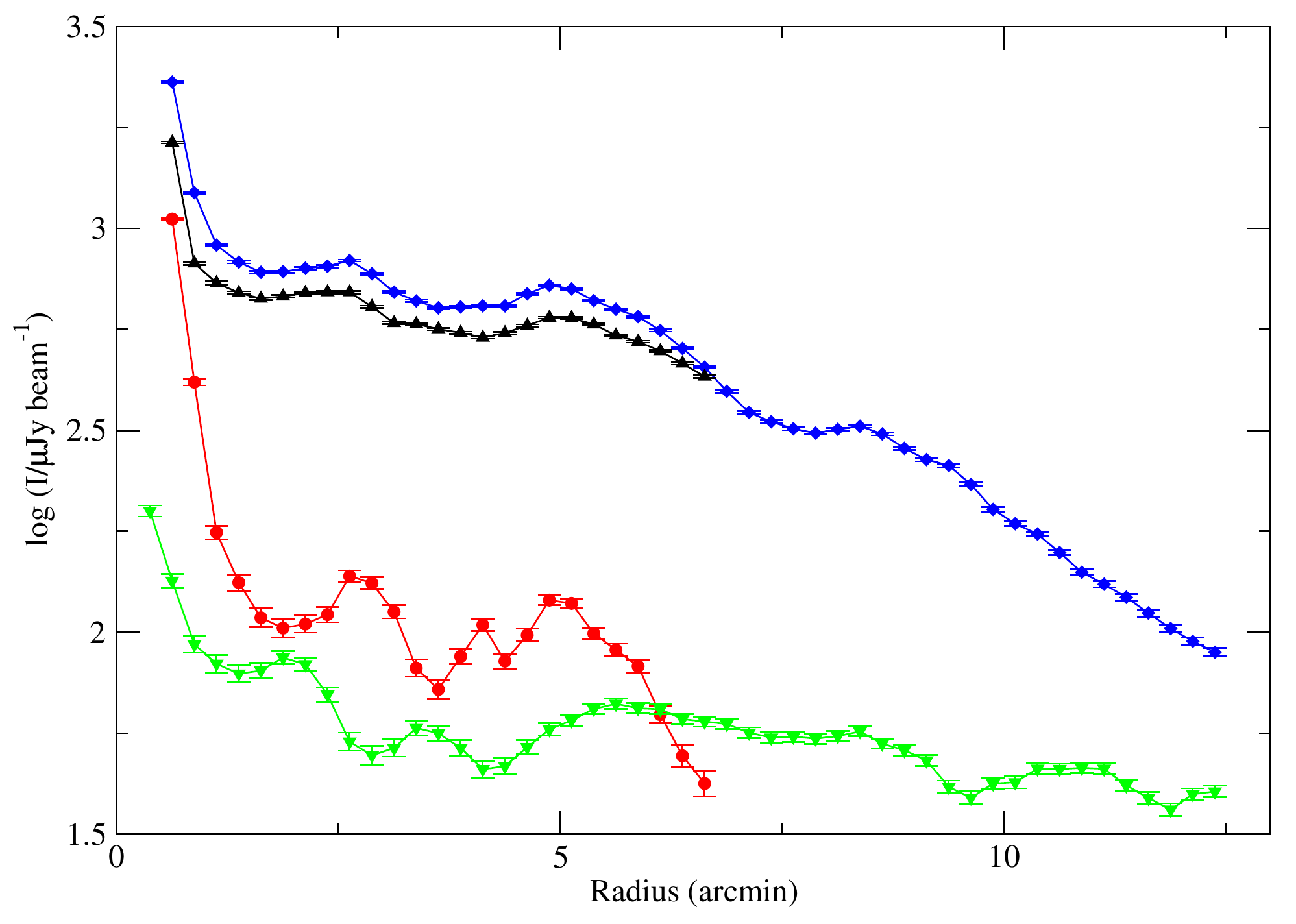}
\caption{{\it Left:\/} Radial distributions of total intensity at a resolution of 25\arcsec\
at \wave{20.1} (upper curve) and at \wave{6.2} (lower curve), determined from the averages in rings
of 0.5\arcmin\ radial width in the plane of the galaxy ($i=31\degr$, $PA=37\degr$).
{\it Right:\/} The radial distributions at \wave{6.2}, from top to bottom: total intensity (blue diamonds),
nonthermal intensity (black upwards triangles), thermal intensity (red circles), and polarized intensity
(green downwards triangles) at a resolution of 25\arcsec, determined from the averages in rings of 0.5\arcmin\
radial width in the plane of the galaxy. The thermal and nonthermal intensities are reliable up to about
7\arcmin\ radius, where the spectral index could be determined with sufficient accuracy (see Fig.~\ref{spec}).
The error bars are determined by the rms noise values in each ring. At the assumed distance, 1\arcmin\ corresponds to about 1\,kpc.
 }
\label{radial}
\end{figure*}

The radial distributions of total radio intensity at \wave{6.2} and \wave{20.1} (Fig.~\ref{radial} left)
reveal breaks beyond about a 5\arcmin\ radius, similar to the break radius of the distribution of thermal intensity
(Fig.~\ref{radial} right) at about 6\arcmin\ radius and that of total neutral gas at about 7\arcmin\ radius
\citep{crosthwaite01}, where the star-formation rate drops to a lower level.
Inside of the break radius, in the inner disk, the distributions of total radio intensities are
flat, with two maxima related to the main spiral arms, and can be described
as smoothed versions of the distribution of thermal intensity (Fig.~\ref{radial} right).
Breaks in the radial distributions of total radio intensity around the break radius of the
star-formation rate were also found in M~33 \citep{taba07a} and in M~51 \citep{mulcahy14}.

\begin{table*}           
\caption{Exponential radial scalelengths $l$ (in arcminutes) of various components of radio intensities
in IC~342. ``Inner'' refers to the radial range 1\arcmin -- 6\arcmin, ``outer'' to 6\arcmin -- 12\farcm5.
}
\centering
\begin{tabular}{lcccccc}
\hline
$\lambda$ (cm) & Total (inner) & Total (outer) & Synchrotron (inner) & Synchrotron (outer) & Polarized synchrotron (outer)\\
\hline
6.2  & $15.5\pm2.6$  & $3.6\pm0.1$   & $16\pm3$           & $3.6\pm0.1$         & $13\pm1$     \\
20.1 & $17.6\pm2.8$  & $6.1\pm0.2$   & $16\pm3$           & $6.1\pm0.2$         & $13\pm1$     \\
\hline
\label{tab:length}
\end{tabular}
\end{table*}

The smooth radial distributions of total and nonthermal intensities can be described by exponential functions
(Fig.~\ref{radial}), while the thermal and polarized intensities fluctuate strongly in the inner disk
(1\arcmin -- 6\arcmin\ radius). The exponential scalelengths $l$ are given in Table~\ref{tab:length}.
As a result of the method that separates thermal from nonthermal emission (Sect.~\ref{sect:thermal}), the nonthermal
intensities at \wave{6.2} and \wave{20.1} have identical scalelengths and only differ in amplitude.
In the outer disk (6\arcmin -- 12\farcm5 radius), the nonthermal and thermal intensities cannot be measured
because the accuracy of the spectral index is not sufficient for a reliable separation.
The scalelength $l_\mathrm{syn}$ of the nonthermal intensity
in the outer disk can be assumed to be the same as for the total intensity because the thermal contribution
is small in this region. The ratio of scalelengths between \wave{20.1} and \wave{6.2} in the outer disk
is $1.7\pm0.1$, which is probably the result of frequency-dependent propagation of CREs
(Sect.~\ref{sect:prop}).

Many spiral galaxies do not exhibit breaks in the radial distribution of total radio intensity, so that
for a comparison, the average exponential scalelengths of IC~342 in the ring 1\arcmin -- 12\farcm5
of $4.9\pm0.2$\,kpc at \wave{6.2} and $7.5\pm0.3$ at \wave{20.1} are used.
These are larger than those in the disks of other spiral galaxies: e.g. $3.0\pm0.2$\,kpc at \wave{6.3}
for M~81 \citep{beck85}, $3.9\pm0.7$\,kpc at \wave{20.5} (nonthermal) for NGC~6946 \citep{walsh02}
and 1--5\,kpc at various wavelengths in several other spiral galaxies \citep{klein81,harnett87,basu13}.
A large optical size $r_{25}$ cannot account for the large scalelengths in IC~342 because $r_{25}$ is
similar for IC~342 (11\,kpc), M~81 (12\,kpc), and NGC~6946 (9\,kpc).
The radio continuum disk of IC~342 appears to extend farther beyond the optical disk than that of
many other spiral galaxies.\footnote{A distance of 3.5\,Mpc is assumed here, while a distance of 1.8\,kpc
as derived by \citet{mccall89} would give smaller scalelengths. However, recent papers converge to the
distance of about 3.5\,Mpc.}

\subsection{Linearly polarized intensity}
\label{sect:pol}

\begin{figure*}[htbp]
\includegraphics[width=0.45\textwidth]{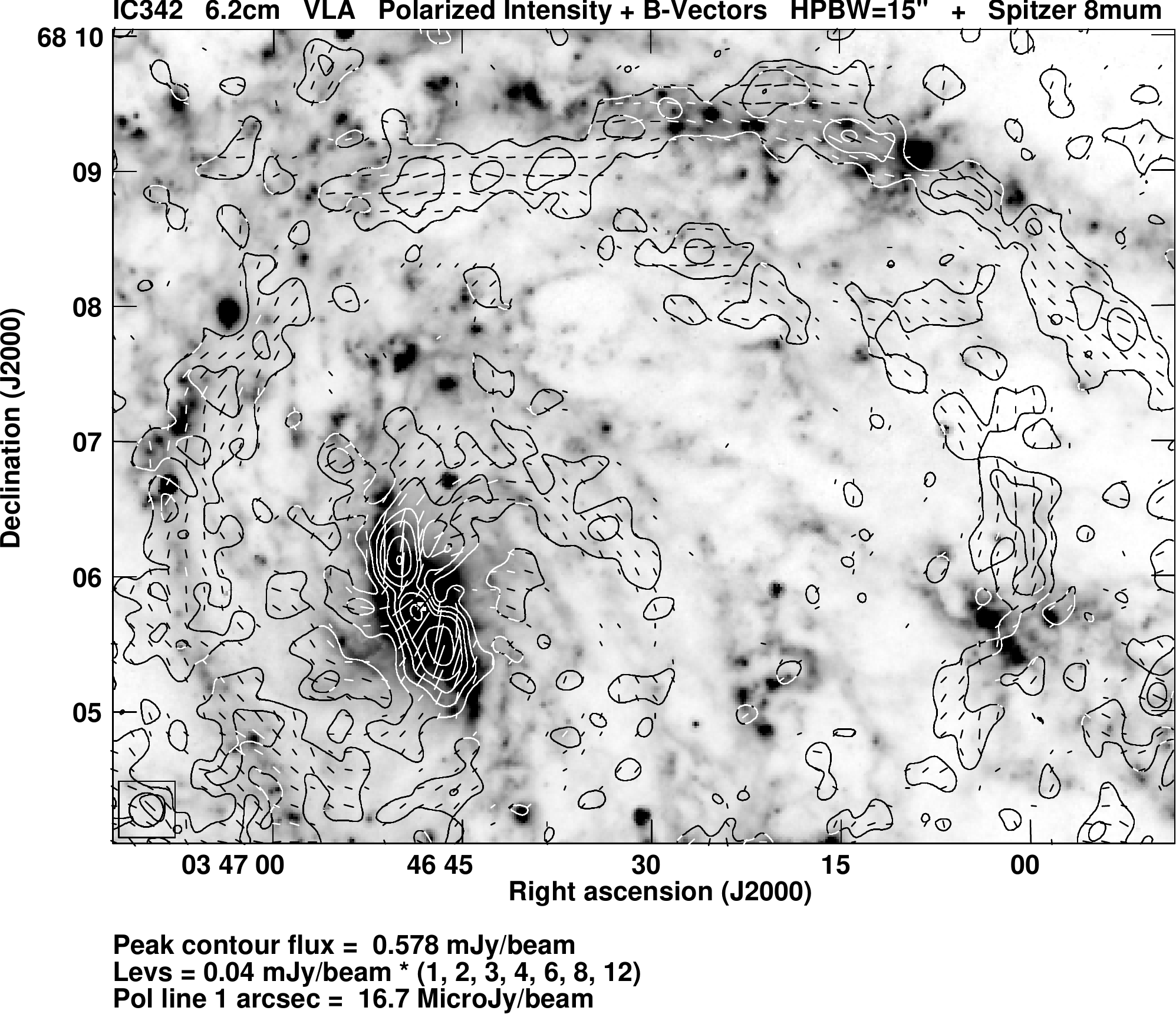}
\hfill
\includegraphics[width=0.45\textwidth]{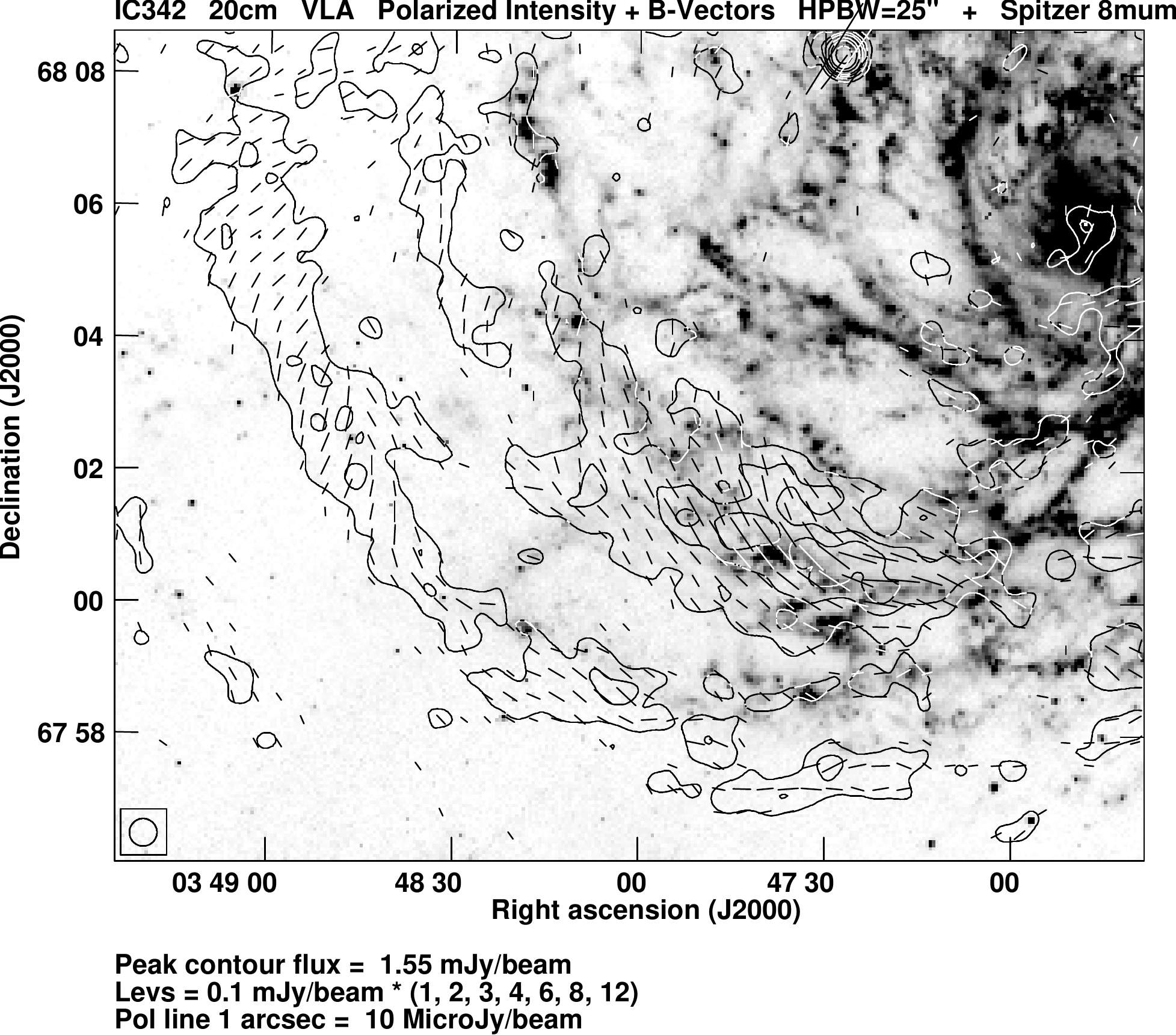}
\caption{ {\it Left:\/} Polarized emission (contours) and observed $B$ vectors ($E$+90\degr) of the central
and north-western regions of IC~342 at \wave{6.2} (VLA only) at 15\arcsec\ resolution, overlaid onto a greyscale
presentation of the infrared map at $8\,\mu$m from the IRAC camera of the SPITZER telescope at 2\arcsec\
resolution \citep{fazio04}.
{\it Right:\/} Polarized emission (contours) and observed $B$ vectors ($E$+90\degr) of the central and
north-western regions of IC~342 at \wave{20.1} (VLA only) at 25\arcsec\ resolution, overlaid onto a greyscale
presentation of the same infrared map at $8\,\mu$m. Faraday rotation and depolarization are strong at
this wavelength. The northern and western parts of the galaxy are not shown here because the emission is
mostly depolarized.}
\label{spitzer}
\end{figure*}

\begin{figure*}[htbp]
\includegraphics[width=0.45\textwidth]{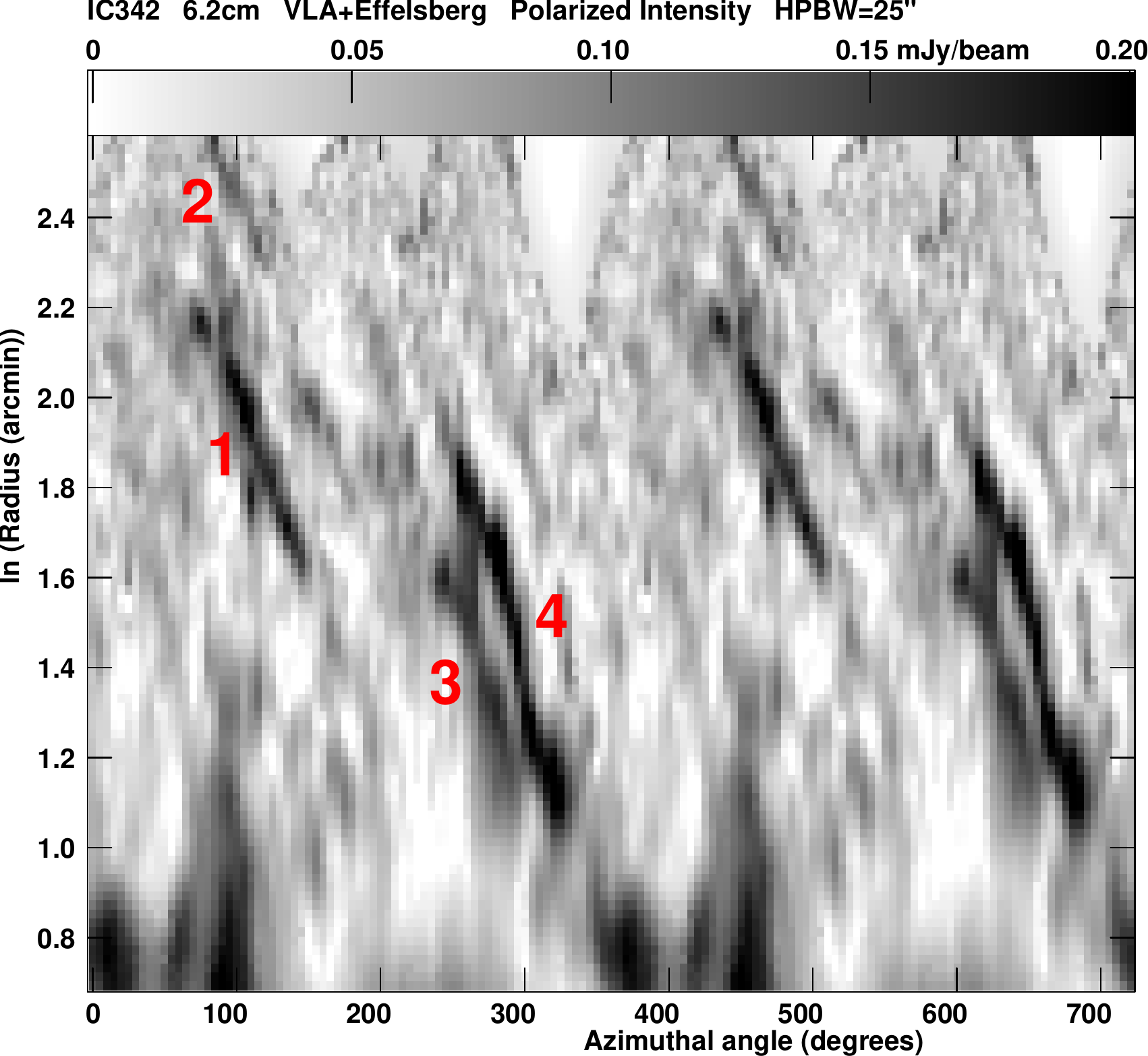}
\hfill
\includegraphics[width=0.48\textwidth]{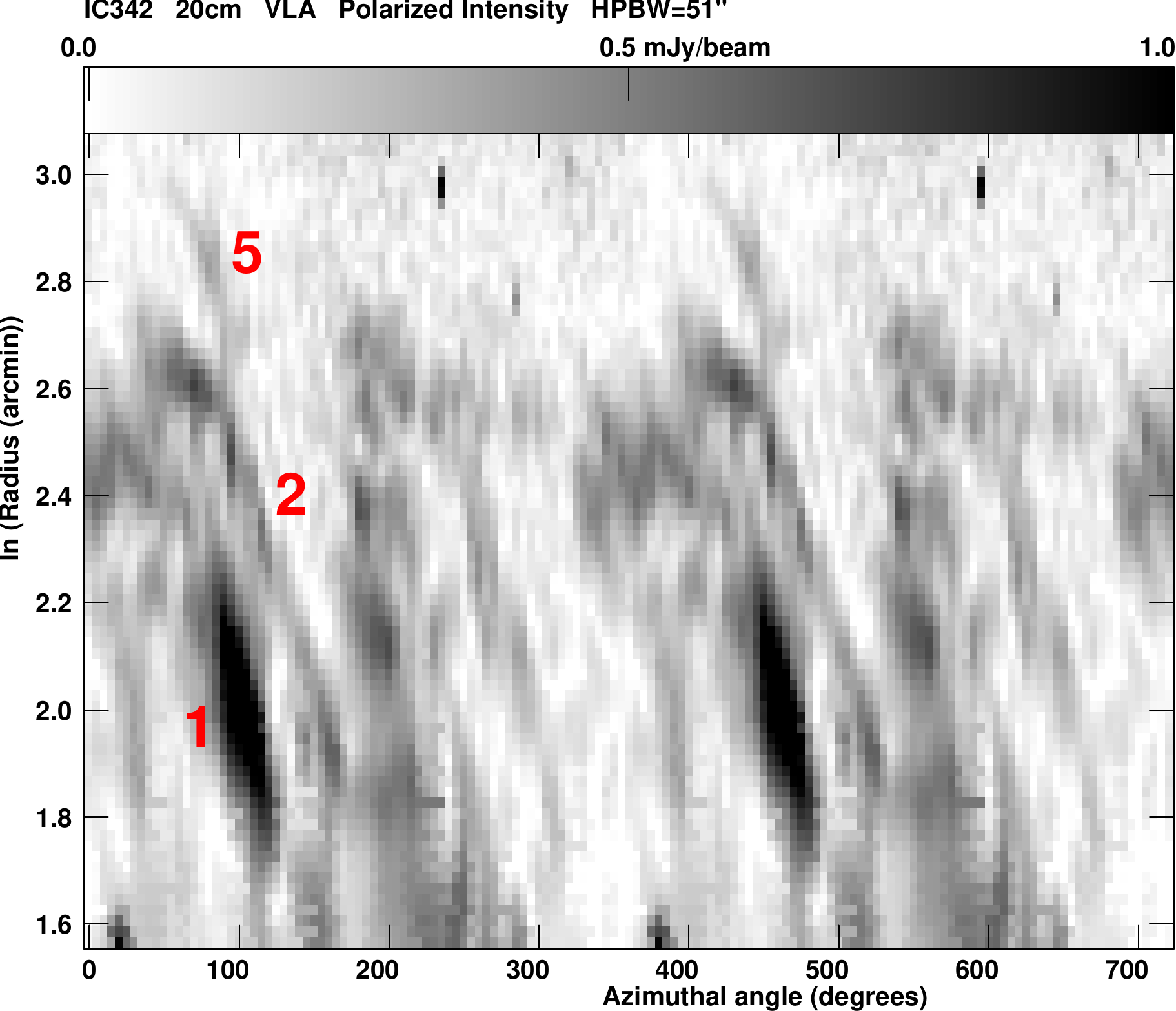}
\caption{{\it Left:\/} Polarized emission at \wave{6.2} (VLA+Effelsberg) at 25\arcsec\ resolution
in polar coordinates (azimuthal angle in degrees, measured counterclockwise from the north-eastern major axis
in the galaxy plane, and $ln$ of radius in arcminutes). The range of azimuthal angles is plotted twice for
better visibility of the spiral arms. {\it Right:\/} Same for the polarized emission at \wave{20.1}
(VLA) at 51\arcsec\ resolution. Numbers refer to the spiral arms listed in Table~\ref{tab:pitch}.
}
\label{spiral}
\end{figure*}

\begin{figure*}[htbp]
\includegraphics[width=0.42\textwidth]{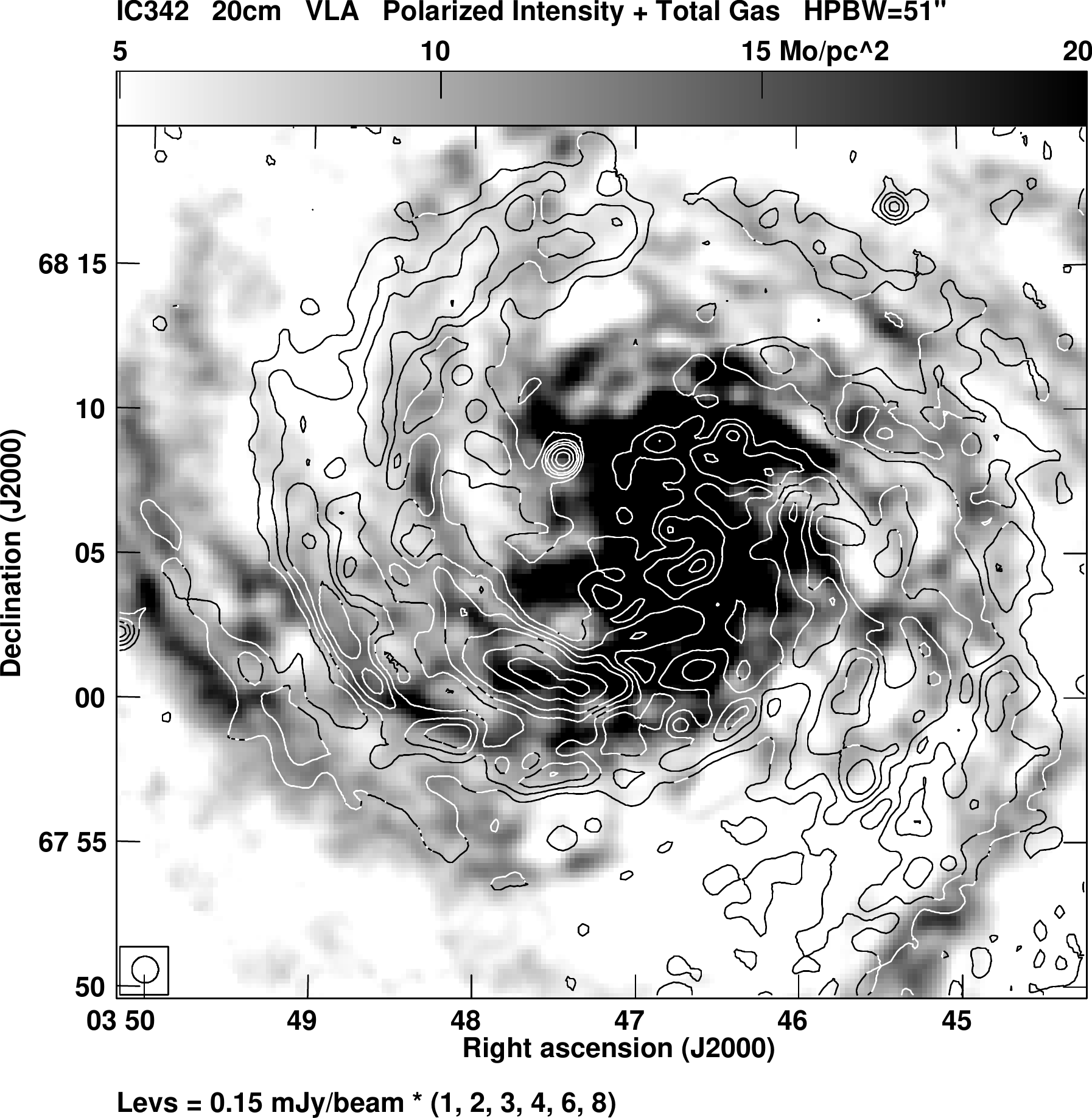}
\hfill
\includegraphics[width=0.52\textwidth]{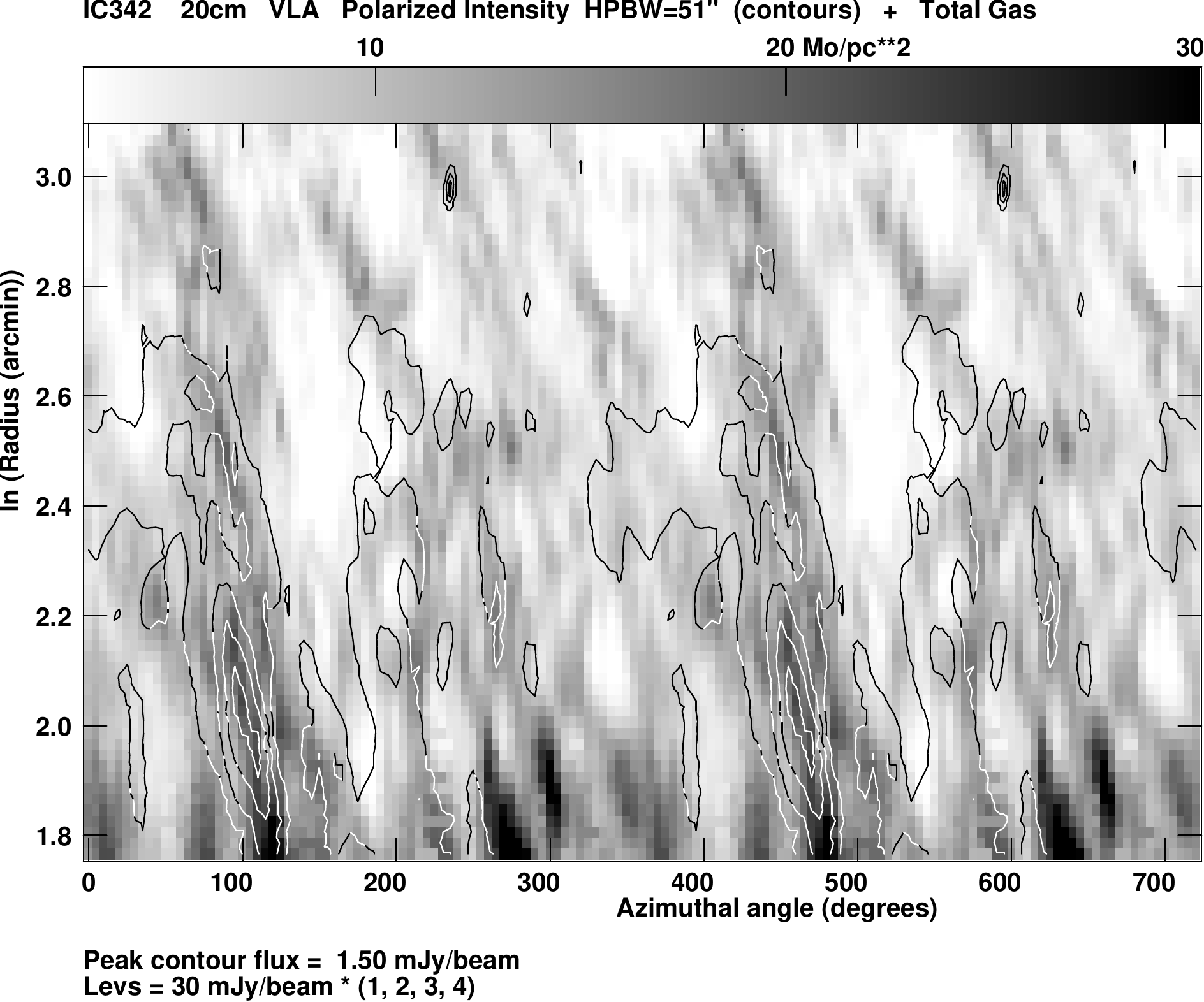}
\caption{{\it Left}: Polarized emission (contours) at \wave{20.1} at 51\arcsec\ resolution, overlaid on a
greyscale map of the surface density of the total neutral gas at 55\arcsec\ resolution \citep{crosthwaite01}
in sky coordinates. {\it Right}: The same quantity plotted in polar coordinates (azimuthal angle measured
counterclockwise from the north-eastern major axis in the galaxy plane and $ln$ of the radius in arcminutes);
the range of azimuthal angles is plotted twice in the right panel for better visibility of the spiral arms.}
\label{gas}
\end{figure*}


The polarized emission in the three Effelsberg maps at the three shortest wavelengths
(Figs.~\ref{cm28eff}--\ref{cm11eff}, right panels) emerges from a spiral configuration of the large-scale
ordered field. Depolarization within the large Effelsberg beams causes the central holes. In the most
sensitive maps at \wave{6.2} and \wave{11.2}, a diffuse disk of polarized emission extends to a radius
of about 20\arcmin\ at the level of five\,times the rms noise, so much further than in the
previous maps by \citet{graeve88}. This extent is greater than seen in total emission (Figs.~\ref{cm6eff}
and \ref{cm11eff} left) because the rms noise values $\sigma$, composed of instrumental noise and the confusion
noise from unresolved background sources, are lower in polarized emission than in total emission
(Table~\ref{tab:obs}). The $3\,\sigma$ detection limit of $200\,\mu$Jy per 180\arcsec\
Effelsberg beam at \wave{6.2} corresponds to $4\,\mu$Jy per 25\arcsec\ beam, so much lower than for the
VLA image of $36\,\mu$Jy. Deep single-dish imaging in polarization allows us to detect weak ordered fields
(see Sect.~\ref{sect:mf}).

Since the polarized foreground of the Milky Way is smooth on the scales of the telescope beams, it
was mostly removed by subtracting linear baselevels from the maps in Stokes $Q$ and $U$.
Only the Effelsberg polarization map at \wave{21.4} (Fig.~\ref{cm21eff} right) is affected by foreground
emission. At this wavelength, the variations in the foreground Faraday rotation measure lead to splitting
of the emission in $Q$ and $U$ into patches with spatial scales smaller than the map size, which are
not removed by subtracting linear baselevels.

At the high resolution of the VLA maps, all four main spiral arms of IC~342 seen in total intensity,
which is a signature of the total field (Sect.~\ref{sect:total}), are partly polarized, with different degrees
of polarization (see below). The polarized emission, a signature of the ordered magnetic field in the
sky plane, follows that of the total emission of the narrow northern and the two broad southern arms
(Fig.~\ref{cm6}), indicating that a fraction of the isotropic turbulent field is compressed or sheared
and becomes anisotropic turbulent.

The dominant structures in the VLA polarization maps at \wave{3.5} and \wave{20.1} and in the combined
VLA+Effelsberg map at \wave{6.2} (Figs.~\ref{cm3}--\ref{cm20b} right panels, Figs.~\ref{spitzer} and~\ref{kitt})
are the well-defined spiral arms. The longest one on the eastern side (no.~1 in Figs.~\ref{cm6} and
\ref{cm20b}, right panels) extends from the south (at about 5\arcmin\ radius) to the far north (at about
12\arcmin\ radius) over at least 30\arcmin\ in length. Two other spiral arms with at least 20\arcmin\
length are seen in the south-east at larger radii, running almost parallel to the inner one (nos.~2 and 5 in
Fig.~\ref{cm20b} right). All three polarization spiral arms are related to spiral structures in total neutral
gas (Fig.~\ref{gas}), though partly displaced inwards (see Sect.~\ref{sect:features}).

Another prominent polarization arm seen at \wave{6.2} (Figs.~\ref{cm6} right and \ref{spitzer} left) is the
inner spiral arm extending from the east (at about 2\arcmin\ distance from the galaxy centre) to the north-west
(between about 3\arcmin\ and 7\arcmin\ distance from the centre). Its brightest sections
are also seen at \wave{3.5} (Fig.~\ref{cm3} right), but not at \wave{20.1} (Figs.~\ref{cm20a} and \ref{cm20b} right)
due to Faraday depolarization.

The narrow arm section located close (east -- north-east) to the central region is different from all other arms:
while clearly resolved in total intensity, it is not resolved in polarized emission at \wave{6.2}
(Fig.\ref{cm6} right) and at \wave{3.5} even at the best available resolution of 12\arcsec, so that its
intrinsic width (corrected for beam smearing) is smaller than 200\,pc (at the assumed distance of 3.5\,Mpc).
Furthermore, the location of the ridge line of polarized emission at \wave{3.5} and \wave{6.2} is shifted
inwards (towards the centre) with respect to that of the total radio and infrared emission
(Fig.~\ref{spitzer} left) by about 10\arcsec\ (170\,pc).\footnote{The location of a feature observed with
a high signal-to-noise ratio $R$ can be measured with an accuracy of $\Theta/(2\,R)$, where $\Theta$ is the
half-power with of the telescope beam. Here, $R$ is about 10 in polarization and even higher in total
intensity, so that the measured shift is highly significant.}
Such a displacement is a probable signature of a density wave that first compresses the field and then
induces star formation and turbulence. A similar displacement has also
been detected in the density-wave spiral galaxy M~51 \citep{patrikeev06}.

The northern section of the polarized spiral arm is split into two parts. The ridge line of its
northern part (no.~4 in Fig.~\ref{cm6} right) oscillates around the infrared spiral arm inwards and outwards with
shifts of up to about 500\,pc (Fig.~\ref{spitzer} left), which is indicative of a helical field (Sect.~\ref{sect:parker}).
It is also detected in polarization at \wave{3.5} (Fig.~\ref{cm3} right), but not at \wave{20.1} where
Faraday depolarization is strong. Wavelength-dependent Faraday depolarization could originate in the dense
magneto-ionic gas of the spiral arm \citep{krause93}, enhanced by depolarization in the halo field on the
north-eastern (receding) side of the major axis (see Sect.~\ref{sect:dp}).

The southern part of the polarized northern filament (no.~3 in Fig.~\ref{cm6} right) is located in the
{\em \emph{interarm}}\ region around
RA, DEC (J2000) $\simeq 03^\mathrm{h}\ 46^\mathrm{m}\ 20^\mathrm{s}$, +68\degr\ 08\arcmin,
resembling a magnetic arm, though its length of about 5\,kpc is much less than that of the magnetic
arms in NGC~6946 (see discussion in Sect.~\ref{sect:ma}). It is also detected in the low-resolution map
at \wave{20.1} (Fig.~\ref{cm20b} right) because Faraday depolarization is smaller in interarm regions.


On the western -- south-western side of the galaxy another three spiral filaments can be traced at \wave{20.1}
(Fig.~\ref{cm20b} right, without numbers), but they are shorter and less well defined than on the eastern side.
The two inner ones are also visible at \wave{6.2} (Fig.~\ref{cm6} right). The outermost filament is located
at the inner edge of a filament in total neutral gas (Fig.~\ref{gas} left). Optical data suggest that the
south-western region is affected by tidal interaction with the Local Group \citep{buta99}, which may have
disturbed the formation of long polarized filaments.

The degrees of polarization of the nonthermal emission at \wave{6.2} vary between about 15\% in the
inner southern and eastern arms and in the northern arm, about 20\% in the northern magnetic arm and
more than 30\% in the outer southern and south-eastern arms, with a peak value of about 60\% in the
outermost arm (no.~2 in Fig.~\ref{cm6} right). The degrees of polarization at \wave{3.5} are similar.
These results indicate that the magnetic field structure in the spiral arms of IC~342 is different
from that in NGC~6946 \citep{beck07}, where the typical degrees of polarization are lower in the spiral arms
(between 5\% and 10\%) but higher in the magnetic arms (35\% on average, up to 45\% locally).
In IC~342 the field is less turbulent in the spiral arms than in those of NGC~6946. The
northern magnetic arm of IC~342 is shorter and its field is less ordered (more turbulent) than in that of
NGC~6946 (see Sect.~\ref{sect:ma}).

\begin{figure*}[htbp]
\vspace{0.7cm}
\includegraphics[width=0.43\textwidth]{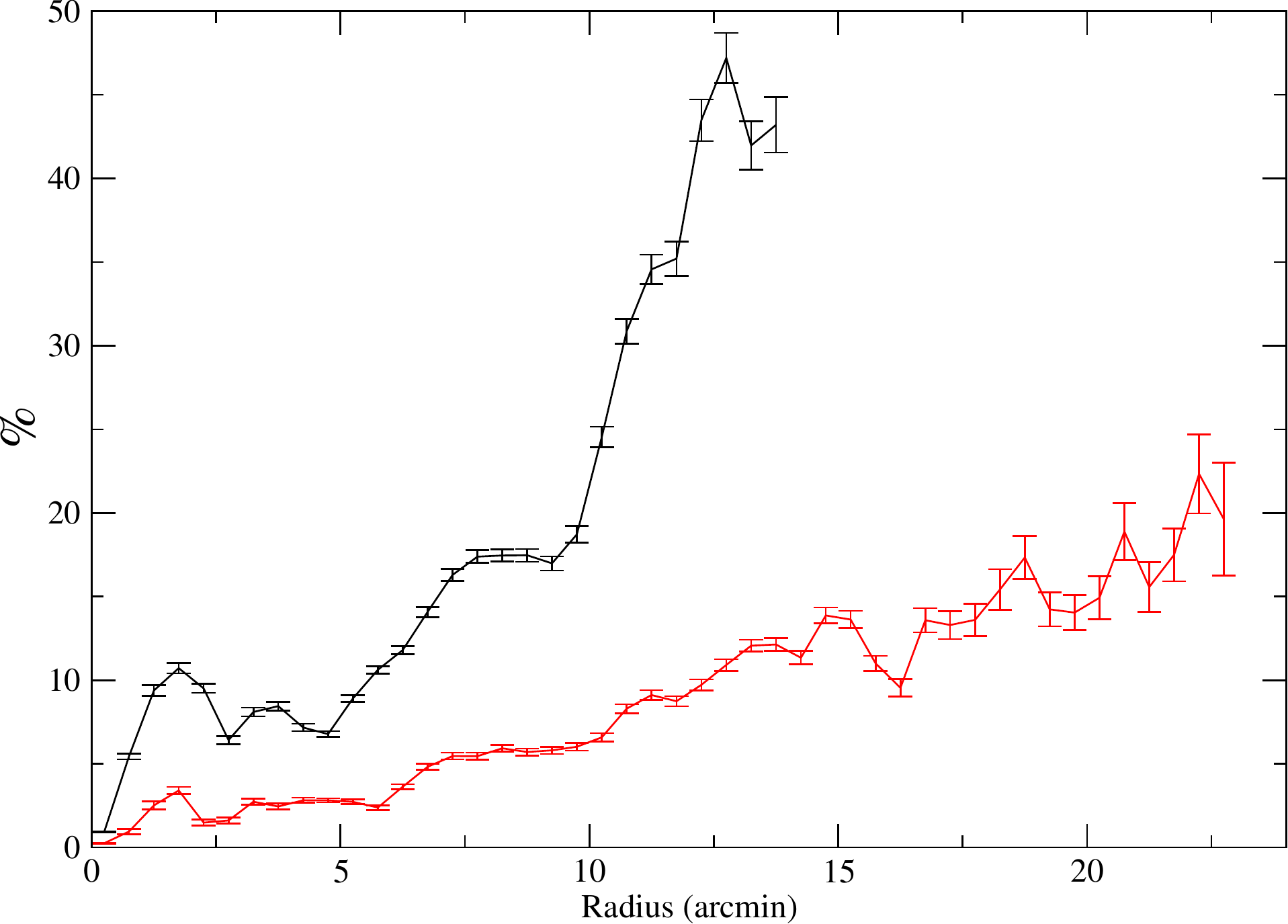}
\hfill
\includegraphics[width=0.43\textwidth]{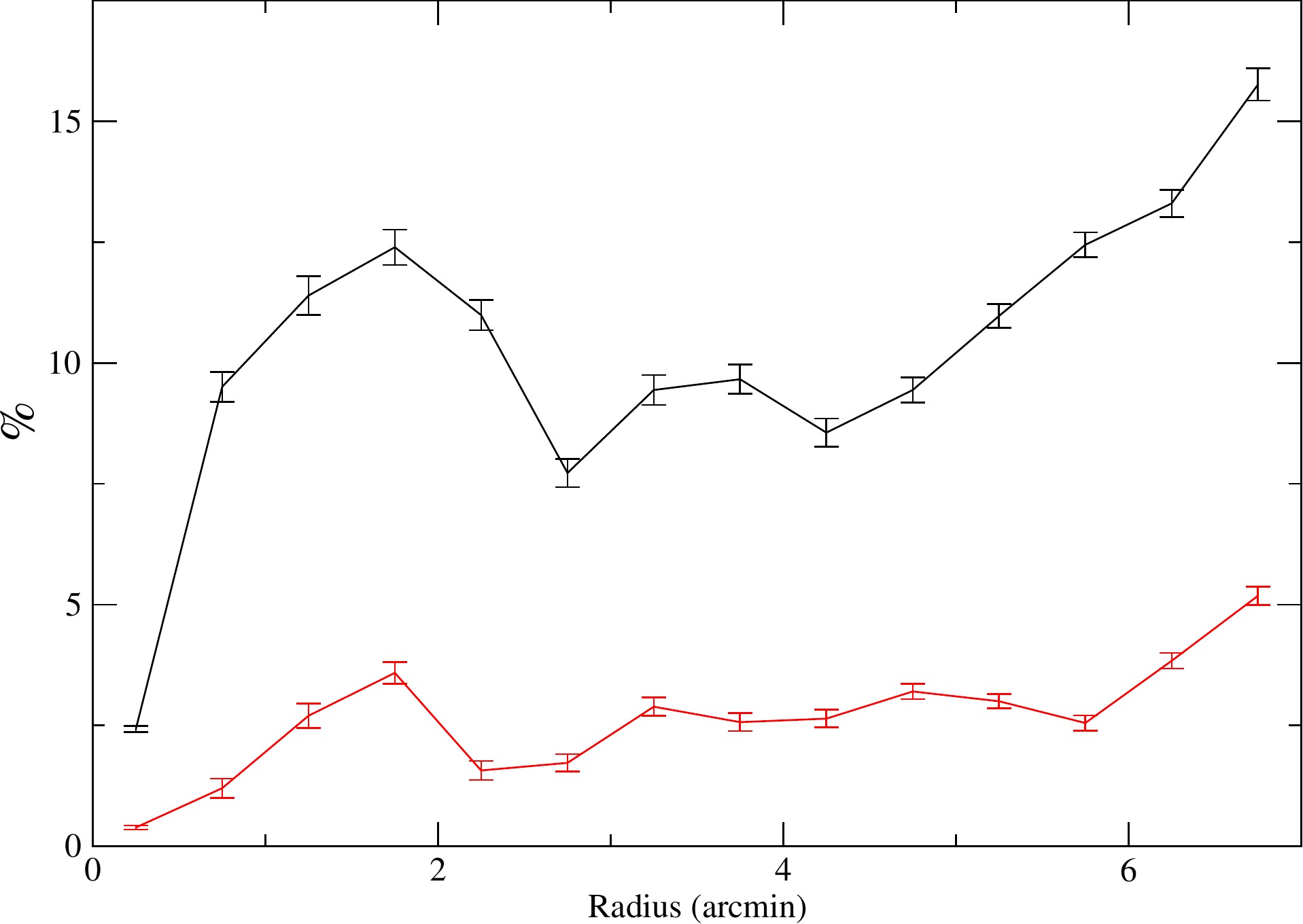}
\caption{{\it Left:\/} Degree of polarization at \wave{6.2} (upper curve) and \wave{20.1}
(lower curve) at a resolution of 25\arcsec, determined from the average total and polarized intensities in rings of 0.5\arcmin\ width in the plane of the galaxy ($i=31\degr$, $PA=37\degr$), as a function of radius.
The polarization degree at \wave{6.2} becomes unreliable beyond 15\,arcmin\ radius owing to the limited map size.
{\it Right:\/} Same for the average degree of synchrotron polarization. Values beyond 7\arcmin\ radius are not
shown because of the limitation of subtracting the thermal emission (see Sect.~\ref{sect:thermal}).
}
\label{perc}
\end{figure*}

The degrees of polarization of the nonthermal emission $p_{syn}$ at \wave{20.1} are low, about 3\% in the eastern
arm, 5\% in the northern arm, and 10\% in the outer south-eastern arms. The average degree of polarization in
rings (Fig.~\ref{perc}) is also lower at \wave{20.1} than at \wave{6.2}, a sign of Faraday depolarization
(Sect.~\ref{sect:dp}). Here, $p_{syn}$ increases with distance from the galaxy centre because towards the outer galaxy
the degree of field order increases (Fig.~\ref{ord}) and Faraday depolarization decreases (Fig.~\ref{dp}).

Spiral arms with constant pitch angles are linear structures in polar coordinates $[ln(r)$, $\phi]$,
where $r$ is the radius and $\phi$ the azimuthal angle. Figure~\ref{spiral} shows the
transforms of Figs.~\ref{cm6} (right) and~\ref{cm20b} in polar coordinates.
In Fig.~\ref{spiral} (left) the spiral arms nos.~3 and 4 form adjacent features, with pitch angles
of -23 to -43\degr\ and -27\degr\ (Table~\ref{tab:pitch}).
The average degrees of polarization at \wave{6.2} (at 25\arcsec\ resolution) are 14\% in the eastern and
northern arms and 18\% in the magnetic arm. After subtracting the thermal emission from the total emission,
the polarization degrees of synchrotron emission become similar (16\% and 18\%, respectively).

Several of the polarization spiral arms discussed above show little or no similarity to spiral arms
of total intensity. There must therefore be mechanisms for generating ordered fields other than
compression or shear, probably mean-field dynamo action. Another major difference is that the bright
diffuse disk seen in total emission lacks polarized emission. This indicates that isotropic
turbulent magnetic fields are responsible for the diffuse emission seen in total intensity,
while the ordered field is mostly concentrated in spiral filaments. Because most of the polarized emission is
already detected with the VLA alone, the diffuse disk of polarized emission seen in the
Effelsberg maps at \wave{6.2} and \wave{11.2} consists of unresolved polarization spiral arms.

The properties of the different types of polarization arms are summarized and discussed in
Sect.~\ref{sect:features}.

\subsection{Magnetic field strengths}
\label{sect:mf}

\begin{figure}[htbp]
\vspace{0.7cm}
\centerline{\includegraphics[width=0.45\textwidth]{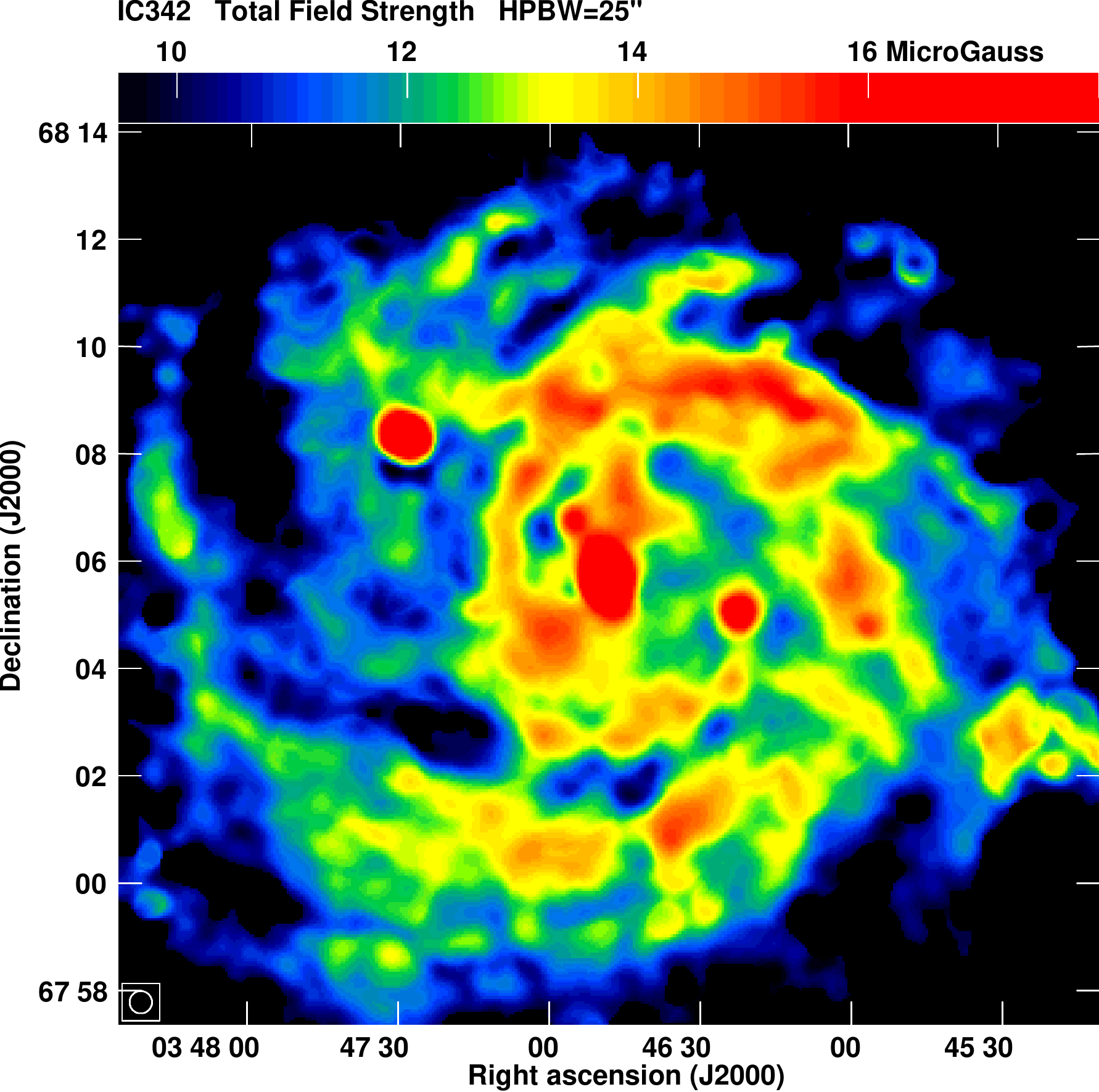}}
\caption{Total magnetic field strength in IC~342, derived from the nonthermal intensity at \wave{6.2}
at 25\arcsec\ resolution and assuming equipartition between the energy densities of total magnetic fields
and total cosmic rays.}
\label{mf}
\end{figure}

The strength of the total magnetic field $B_\mathrm{tot}$ and its ordered component $B_\mathrm{ord}$ on the sky
plane is derived from both the total synchrotron intensity and its degree of linear polarization
(Sect.~\ref{sect:pol}), if equipartition between the energy densities of total magnetic fields and
total cosmic rays is valid. In addition, the ratio $K$ between the number densities of cosmic-ray protons and
electrons, the pathlength $L$ through the synchrotron-emitting medium, and the synchrotron spectral index
$\alpha_\mathrm{syn}$ need to be known \citep{beck+krause05}. Here, $K=100$, a pathlength through the thick disk of
$L=1$\,kpc/cos(i) (where $i=31\degr$ is the galaxy's inclination), and $\alpha_\mathrm{syn}=-1.0$
(constant over the galaxy) are assumed. The distribution of total field strengths is shown in Fig.~\ref{mf}.

The total field strength varies from $30\muG$ in the inner region to $13\muG$ at a radius
of 5\arcmin, to $11\muG$ at a radius of 10\arcmin\ and to $8\muG$ at the disk edge in Fig.~\ref{mf}.
The average total field strength is $15\muG$ within a 3\arcmin\ radius and $13\muG$ within a 6\arcmin\ radius.
The ordered field decreases slowly from $5\muG$ near the centre
to $4\muG$ at a 5\arcmin\ radius, but increases again to $5\muG$ at 8\arcmin\ radius.
Typical values of the total field strength in the main optical spiral arms are $15\muG$, those of
the ordered field $5\muG$. The strengths of the ordered field in the northern magnetic arm
is $6\muG$. The strongest ordered fields of $8\muG$ are found in the outer southern and
south-eastern arms. The average strengths of the total and ordered fields within a 7\arcmin\ radius
are $13\muG$ and $4\muG$, respectively (Table~\ref{tab:flux}).

The uncertainty in field strength is given by the uncertainties of the assumptions.
The largest uncertainties are in the values assumed for the pathlength $L$
and the ratio $K$. Changing one of these values by a factor $a$ changes the field
strength by a factor of $a^{-1/(1-\alpha_\mathrm{syn})}$. Even a large uncertainty of $a=2$
would decrease the field strength by only 16\%.

The total magnetic field strength derived from synchrotron intensity is overestimated if the field fluctuates
along the line of sight \citep{beck03}. If the amplitude of the fluctuations in field strength is similar to its
mean value, the field is too strong by a factor of about 1.4. Such extreme fluctuations may occur in
turbulent star-forming regions, but hardly in the average ISM.

The assumption of a constant synchrotron spectral index $\alpha_\mathrm{syn}$ in galaxies is
only valid on average, because the synchrotron spectrum is known to be flatter in spiral arms and steeper
between the arms \citep{taba07b}. This means that the magnetic field strength is underestimated in the
arms and overestimated in the interarm regions. A consideration would require an improved separation
of thermal and nonthermal emission, which is not expected to modify the results significantly.
An extreme variation in $\alpha_\mathrm{syn}$ by $\pm0.3$ would change the field strengths by about $\pm2\muG$.

The results for magnetic field strengths and magnetic energy densities in this paper are based on the
assumption of energy density equipartition. The general concept of energy equipartition
has often been questioned. For example, a correlation analysis of radio continuum maps from the Milky Way
and from the nearby galaxy M~33 by \citet{stepanov14} showed that equipartition does not hold on small
scales, which is understandable in view of the propagation length of cosmic-ray electrons
(Sect.~\ref{sect:prop}). Using the radio--IR correlation, \citet{basu12} and \citet{basu13} argue that
the equipartition assumption is valid on $\ge1$\,kpc scales.
Further arguments for equipartition on large scales come from the joint analysis of radio continuum and
$\gamma$--ray data, allowing an independent determination of magnetic field strengths, such as in the
Large Magellanic Cloud (LMC; see \citep{mao12} and in M~82 \citep{yoast13}.

Another limitation of the equipartition condition emerges from energy losses of cosmic-ray electrons.
Energy equipartition is valid between the total magnetic fields and the total cosmic rays that are
dominated by protons. Synchrotron radio waves are emitted by cosmic-ray electrons (CREs) that herewith lose
energy. Further loss processes are the inverse Compton effect with photons of the galaxy's radiation field and the
CMB, bremsstrahlung and ionization losses with atoms of the neutral gas, and adiabatic losses in expanding outflows.
The energy losses of cosmic-ray protons are much greater. As a result, the CR proton-to-electron ratio $K$
increases away from CR sources, and the magnetic field strengths are underestimated.

The average surface brightness of radio synchrotron emission (in mJy/arcsec$^2$) of IC~342 is less than
half of that in the spiral galaxy NGC~6946 \citep{beck07}. As a result,
the average total equipartition magnetic field $B_\mathrm{tot}$ within 7\,kpc radius of $13\muG$
(Table~\ref{tab:flux}) is about 20\% weaker in IC~342 than in NGC~6946, and
the average ordered field of $4.3\muG$ is even about 30\% weaker in IC~342.

The strength of the total magnetic field $B_\mathrm{tot}$ and the star-formation rate per surface area
$\Sigma_\mathrm{SFR}$ are related as $B_\mathrm{tot} \propto \Sigma_\mathrm{SFR}^{\,\,\,\,0.14 ... 0.3}$ \citep{taba13a,heesen14}. However, $\Sigma_\mathrm{SFR}$, derived from IR and H$\alpha$ data,
is about 10\% smaller in IC~342 than in NGC~6946 \citep{calzetti10}, so that $B_\mathrm{tot}$
is expected to be only about 1\%\ to 3\% smaller in IC~342, which disagrees with the observations.
This indicates that the relation between $B_\mathrm{tot}$ and $\Sigma_\mathrm{SFR}$ is not universal,
but it varies between galaxies. Parameters other than $\Sigma_\mathrm{SFR}$ determine the magnetic field
strength, such as the efficiency of magnetic field amplification or the propagation of cosmic-ray electrons
(Sect.~\ref{sect:prop}). Still, IC~342 does not deviate from the global radio--IR correlation of integrated
flux densities (Sect.~\ref{sect:flux}).

This paper presents some of the most sensitive radio continuum maps of any spiral galaxy obtained so far.
The faintest total intensity detected in IC~342 with the VLA at \wave{6.2} is about $\simeq60\,\mu$Jy per
25\arcsec\ beam (3\,times the rms noise) (Fig.~\ref{cm6} left),
while the faintest total intensity of IC~342 detected with the Effelsberg telescope is $\simeq1.5$\,mJy per
3\arcmin\ beam (Fig.~\ref{cm6eff} left). These values correspond to total field strengths
of about $7\muG$ and $6\muG$, respectively.
The detection of polarized intensity of $\simeq0.2$\,mJy per 3\arcmin\ beam with the Effelsberg
telescope (Fig.~\ref{cm6eff} right) corresponds to a strength of the ordered field of $2.5\muG$ in IC~342
(assuming a degree of polarization of 20\%). Similarly weak emission was detected with
Effelsberg observations of other galaxies (e.g. NGC~5907 \citep{dumke00}).

\subsection{Degree of field order}
\label{sect:order}

The degree of synchrotron polarization $p_n$ is a measure of the ratio $q$ of the field strength
of the ordered field in the sky plane and the
isotropic turbulent field, the \emph{\emph{degree of field order}}, $q=B_\mathrm{ord}/B_\mathrm{tur}$.
The field observed in polarization can be a regular field or an anisotropic turbulent field
(see footnote~(2)). For equipartition between the energy densities of cosmic rays and magnetic field,
Eq.~(4) from \citet{beck07} applies. If cosmic rays are uncorrelated with the magnetic field or constant
in energy density, Eq.~(2) from \citet{beck07} should be used. The equipartition case yields about 15\% lower
values of $q$ than the non-equipartition case.

\begin{figure}[htbp]
\centerline{\includegraphics[width=0.45\textwidth]{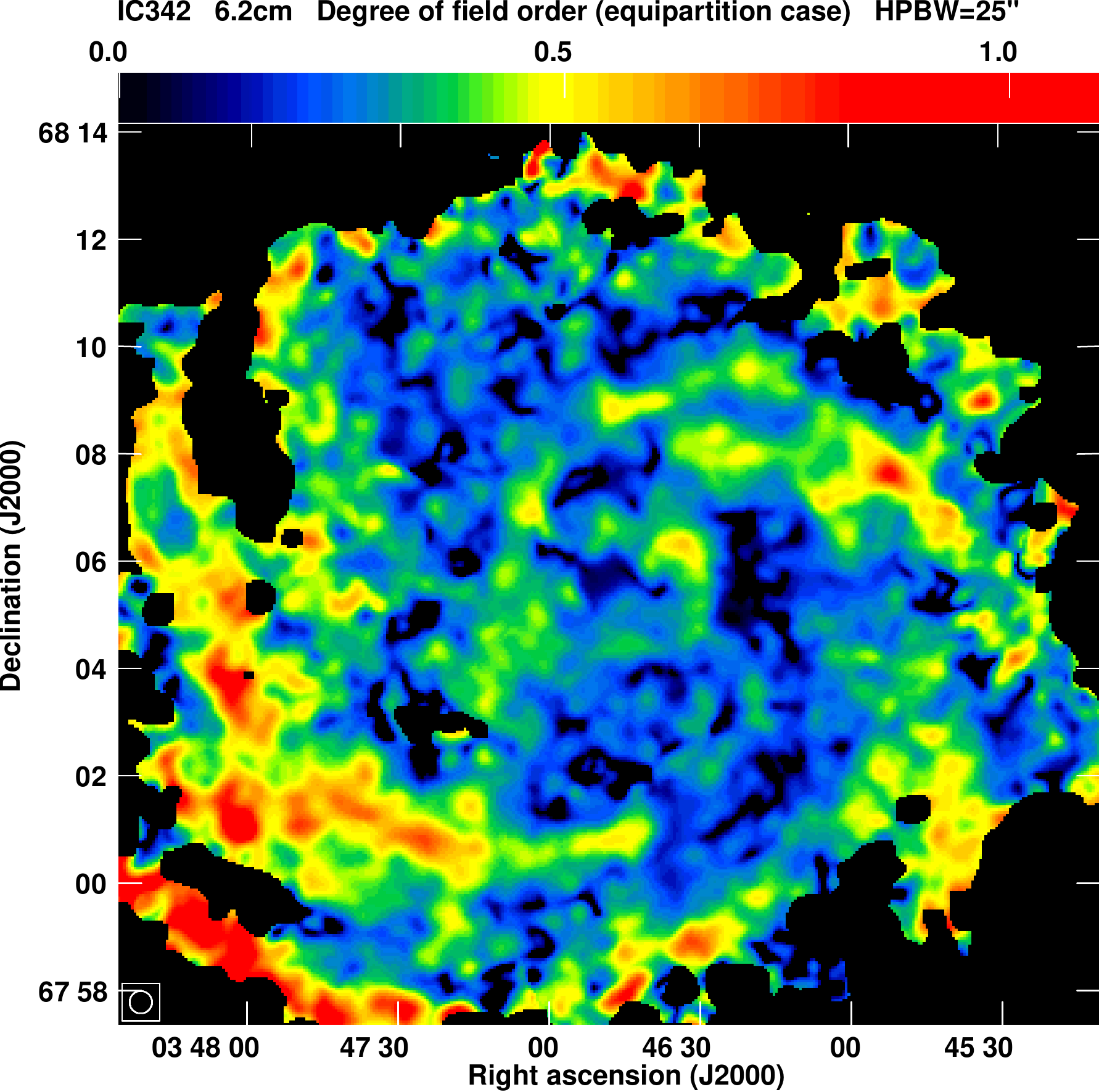}}
\caption{Degree of field order $q$, the ratio of the
ordered field strength in the sky plane to the isotropic turbulent
field strength, derived from the fractional polarization of the
nonthermal intensity of IC~342 at \wave{6.2} at 25\arcsec\
resolution and assuming equipartition between the energy densities of cosmic rays
and magnetic field. $q$ is computed only at points where the total
nonthermal intensities at \wave{6.2} are larger than 10\,times the rms noise.}
\label{ord}
\end{figure}

According to Fig.~\ref{ord}, the degree of field order $q$ is about 0.3 in the inner galaxy and increases to
about 0.4 in the outer parts. In the polarized filaments east and north of the centre, the ordered field can
be locally half as strong as the isotropic turbulent field. In the highly polarized outermost arms in the
south-east, the ordered field in the sky plane becomes comparable in strength to that of the isotropic turbulent one.
Mean-field dynamo models predict a ratio $q$ of 0.3--0.5 \citep{gressel08}, which increases with
decreasing star-formation rate. This is in good agreement with the observations presented here.

The average degree of field order $<q>$ within 7\arcmin\ radius is $0.35\pm0.03$ (Table~\ref{tab:flux}), similar
to that in M~33, but less than in M~31, M~51, and NGC~6946 \citep{taba13b}. A low degree of field order
predicts a small diffusion coefficient $D_\parallel$ of CREs along the ordered field
\citep[e.g.][Eq.~3.41]{shalchi09} (see discussion in Sect.~\ref{sect:prop}).

\subsection{Global properties of the radio emission}
\label{sect:flux}

\begin{table*}           
\caption{Integrated radio properties of IC~342}
\centering
\begin{tabular}{lcccccccc}
\hline
$\lambda$ (cm)  & $S_\mathrm{tot}$ (mJy) & $S_\mathrm{pol}$ (mJy) & $<p>$ (\%) & $<p_\mathrm{syn}>$ (\%) & $B_\mathrm{tot}$ ($\muG$) & $B_\mathrm{tur}$ ($\muG$) & $B_\mathrm{ord}$ ($\muG$) & $<q>$\\
\hline
6.2 & $782\pm66$ & $\,\,90\pm20$ & $11.5\pm2.7$ & $10.6\pm2.1$ & $12.9\pm0.2$  & $12.2\pm0.2$  & $4.3\pm0.4$ & $0.35\pm0.03$\\
20.1 & $2550\pm170$ & $101\pm10$ & $\,\,4.0\pm0.5$ & $\,\,2.9\pm0.3$ & $12.9\pm0.2$ & -- & -- & -- \\
\hline
\label{tab:flux}
\end{tabular}
\end{table*}

Table~\ref{tab:flux} gives the flux densities of the total and polarized emission of IC~342 at \wave{6.2}
and \wave{20.1}, integrated in rings in the galaxy's plane to the optical radius of $r_{25}=11$\arcmin\
(about 11\,kpc), and the resulting value of the average degree of polarization $<p>$.
A bright background source (no.~2 in Table~\ref{tab:sources}) was subtracted before integration.
The average degree of synchrotron polarization $<p_\mathrm{syn}>$ refers to the radial range $r\le7\arcmin$
for which the map of synchrotron intensity (Fig.~\ref{th} right) is complete.
The errors in the integrated flux densities are dominated by the uncertainties of the general zerolevel
far away from the galaxy.

The average values of the total equipartition field $B_\mathrm{tot}$, the turbulent field $B_\mathrm{tur}$,
the ordered field $B_\mathrm{ord}$, and the average degree of field order $<q>=B_\mathrm{ord}/B_\mathrm{tur}$
in Table~\ref{tab:flux} are derived from the average synchrotron intensity and $<p_\mathrm{syn}>$ at \wave{6.2}
in the radial range $r\le7\arcmin$.
Because the average synchrotron intensities at
\wave{6.2} and \wave{20.1} are related due to the method of subtracting the thermal emission, they
yield the same value for $B_\mathrm{tot}$. $<p_\mathrm{syn}>$ at \wave{20.1} is strongly affected by Faraday
depolarization and cannot be used to measure $B_\mathrm{ord}$.

The flux densities at \wave{6.2}, integrated to 15\arcmin\ radius, are $887\pm115$\,mJy for the total and
$131\pm34$\,mJy for the polarized intensity, which are the same within the error bars as those measured by \citet{graeve88}.

IC~342 is similar to the spiral galaxy NGC~6946 in many aspects. The infrared ($24\,\mu$m) and radio (20\,cm)
flux densities of NGC~6946 integrated to $r_{25}=9$\,kpc are 246~Jy \citep{dale12} and 1.44~Jy \citep{taba13a}, respectively. The ratio of 170 is in the typical range for most spiral galaxies \citep{murphy08}.
The corresponding values for IC~342 are 448~Jy \citep{dale12} and 2.55~Jy (Table~\ref{tab:flux}), which yields
a ratio of 175. Both galaxies follow the global radio--IR correlation. The lower radio surface brightness
of IC~342 (Sect.~\ref{sect:mf}) appears to be compensated by the larger size of IC~342, which is
$r_{25}\simeq11$\,kpc compared to $r_{25}\simeq9$\,kpc of NGC~6946.

\subsection{Pitch angles of the spiral magnetic field and the spiral structures}
\label{sect:pitch}

To determine the intrinsic magnetic pitch angle \footnote{The pitch angle is positive (negative) for a spiral
that is winding outwards in the counterclockwise (clockwise) direction.},
the observed $B$ vector is corrected for Faraday rotation,
transformed into the galaxy's plane and the position angle of the local circumferential orientation subtracted.
The intrinsic orientation of the ordered magnetic field in the sky plane can be determined from a map of
polarization angles at a sufficiently short wavelength where Faraday rotation does not play a role or
from a map of polarization angles at (at least) two wavelengths, corrected for Faraday rotation.

\begin{figure}[htbp]
\vspace{0.5cm}
\hspace{0.4cm}
\centerline{\includegraphics[width=0.45\textwidth]{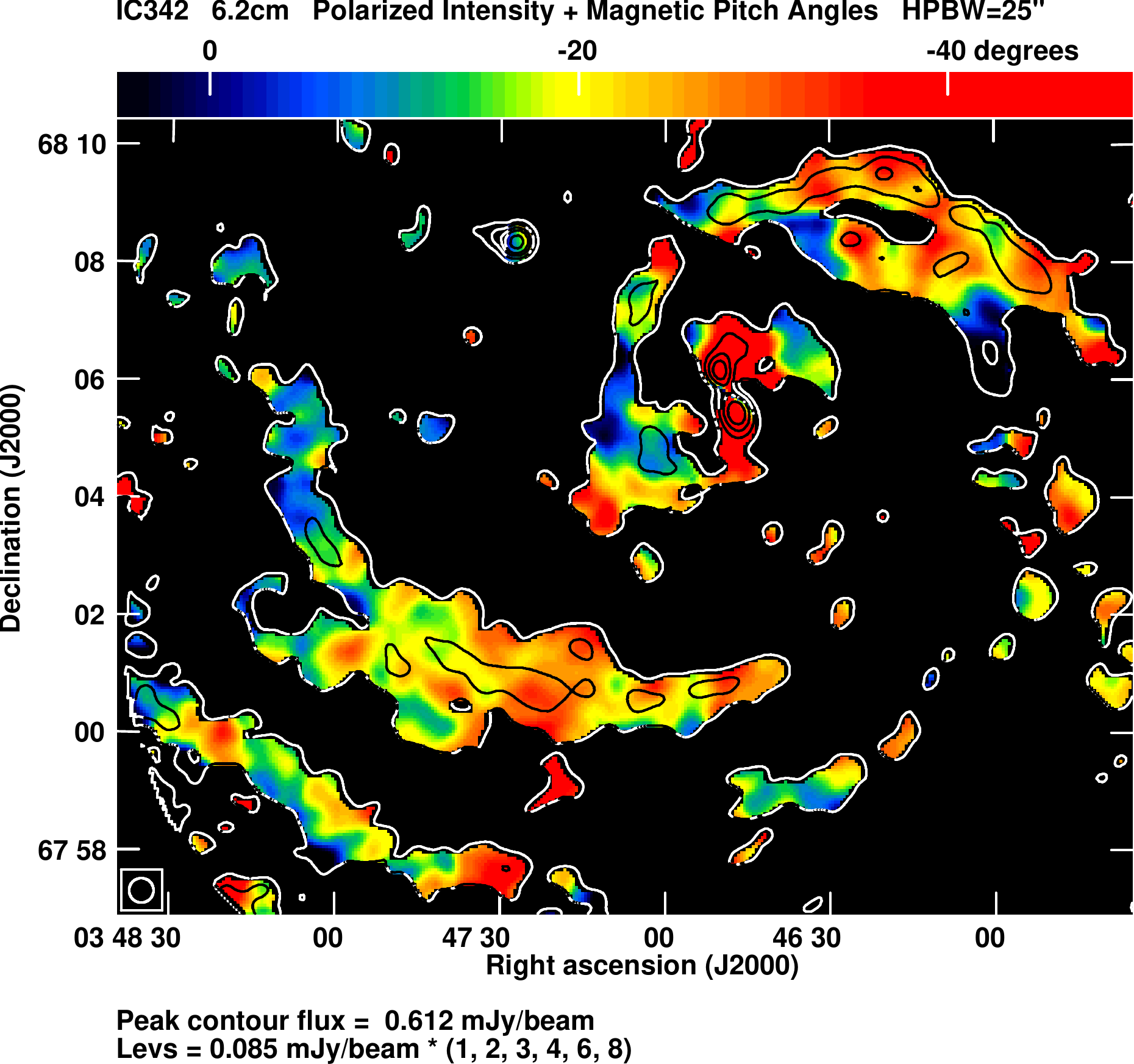}}
\caption{Pitch angle (greyscale) of the apparent magnetic field (not corrected for Faraday rotation) in the
plane of IC~342 at \wave{6.2} at 25\arcsec\ resolution, determined at pixels where the polarized intensities at
\wave{6.2} are larger than 7\,times the rms noise, so that the maximum error due to noise is
$\pm (1/14)$\,rad\,$\simeq 4\degr$. Contours show the polarized intensity at \wave{6.2}.
}
\label{pitch}
\end{figure}

\begin{figure}[htbp]
\vspace{0.7cm}
\centerline{\includegraphics[width=0.42\textwidth]{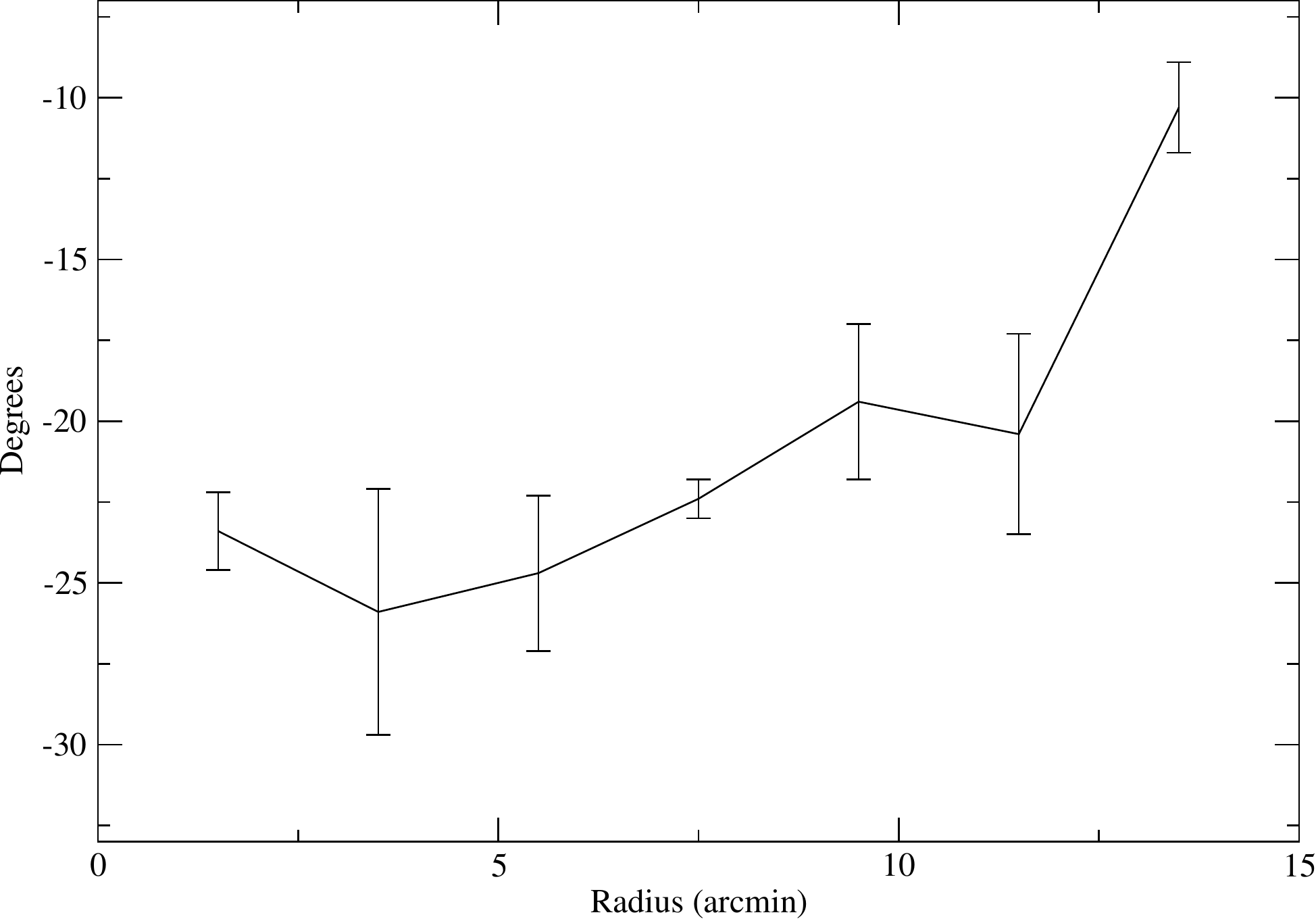}}
\caption{Average magnetic pitch angle in the galaxy's plane and its radial variation in IC~342.
The errors are determined from the standard deviations in Stokes $Q$ and $U$ in each ring.}
\label{pitch_radial}
\end{figure}

\begin{table*}           
\caption{Average pitch angles of spiral arm structures of polarized intensity and magnetic pitch angles of the ordered
field}
\centering
\begin{tabular}{llllll}
\hline
$\lambda$ (cm) & No. & Azimuthal range (\degr) & Radial range (\arcmin) & Structure pitch angle (\degr) & Magnetic pitch angle (\degr)\\
\hline
6.2  & 1 & 70-150  & 4.8-7.9  & $-26\pm1$ & $-23\pm8$ \\
            & 2 & 80-120  & 9.7-13.7 & $-22\pm2$ & $-17\pm12$ \\
            & 3 & 230-290 & 2.7-5.2  & $-27\pm1$ & $-23\pm7$ \\
            & 4 & 240-330 & 2.7-7.4  & $-23$ ... $-43$ & $-27\pm7$ \\
20.1 & 1 & 80-130  & 5.7-10   & $-42\pm3$ \\
            & 2 & 80-170  & 6.3-13   & $-25\pm1$ \\
            & 5 & 60-100  & 14-20    & $-24\pm3$ \\
\hline
\label{tab:pitch}
\end{tabular}
\end{table*}

The Effelsberg map at \wave{2.8} (Fig.~\ref{cm28eff}) has signal-to-noise ratios that are too low
to provide reliable maps of intrinsic magnetic field orientations.
At \wave{3.5} the signal-to-noise ratios of the polarized intensity are high enough to measure magnetic
pitch angles in restricted regions of the northern and eastern spiral arms and in the central region.
The local extrema of Faraday rotation of about $\pm200\radm$ (Fig.~\ref{rm} left) correspond to small
rotation angles of up to $\pm14\degr$ at \wave{3.5}.
The intrinsic magnetic pitch angle shows systematic variations between $-20\degr$ and
$+40\degr$ along the northern arm, is roughly constant (about $-20\degr$) in the eastern arm (Fig.~\ref{cm3} right),
and obtains high absolute values in the central region (Fig.~\ref{central}).

The \wave{6.2} polarized intensities at 25\arcsec\ resolution have higher signal-to-noise ratios.
Faraday rotation angles are stronger (up to $\pm45\degr$) than at \wave{3.5}, but cannot be corrected
over the galaxy disk with help of a second frequency, because
the polarized emission at \wave{3.5} is too weak, and the polarized emission at \wave{20} is affected
by Faraday depolarization, so that the polarization angles at \wave{20} refer to the upper disk and halo.
The apparent $B$ vectors at \wave{6.2} (Fig.~\ref{cm6}), which have not been corrected for Faraday rotation,
are transformed into the galaxy's plane. The position angle of the local circumferential orientation is
subtracted at each point, resulting in a map of magnetic pitch angles (Fig.~\ref{pitch}).

The ordered fields in the main spiral arms in the north -- north-west (no.~4 in Fig.~\ref{cm6} right) and
south-east (no.~1) have similar average magnetic pitch angles of $\approx -25\degr$.
(Average pitch angles are still useful because the average Faraday rotation over a spiral arm is small.)
The central region shows very large pitch angles of $\approx -60\degr$ in the northern lobe and $\approx -40\degr$
in the southern lobe owing to the bar (see Fig.~\ref{central}). The outer arm in the south-east (no.~2) has a
small magnetic pitch angle of $\approx -17\degr$.

The intensities in Stokes $Q$ and $U$ at \wave{6.2} are averaged in sectors of 10\degr\ width in azimuthal
angle and 2\arcmin\ in radius in the plane of the galaxy. The average polarization angles are transformed into
magnetic pitch angles with respect to the local azimuthal orientation in the galaxy plane and then averaged
along each ring. The average pitch angle $\psi$ of the ordered field increases from about $-23\degr$ at
1.5\arcmin\ radius to about $-10\degr$ at 13\arcmin\ radius (Fig.~\ref{pitch_radial}). The average Faraday rotation
measure in the azimuthal averages is that of the Galactic foreground of $-10\pm2\radm$ (Sect.~\ref{sect:rm}),
which corresponds to a rotation of about $-2\degr$ at \wave{6.2} and hence does not significantly affect the pitch
angles in Fig.~\ref{pitch_radial}.

\begin{figure*}[htbp]
\vspace{0.7cm}
\includegraphics[width=0.42\textwidth]{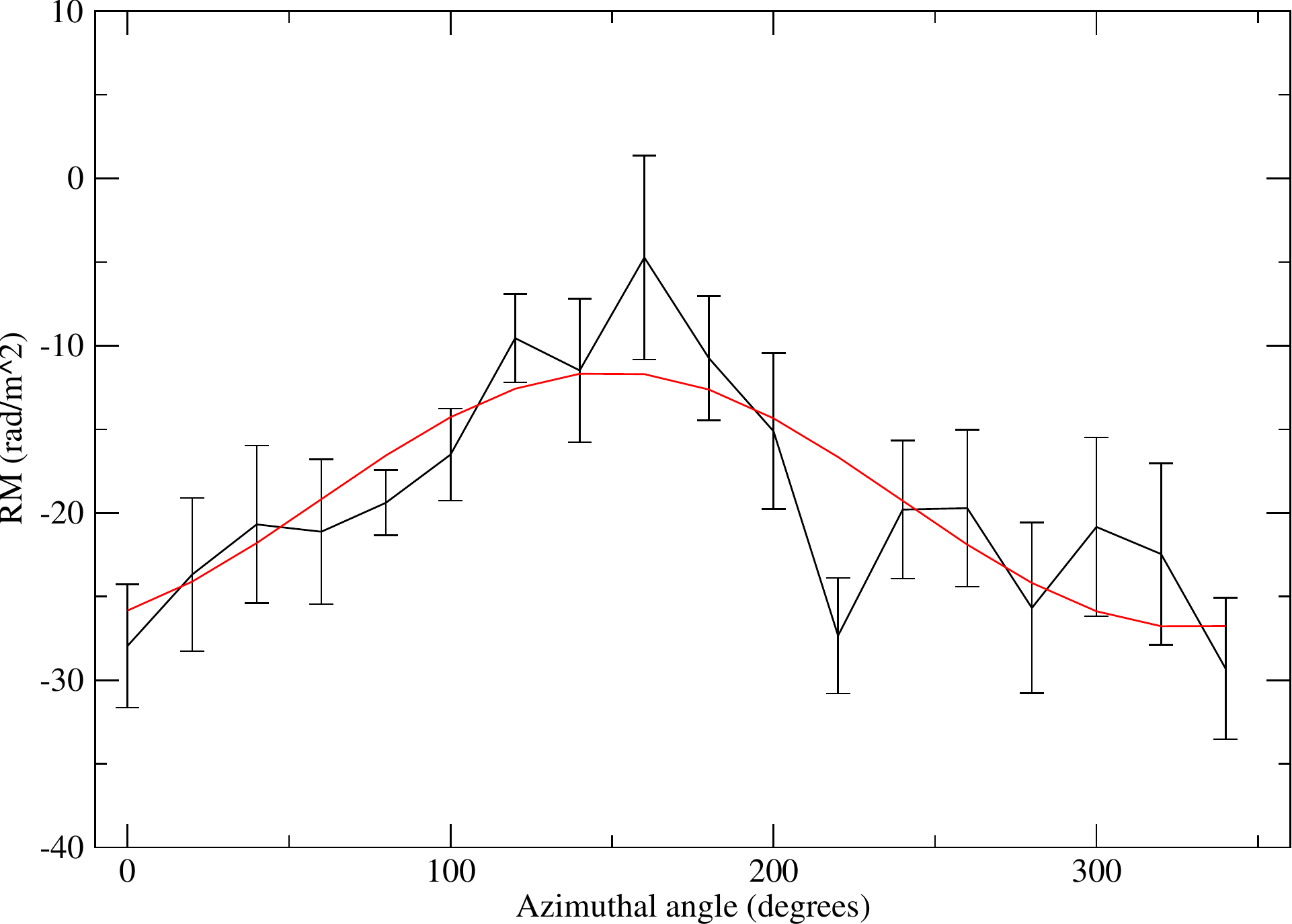}
\hfill
\includegraphics[width=0.42\textwidth]{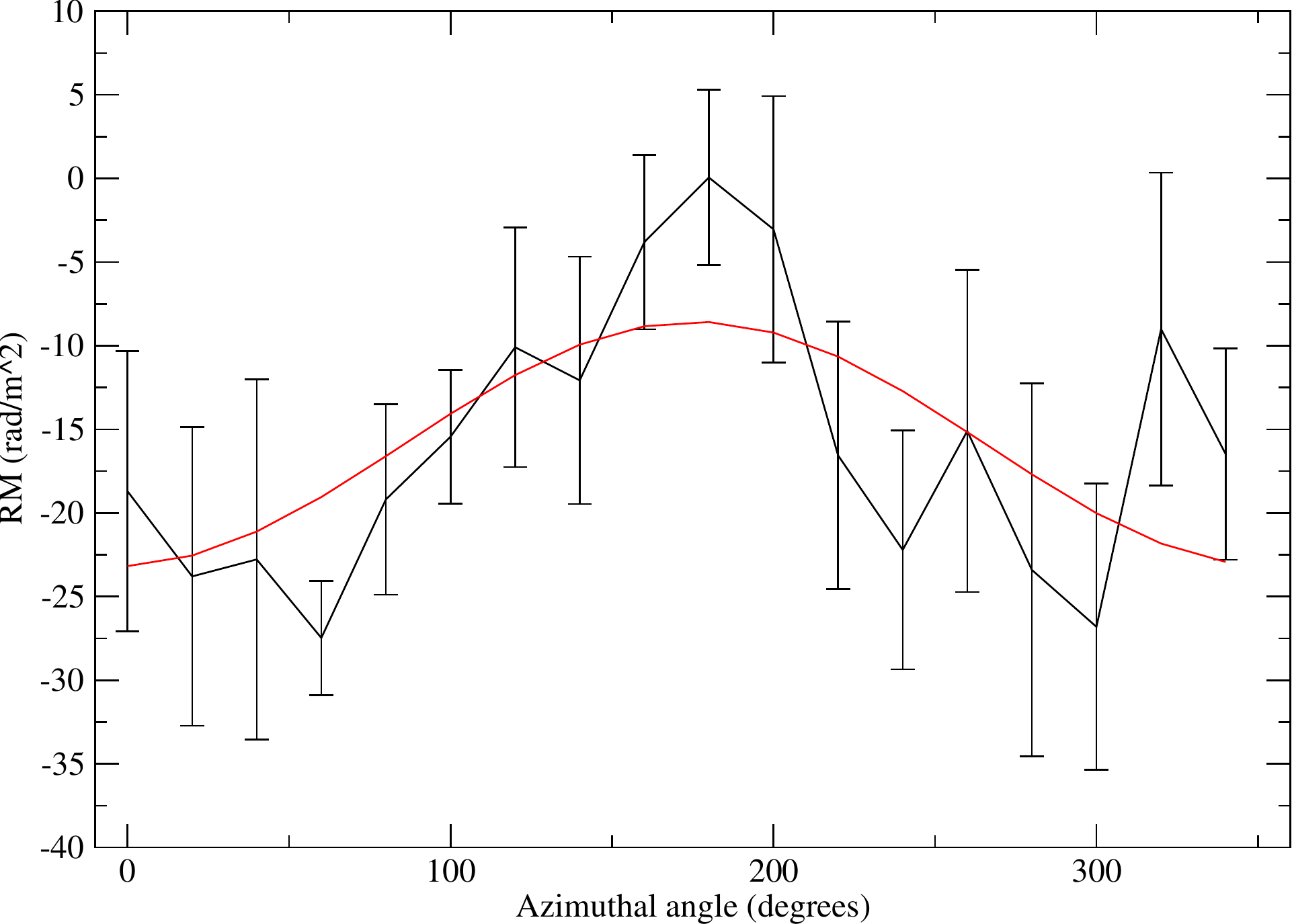}
\caption{{\it Left:\/} Faraday rotation measures $RM$ between \wave{6.2} and \wave{11.2} (Effelsberg only)
at 5\arcmin\ resolution, averaged in sectors of 10\degr\ azimuthal width in the plane of the galaxy
(counted counterclockwise from the north-eastern major axis) for the ring 7\farcm5\ -- 12\farcm5.
The error bars are computed from the standard deviations of the polarization angles at the two wavelengths.
The parameters of the fitted line are given in Table~\ref{tab:rm}.
{\it Right:\/} Same for the ring 12\farcm5\ -- 17\farcm5.
}
\label{rmeff}
\end{figure*}

By fitting the $RM$ variation between \wave{6.2} and \wave{11} with azimuthal angle in the galaxy's plane,
\citet{graeve88} found average pitch angles of the axisymmetric spiral (ASS) field of $-20.5\degr\pm1.7\degr$
in the ring $5\farcm5 - 10\farcm0$ and $-18.0\degr\pm1.9\degr$ in the ring
$10\farcm0 - 14\farcm4$. The $RM$ variation between \wave{6.2} and \wave{20} gave $-20\degr\pm4\degr$
and $-16\degr\pm11\degr$ in the same rings \citep{sokoloff92}. These values are in good
agreement with Fig.~\ref{pitch_radial}. The pitch angles of the azimuthal $RM$ variation derived from
the new Effelsberg maps (Fig.~\ref{rmeff}) are also consistent with Fig.~\ref{pitch_radial}.

In the thin-disk approximation of the mean-field dynamo, the magnetic pitch
angle \footnote{The magnetic pitch angle is not related to the pitch angle of the spiral structures
of the ordered field, though both numbers are generally observed to be similar.} $\psi$
is given by $\psi = - (R_{\alpha}/R_{\Omega})^{1/2}$ where $R_{\alpha}$ and $R_{\Omega}$ are the dynamo
numbers \citep{shukurov05}. Simplified estimates for the dynamo numbers and a flat rotation curve,
which is valid beyond a radius of about 7\arcmin\ in IC~342 \citep{newton80a},
give $\psi = - d/h$ where $d$ is the turbulence scale (about 50\,pc, \citet{fletcher11}),
and $h$ is the scale height of the ionized gas. The radial variation of $\psi$ in Fig.~\ref{pitch_radial}
indicates that the scale height $h$ is approximately constant until 12\arcmin\ radius and increases by
about a factor of two between 12\arcmin\ and 14\arcmin\ radius. Such flaring is in good agreement
with the results for the scaleheights of $\HI$ disks in spiral galaxies \citep{bagetakos11}, suggesting that
the disk of ionized gas also flares. However, there has been no such hint from H$\alpha$ observations of
edge-on spiral galaxies so far \citep{hoopes99}. Alternatively,
the magnetic pitch angle may be affected by gas flows, for instance, by outflows that become weaker towards the outer
disk and can decrease the effective dynamo number \citep{shukurov06,sur07}.

The pitch angles of the spiral structures of the ordered magnetic field are measured by the slope
of the linear features in the map of polarized intensity plotted in polar coordinates (Fig.~\ref{spiral}),
where five spiral arms can be distinguished.
The average structural pitch angles are given in Table~\ref{tab:pitch}.
Their absolute values decrease with increasing radius.
Arm no.~1 at \wave{20.1} is the extension of arm no.~1 observed at \wave{6.2}, but with a smaller
pitch angle. Arm no.~2 has a similar pitch angle at both wavelengths.
The magnetic arm (no.~3) has a similar pitch angle as those of the other arms.
The adjacent main inner spiral arm (no.~4) has a smaller pitch angle of about $-43\degr$ in the central part
(Fig.~\ref{spiral} left), while the inner and outer parts of this arm have larger pitch angles of about
$-23\degr$.

The average magnetic pitch angles and their dispersions given in the last column of Table~\ref{tab:pitch}
are determined at \wave{6.2} (Fig.~\ref{pitch}).
Magnetic pitch angles cannot be measured at \wave{20.1} due to strong Faraday rotation at this wavelength.

Table~\ref{tab:pitch} shows that the orientation
of the magnetic field at \wave{6.2} is on average well aligned with the polarization spiral arms. The variations
within each structure (standard deviations of $7\degr\ - 12\degr$) are larger than the noise fluctuations of
$\Delta\psi=0.5$\,rad$/R \le 4\degr$ for signal-to-noise-ratios of $R \ge 7$, as used in Fig.~\ref{pitch}.
This indicates intrinsic fluctuations in Faraday rotation and/or in the orientations of the magnetic field.

\subsection{Faraday rotation}
\label{sect:rm}

\begin{table}           
\caption{Fit results to Fig.~\ref{rmeff}: $RM_\mathrm{fg}$ is the $RM$ of the Galactic foreground,
$RM_\mathrm{max}$ is the amplitude of the $RM$ variation and $\phi_0$ the phase.}
\centering
\begin{tabular}{cccc}
\hline
Ring & $RM_\mathrm{fg}$ & $RM_\mathrm{max}$ & $\phi_0$ \\
\hline
7\farcm5\ -- 12\farcm5 & $-19\pm4\radm$ & $8\pm2\radm$ & $-30\pm10\degr$ \\
12\farcm5\ -- 17\farcm5 & $-16\pm4\radm$ & $6\pm2\radm$ & $-4\pm14\degr$ \\
\hline
\label{tab:rm}
\end{tabular}
\end{table}

The Effelsberg maps in Stokes parameters $Q$ and $U$ at \wave{6.2} and \wave{11.2} are smoothed to a common beam
of 5\arcmin. Faraday rotation measures ($RMs$) between the polarization angles at these two wavelengths are computed
in rings of 5\arcmin\ radial
width in the plane of the galaxy (Fig.~\ref{rmeff}). The points in each ring are fitted by a function predicted
for an axisymmetric spiral (ASS) field, the basic mode generated by mean-field dynamo action
\citep{krause87,graeve88,krause89a}, namely $RM = RM_\mathrm{fg} \, - \, RM_\mathrm{max} \,\, cos(\phi - \phi_0)$,
where $RM_\mathrm{fg}$ is the $RM$ of the
Galactic foreground, $RM_\mathrm{max}$ is the amplitude of the $RM$ variation, $\phi$ the azimuthal angle, and
$\phi_0$ the phase. For the axisymmetric spiral (ASS) field pattern, $\phi_0$ is identical to the average
pitch angle $\psi$ of the field in the corresponding ring. The results of the fits for the outer two rings are given
in Table~\ref{tab:rm}. The data points for the ring 2\farcm5\ -- 7\farcm5 do not show a significant cosine
variation.
These results are generally consistent with the previous results at the same wavelengths for the radial rings
5\farcm5\ -- 10\farcm0 and 10\farcm0\ -- 14\farcm4 by \citet{graeve88}, but have smaller errors than the previous
data. The values of $RM_\mathrm{max}$ in Table~\ref{tab:rm} are significantly smaller than those in
\citet{graeve88}.

The inclination-corrected amplitude of the ASS field of $RM_0 = RM_\mathrm{max} / \, tan(i)$ is $13\radm$ and $11\radm$
in the two rings of IC~342, which is several times smaller than in M~31 \citep{fletcher04} and M~51 \citep{fletcher11}
and about ten times smaller than in NGC~6946 \citep{ehle93}. The
$RM_0$ is related to the strength of the ASS field $B_0$ as $RM_0 = 0.81\,B_0 <n_\mathrm{e}> h$.
Assuming an average electron density along the line of sight of $<n_\mathrm{e}> \, \simeq 0.03\ccm$ and a scale
height of the ionized gas of $h\simeq 1000$\,pc, $B_0 \simeq 0.5\muG$ and $\simeq 0.4\muG$, similar to the values
for M~33, but much lower than in NGC~6946 \citep{ehle93} and several other spiral galaxies \citep{vaneck15}.

The weak ASS-type field and the pitch angles of about $-20\degr$ and $-10\degr$ at 10\arcmin\ and 15\arcmin\
(Fig.~\ref{pitch_radial}) are inconsistent with the relation between these quantities found by
\citet{vaneck15} (their Fig.~9d), so that the significance of this relation needs further investigation.

The values of
$RM_\mathrm{fg}$ from Table~\ref{tab:rm} are not well constrained and can be determined at longer wavelengths
with higher accuracy (see below). The spiral pitch angles $\phi_0$, with large errors, are consistent with
those derived independently from the polarization vectors (Fig.~\ref{pitch_radial}), supporting an ASS-type
field pattern.

\begin{figure*}[htbp]
\includegraphics[width=0.43\textwidth]{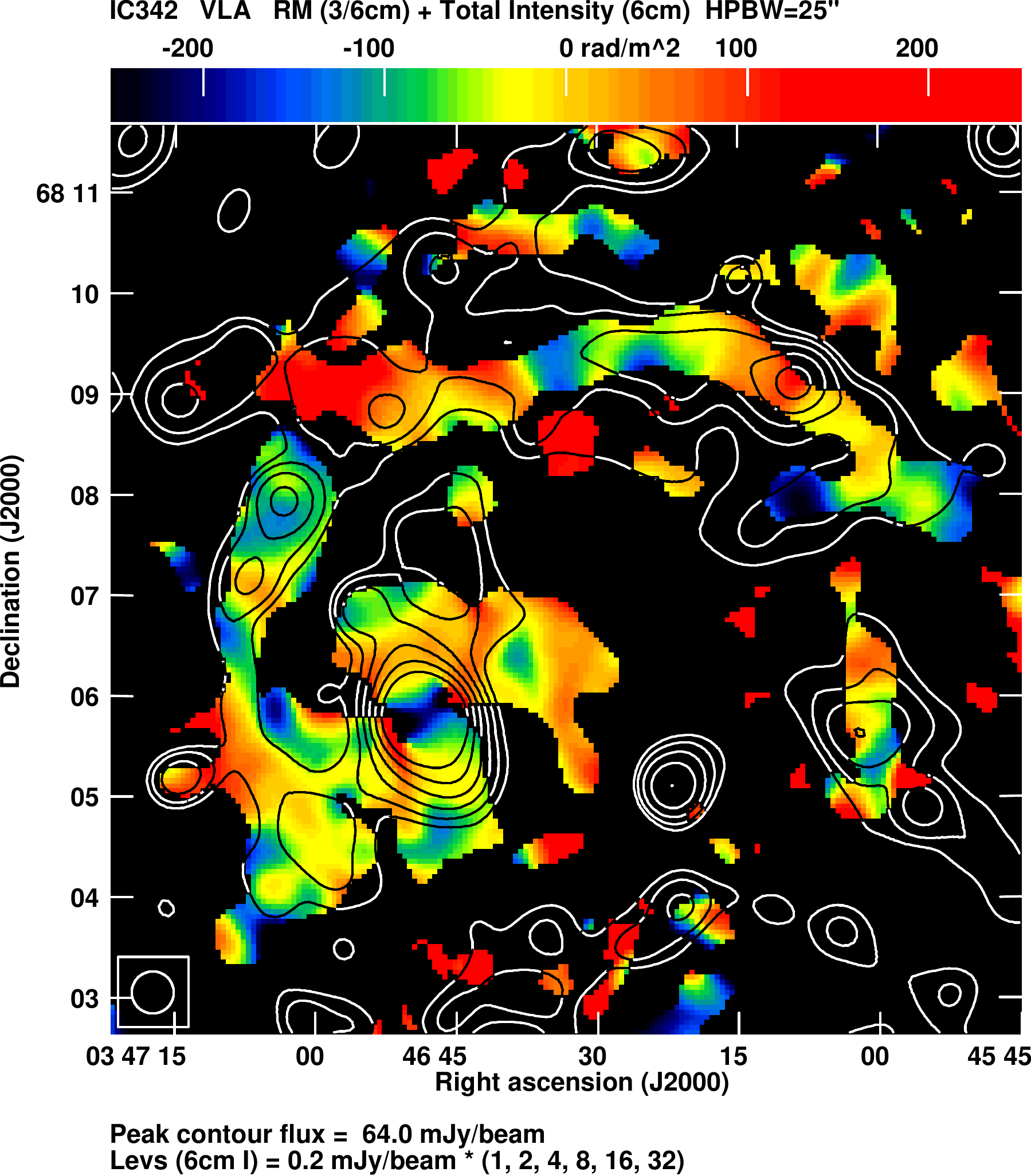}
\hfill
\includegraphics[width=0.49\textwidth]{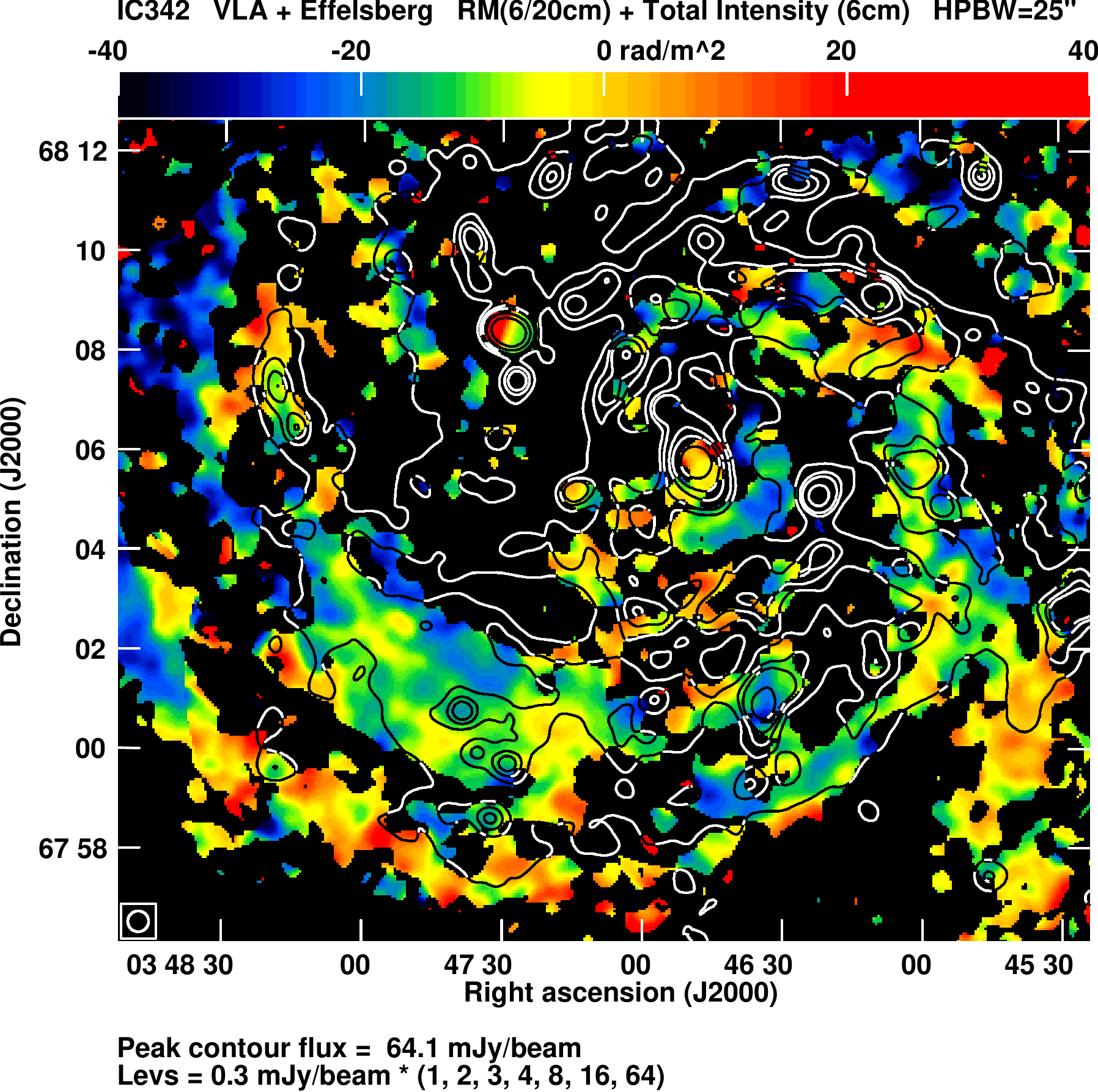}
\caption{{\it Left:\/} $RMs$ between \wave{3.5} and \wave{6.2} (VLA only) at 25\arcsec\ resolution
in the central and northern regions of IC~342 at points
where the polarized intensities at both wavelengths exceeds 3\,times
the rms noise (taken from that in $Q$ and $U$). Contours show
the total intensity at \wave{6.2}.
{\it Right:\/} Faraday rotation measures $RM$ between
\wave{6.2}  (VLA+Effelsberg) and \wave{20.1} (VLA) at 25\arcsec\ resolution at points
where the polarized intensities at both wavelengths exceeds 3\,times the rms noise.
Contours show the total intensity at \wave{6.2}.
}
\label{rm}
\end{figure*}

Using the VLA-only data that do not include the emission on large scales, a map of $RMs$ (Fig.~\ref{rm} left)
is computed from the maps of polarization angles at \wave{3.5} and \wave{6.2}.
A second $RM$ map (Fig.~\ref{rm} right) is obtained from the polarization angles at \wave{6.2}
from the combined VLA+Effelsberg data and the \wave{20.1} VLA data, both of which also
include the large-scale emission. Ambiguities in $RM$ values (owing to the $\pm n \cdot \pi$ ambiguity of
polarization angles) of $\pm n \cdot 1230\radm$ between \wave{3.5} and \wave{6.2} and $\pm n \cdot 86\radm$
between \wave{6.2} and \wave{20.1} are larger than the average values in Fig.~\ref{rm} and can be excluded.

The average value in $RM(6/20)$ of $-10\pm2\radm$ is adopted as the foreground $RM_\mathrm{fg}$ from our Galaxy.
It agrees with the value given by \citet{taylor09} and \citet{oppermann12} at the Galactic coordinates of IC~342
(l=138.2\degr, b=+10.6\degr) and with the results by \citet{graeve88} and \citet{krause89a}.
This value has to be subtracted to obtain $RM_i$ intrinsic to IC~342.

The distribution of $RM(3/6)$ (Fig.~\ref{rm} left) at 25\arcsec\ (about 400\,pc) resolution
does not show the large-scale pattern of the ASS field seen in Fig.~\ref{rmeff},
because emission at scales beyond about 2\,kpc is missing in these data and the ASS field is very weak.
Typical $RMs$ on smaller scales seen in Fig.~\ref{rm} (left) are about ten times stronger than
the amplitude $RM_\mathrm{max}$ of the underlying ASS field (Table~\ref{tab:rm}). The variations in $RM(3/6)$
are probably the result of local gas motions and instabilities. In particular, $RM(3/6)$ varies
systematically along the northern arm, indicating helically twisted field loops
(Sect.~\ref{sect:parker}).

This result may explain why \citet{vaneck15} could not find compelling evidence of any mean-field
dynamo action from the current polarization data of nearby spiral galaxies: The regular field with an ASS pattern
contributes only a small fraction to the ordered field, and the observational signatures of mean-field dynamo
action are hidden by other processes shaping the field on scales of a few 100\,pc.


The values of $RM(6/20)$ in IC~342 (Fig.~\ref{rm} right) are generally lower than $RM(3/6)$, as expected for
strong Faraday depolarization at \wave{20.1} (Sect.~\ref{sect:dp}). At this wavelength, emission from the
inner disk is almost completely depolarized, leaving emission from the upper disk and the halo,
as in M~51 \citep{fletcher11}.

The noise in the maps of polarized intensity accounts
for $RM$ fluctuations of $\Delta RM = (1/R_1^2 + 1/R_2^2)^{1/2} \, / \, (2\,|\, \lambda_2^2-\lambda_1^2 \,|\,)$,
where $R_1$ and $R_2$ are the signal-to-noise ratios of the polarized intensities at $\lambda_1$ and $\lambda_2$.
The rms noise in $PI$ is taken to be the mean rms noise in $Q$ and $U$.
The noise error $\Delta RM$ of $RM$ between \wave{3.5} and \wave{6.2} (Fig.~\ref{rm} left)
is dominated by the fluctuations caused by the low signal-to-noise ratios $R$ at \wave{3.5};
it increases from $30\radm$ in the brightest part of the northern arm,
to $50\radm$ on average over the northern and eastern arms and to $92\radm$ at
the cutoff limit of $R=3$. This agrees with the measured $RM$ dispersion of $\simeq50\radm$
in the northern arm.

The noise error $\Delta RM$ of $RM$ between \wave{6.2} and \wave{20.1} (Fig.~\ref{rm} right)
increases from $1.5\radm$ in the brightest part of the south-eastern arm,
to $2.5\radm$ over the whole south-eastern arm and to $6.4\radm$ at the cutoff limit of $R=3$.
The measured $RM$ dispersion in the south-eastern arm is larger ($\simeq4\radm$) and indicates a
contribution of ISM turbulence and field loops to the $RM$ dispersion.

A detailed analysis of the polarization angles at all four wavelengths will follow in a subsequent paper.

\subsection{Faraday depolarization}
\label{sect:dp}

Faraday depolarization is usually measured by the ratio
$DP$ of the degrees of polarization of the synchrotron intensities at two wavelengths,
where $DP=1$ means no depolarization and $DP=0$ means total depolarization.
The determination of the degree of synchrotron polarization requires subtracting the
thermal emission from the total emission, which is subject to the uncertainty of the synchrotron
spectral index (see Sect.~\ref{sect:thermal}). Furthermore, the noise in the four input maps of total
and polarized intensities at both wavelengths contributes to the uncertainties in $DP$.
To reduce these problems, $DP$ is computed as

\begin{equation}
DP=(PI_1/PI_2) \cdot (\nu_2/\nu_1)^{\alpha_\mathrm{syn}}\, ,
\end{equation}

\noindent where $PI$ is the polarized intensity at frequency $\nu$, and $\alpha_\mathrm{syn}$ is the
synchrotron spectral index that is assumed to be constant ($\alpha_\mathrm{syn}=-1.0$) across the galaxy.
Deviations from the assumed $\alpha_\mathrm{syn}$ affect $DP$ less severely than the uncertainties
of thermal fractions.

\begin{figure}[htbp]
\includegraphics[width=0.43\textwidth]{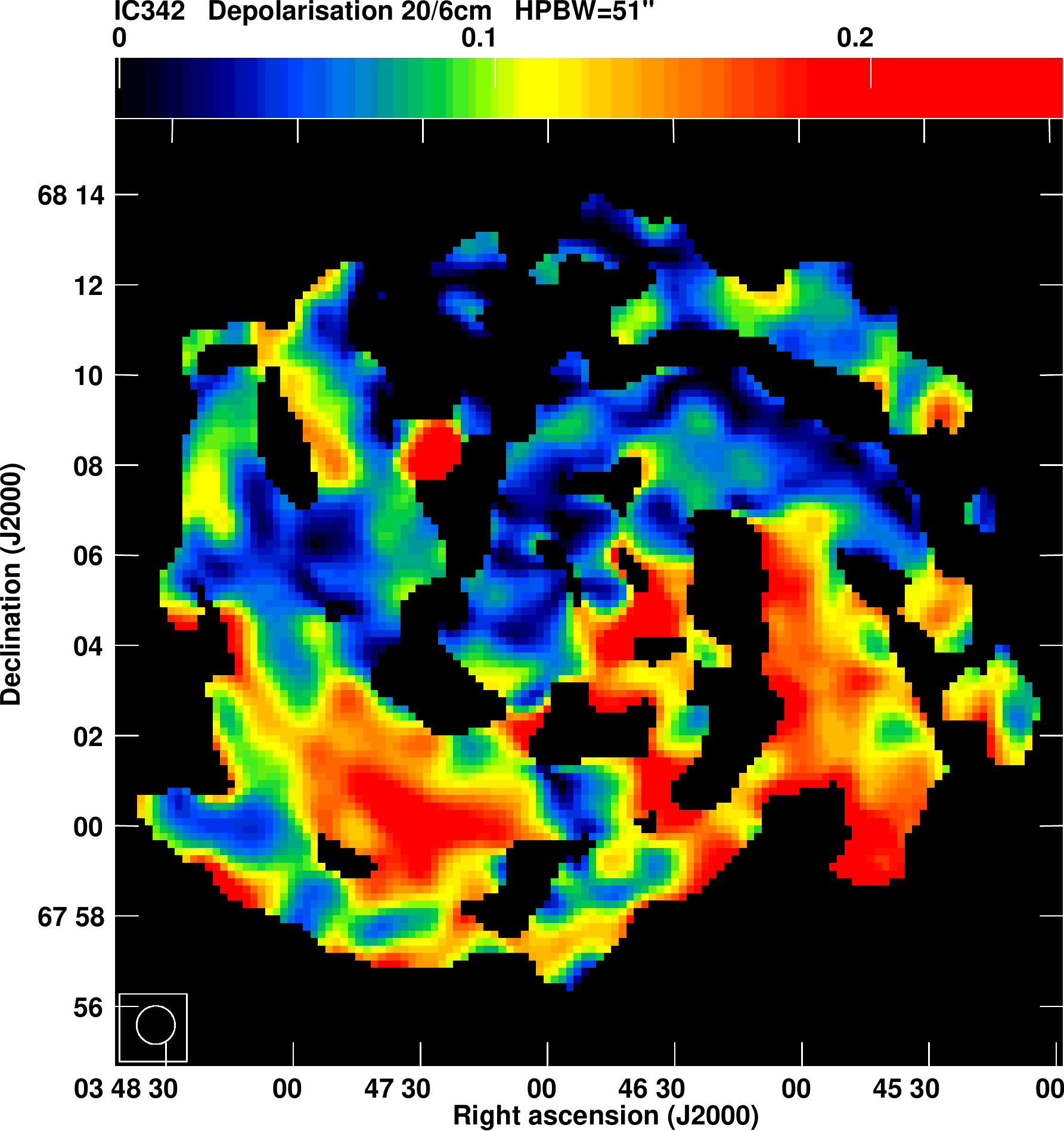}
\caption{Ratio $DP$ between polarized intensities at
\wave{20.1} and \wave{6.2} at 51\arcsec\ resolution, corrected for the average
synchrotron spectral index, as a measure of Faraday depolarization. The ratio
is computed only at points where the polarized intensities at both
wavelengths exceeds 5\,times the rms noise.
} \label{dp}
\end{figure}

The polarization maps at \wave{20.1} and \wave{6.2} at 51\arcsec\ resolution yield a map of
Faraday depolarization (Fig.~\ref{dp}). The
$DP$ is around 0.15 in the southern and western magnetic arms. Lower values around 0.05, i.e. stronger
depolarization, are found in the northern and north-eastern parts of the galaxy, located around the side
of the major axis (at $PA=37\degr$), where the galaxy's rotation is {\em \emph{receding}} (i.e. the radial velocities
are positive). Similar $DP$ asymmetries along the major axis of the projected galaxy disk in the sky plane
were found in many other galaxies, for which the minimum polarized emission at \wave{20} owing to strong depolarization
is always located on the receding side \citep{braun10, vollmer13}. A combination of spiral fields in the disk
and vertical fields in the halo, as observed in edge-on galaxies \citep{krause14} and predicted by
dynamo models (i.e. large-scale helical fields), is able to explain such asymmetries.

Differential Faraday rotation within the emitting layer leads to depolarization that varies with $\lambda$ as a
$sin(x)/x$ function, with $x=2 \, |RM_\mathrm{i}| \, \lambda^2$ \citep{sokoloff98}. At
\wave{20.1} strong $DP$ is expected for $|RM_\mathrm{i}| > 30\radm$, with lines of zero polarization (``canals'')
along level lines with $|RM_\mathrm{i}|=n \, \cdot \, 39\radm$. Because these are not seen in Fig.~\ref{cm20b} (right),
differential Faraday rotation is probably unimportant in IC~342, and turbulent fields are responsible for
Faraday depolarization, as already concluded by \citet{krause89a} and \citet{krause93}.

Internal Faraday dispersion by turbulence in the magneto-ionic ISM along the line of
sight is computed as \citep{sokoloff98}:

\begin{equation}
p\, = \, p_0 \, [\,1-\exp(-2S)\,]\, / \, (2S)\, ,
\end{equation}

\noindent where $S=\sigma_\mathrm{RM}^2 \, \lambda^4$, and $\sigma_\mathrm{RM}$ is the dispersion in intrinsic
rotation measure $RM_\mathrm{i}$. The average value of $DP=0.1$ between \wave{20.1} and \wave{6.2} requires a
dispersion of $\sigma_\mathrm{RM}\simeq55\radm$.

Faraday dispersion in the turbulent ISM is described as
$\sigma_\mathrm{RM}\,=\,0.81\, <n_\mathrm{e}>\, B_\mathrm{r}\, d\, (L/(d\,f))^{0.5}$, where $<n_\mathrm{e}>$ is the
average thermal electron density of the diffuse ionized gas along the line of sight (in $\ccm$), $B_\mathrm{r}$
the random field strength (in $\muG$), $L$ the pathlength through the thermal gas (in \pc), $d$ the turbulent
scale (in \pc), and $f$ the volume filling factor of the Faraday-rotating gas. Faraday dispersion occurs only if
$d \ll \Theta$ where $\Theta$ is the beamsize at the galaxy's distance ($\Theta \simeq 900$\,pc for the beam
used for Fig.~\ref{dp}).

While the thermal optical and thermal
radio emission is dominated by $\HII$ regions with a small filling factor, Faraday rotation and depolarization
occur in the diffuse ionized medium with a large filling factor. Values of $<n_\mathrm{e}>=0.03\ccm$,
$B_\mathrm{r}=10\muG$, $L=1000\pc$, $d=50\pc$ and $f=0.5$, which are standard values for the ISM of galaxies,
yield the required $\sigma_\mathrm{RM}$.

The intrinsic rotation measures between \wave{6.2} and \wave{20.1} (after subtracting the Galactic
foreground contribution) are generally smaller than those between \wave{3.5} and \wave{6.2}. The disks of
IC~342 and many other galaxies studied so far are not transparent to polarized decimetre radio waves, they are
``Faraday thick''. \citet{berkhuijsen97} modelled the similar situation
in the spiral galaxy M~51 by assuming Faraday dispersion, which depolarizes most of the disk, plus a
foreground layer in the upper disk or in the halo, which has a lower polarized intensity and rotation
measure than those of the full disk.

\subsection{Unresolved sources}
\label{sect:sources}

A compact, bright, and unpolarized source with a flat spectrum located in the northern spiral arm
(at RA, DEC (J2000) $\simeq 03^\mathrm{h}\ 46^\mathrm{m}\ 08\fs 9$, +68\degr\ 09\arcmin\ 07\arcsec)
is seen at all wavelengths, but is particularly prominent in the \wave{3.5} map (Fig.~\ref{cm3} left).
The flux densities are $3.1\pm0.1$\,mJy at \wave{3.5}, $3.4\pm0.1$\,mJy at \wave{6.2} and $3.0\pm0.2$\,mJy
at \wave{20.1}. It is also seen in the infrared maps (Figs.~\ref{wise} and \ref{spitzer}), so can be
identified as a compact star-forming region.

\begin{table*}           
\caption{Polarized background sources in the region around IC~342. Sources 1 and 3 are measured with resolutions
of 10\arcmin\ (\wave{21.4}) or 5\arcmin\ (\wave{6.2} and \wave{11.2}), sources 2A and 2B with resolutions of 15\arcsec\
(\wave{6.2} and \wave{20.1}). Flux densities $S_{\lambda}$ are given in mJy/beam, degrees of polarization in \%
and $RM$ in $\radm$.}
\centering
\small
\begin{tabular}{cccccccccccc}
\hline
No. & RA(J2000) & DEC(J2000) & $S_6$ & $S_{11}$ & $S_{20,21}$ & $p_6$ & $p_{11}$ & $p_{21,20}$ & RM(6/11) & RM(6/20,21) \\
\hline
1  & 03 41 23   & +68 01 00 & $121\pm4$ & $184\pm2$ & $342\pm9$ & $7.5\pm0.3$ & $6.9\pm0.3$ & $4.4\pm0.6$ & $-10.4\pm0.4$ & $-18\pm3$    \\
2A & 03 47 27.3 & +68 08 20 & $10.8\pm0.2$ & -- & $32.1\pm0.4$ & $5.4\pm0.1$ & -- & $4.8\pm0.1$ & -- & $-7\pm1$ \\
2B & 03 47 29.8 & +68 08 24 & $9.0\pm0.1$  & -- & $29.0\pm0.4$ & $3.3\pm0.1$ & -- & $1.6\pm0.1$ & -- & $-35\pm2$ \\
3  & 03 52 21   & +67 58 20 & $163\pm4$ & $280\pm2$ & $574\pm7$ & $8.4\pm0.2$& $7.9\pm0.2$ & $3.7\pm0.4$ & $-24.4\pm0.3$ & $-12.7\pm0.1$ \\
\hline
\label{tab:sources}
\end{tabular}
\end{table*}

Most unresolved sources in the radio images presented here are too bright to be $\HII$ regions or
supernova remnants in IC~342; they are distant QSOs or radio galaxies in the background.
The largest number of sources is detected at \wave{20.1} (Fig.~\ref{cm20b} left). This is
partly due to the large primary beam of the VLA at this wavelength and partly due to the steep synchrotron
spectrum of most distant radio sources.

Four background sources show significant degrees of
polarization between 2\% and 12\% at \wave{20.1} and at 51\arcsec\ resolution.
Only one of these polarized sources is also included in the field observed at \wave{6.2} (Fig.~\ref{cm6}).
It consists of two components (named 2A and 2B in Table~\ref{tab:sources}).
The $RM$ of the western component of $-7 \radm$ is similar to that of the Galactic foreground,
while the $RM \simeq -35 \pm n \cdot 86 \radm$ of the eastern component has a significant internal rotation.
The $RM$ ambiguity cannot be solved with help of the data at other wavelengths because the sources are located
outside of the VLA fields observed at \wave{3.5}. With the large Effelsberg beams at \wave{2.8} and \wave{11.2}
they cannot be distinguished from the diffuse emission.

Polarization from two strong, compact background sources (named 1 and 3 in Table~\ref{tab:sources}) is detected in the
Effelsberg maps at \wave{6.2}, \wave{11.2} and \wave{21.4}. Their $RMs$ are wavelength-dependent and hence partly
of internal origin. Both sources are interesting targets for detailed investigations of future
multi-frequency data to be obtained with JVLA and LOFAR.

\subsection{Magnetic fields in the central region}
\label{sect:central}

\begin{figure}[htbp]
\centerline{\includegraphics[width=0.4\textwidth]{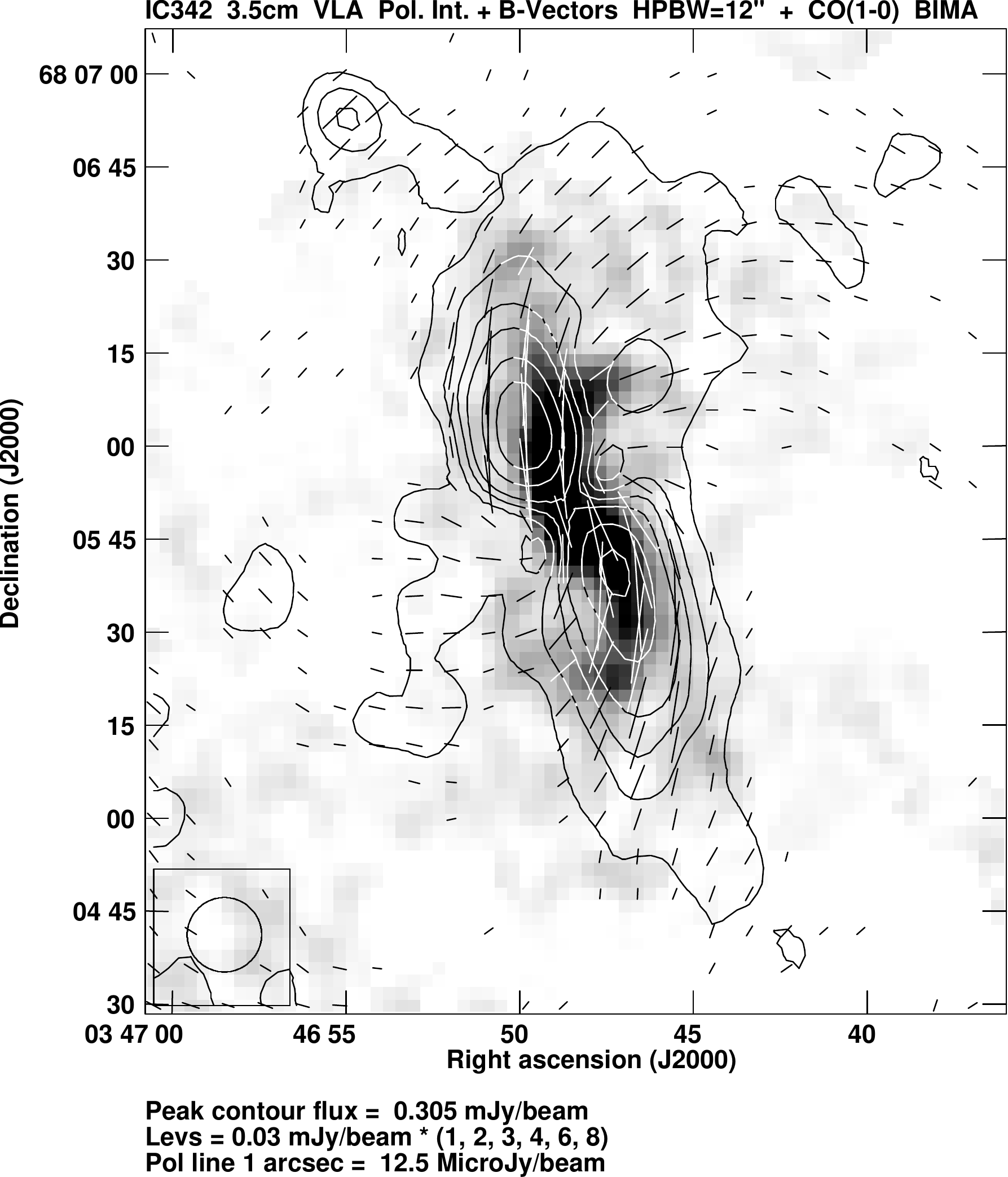}}
\caption{Linearly polarized intensity and observed $B$ vectors ($E$+90\degr) in the central region
at \wave{3.5} at 12\arcsec\ resolution, overlaid on a greyscale presentation of the CO(1-0) emission
at 2.6\,mm from the BIMA survey at $5\farcs6\times5\farcs1$ resolution \citep{helfer03}.}
\label{central}
\end{figure}

The central region of IC~342 hosts a bar of dust and cold gas with a wealth of molecular lines
\citep{eckart90,schulz01,helfer03,meier05,lebron11} and starburst activity \citep{ishizuki90}.
The recent nuclear starburst event is 4--30\,Myr old, and
the present star-formation rate is reduced because fueling the nucleus is currently prevented by feedback
from stellar winds and supernova shock fronts \citep{schinnerer08}. The nuclear region is also bright in
radio continuum \citep{turner83} and in X-rays \citep{bregman93,bauer03}.
In terms of size, dynamical mass, molecular mass, and star-formation rate, the nucleus of IC~342 is a
potential twin of the Galactic centre \citep{meier14}.

While the central bar in total intensity at the highest available resolution (Fig.~\ref{cm3} left)
coincides with the central bar in the CO line emission of molecular gas \citep[e.g.][]{helfer03}, the
polarized emission reveals a ``double-lobe'' structure that is displaced from the CO bar (Fig.~\ref{central}).
The average degree of polarization (uncorrected for thermal emission) is about 8\% in the northern lobe
and about 4\% in the south. The northern peak in polarized intensity is offset by about 5\arcsec\
($\simeq90$\,pc) from the CO ridge towards the east and the southern peak by 2\arcsec\ ($\simeq30$\,pc) towards the
west. Since the galaxy is rotating counter-clockwise, the ordered field is strongest on the preceding side of the bar.
The orientations of the ordered field lines follow the ridge of the bar, but around the bar
the field is oriented at large angles with respect to the bar ridge, forming a spiral pattern with large
pitch angles.

These results resemble the observations of polarized emission in the galaxies NGC~1097 and NGC~1365
hosting large bars \citep{beck05c}. According to numerical modelling of gas in a bar potential
\citep{athanassoula92}, the flow velocity of gas and magnetic fields is faster than the pattern speed of the bar
and enters the bar from the receding side. Field compression is greatest on the preceding side of the bar.
The ordered field in the central region of IC~342 traces the flow pattern that is a down-sized version of
the field pattern
around large bars. The magnetic field may support the outward transport of angular momentum that is required
to fuel the central region with gas.

The average Faraday rotation between \wave{3.5} and \wave{6.2} is positive in the northern ``lobe'' and negative
in the southern one (Fig.~\ref{rm} left). Taking the orientation of the galaxy's disk into account, the radial
component of the regular magnetic field is directed away from the centre of IC~342 on the northern and
the southern sides. This direction is {\em \emph{opposite}}\ to the radial component of the axisymmetric disk field
that points inwards \citep{graeve88,krause89a}, with a {\em \emph{field reversal}}\ between the central region and the
spiral arm region.

Mean-field dynamo models can generate field reversals \citep[e.g.][]{moss12}, but these are not associated
with large jumps in magnetic pitch angles as observed in IC~342. As a result, the magnetic field in the central region
is probably disconnected from the axisymmetric field in the disk. A similar dichotomy between the central and
the outer disk field has also been found in M~31 \citep{giess14}. Independent mean-field dynamos may operate in
the disk and in the central region of a galaxy. The central disk in M~31 is strongly inclined with respect
to the outer disk. The inclination of the bar region of IC~342 is hard to measure, but is probably different
from that of the disk, too \citep{crosthwaite01}.

\section{Discussion}

\subsection{Propagation of cosmic-ray electrons}
\label{sect:prop}

Beyond of 5\arcmin\ radius in the disk of IC~342, there is a rapid decrease in thermal intensity
(Fig.~\ref{radial} right). Thermal emission is a measure of the star-formation rate and of the production
of massive stars, the progenitors of supernova remnants, which are the probable sites of cosmic-ray
acceleration. The total intensity in the outer disk is mostly synchrotron emission and reflects the
distribution of cosmic-ray electrons (CREs) diffusing and radiating in the interstellar magnetic field.
The break in radio synchrotron intensity at about 6\arcmin\ radius (Fig.~\ref{radial} right) indicates
a break in the distribution of cosmic-ray sources. The smooth distribution beyond the break radius can
be explained by diffusion of the CREs of a few kpc before they leave the star-forming disk.
The wavelength-dependent distributions (Fig.~\ref{radial} left) can also be
understood as the result of CRE diffusion, as suggested previously by \citet{basu13}, \citet{taba13b}, and
\citet{berkhuijsen13}.

In the case of propagation by
energy-independent {\em \emph{diffusion}}, the CRE particles are scattered by the irregularities in the
magnetic field. The diffusion length ($l_{\rm d}$) depends on the diffusion coefficient
($D$) and the CRE lifetime ($t_{\rm CRE}$) as $l_{\rm d} \propto (D \, t_{\rm CRE})^{0.5}$.
If, however, CREs propagate with respect to the gas by the {\em \emph{streaming instability}}\ \citep[e.g.][]{kulsrud05},
the propagation length is $l_{\rm s} = {\rm v_A} \, t_{\rm CRE}$, where ${\rm v_A}$ is the Alfv\'en velocity.
It is generally assumed that $B_{\rm tot} \propto \rho^{0.5}$ (where $\rho$ is the gas density), so that
${\rm v_A} \propto B_{\rm tot} \, \rho^{-0.5} \simeq$~const. If the CRE lifetime is limited by synchrotron loss,
$t_{\rm CRE} \propto B_{\rm tot}^{\,\,-1.5} \, \nu^{\,\,-0.5}$,
where $\nu$ is the frequency at which the CREs of energy $E$ radiate in the total field $B_{\rm tot}$
in the sky plane ($\nu \propto B_{\rm tot,\perp} \, E^2$).
If the CRE lifetime is limited by inverse Compton loss,
$t_{\rm CRE} \propto B_{\rm tot}^{\,\,0.5} \, B_{\rm IC}^{\,\,-2} \, \nu^{\,\,-0.5}$,
where $B_{\rm IC} = 3.3 \, (1+z)^2 \muG$ is the equivalent magnetic field strength having the same energy
density as the radiation field.
As a result, we expect $l_{\rm d} \propto \nu^{\,\,-0.25}$ for diffusive propagation
or $l_{\rm s} \propto \nu^{\,\,-0.5}$ for streaming propagation.

The frequency dependence of synchrotron scalelengths in the outer disk of IC~342 (Table~\ref{tab:length}) yields
$l_\mathrm{sync} \propto \nu^{\,\,-0.45\pm0.08}$, which is in good agreement with {\em \emph{streaming propagation}}.
Forthcoming observations at lower frequencies with LOFAR will allow us to test this dependence
with higher accuracy. The streaming propagation length for a given CRE lifetime is longer than that for
diffusive propagation, which may also explain the large scalelengths of the radio disk of IC~342 (Sect.~\ref{sect:radial}) and the low radio surface brightness (Sect.~\ref{sect:mf}).

LOFAR and VLA data of the spiral galaxy M~51 also reveal breaks in the radial distribution of the nonthermal emission
at a radius of about 10\,kpc, beyond which the star-forming activity decreases sharply \citep{mulcahy14},
suggesting a similarity of the underlying physical processes between M~51 and IC~342.
However, the difference in exponential scalelengths at \wave{20} between the inner and the outer disks is
shorter in M~51 than in IC~342. Furthermore, the exponential scalelengths in the outer disk of M~51 have a
frequency dependence that is weaker than in IC~342 and is consistent with diffusive propagation.
\citet{basu13} studied the ratio of scalelengths between \wave{90} and \wave{20} in five spiral galaxies. For two galaxies (NGC~4736 and NGC~6946) the values are consistent with diffusive propagation,
and three cases are unclear.

CRE propagation seems to be different between galaxies, but the reason remains unclear.
M~51 has a weak large-scale regular field but a strong anisotropic turbulent field \citep{fletcher11} that
may support diffusive propagation along the field. A regular magnetic field generated
by a mean-field dynamo like in IC~342 (Fig.~\ref{cm6eff}) may favour streaming propagation. However, the
field is more regular in NGC~6946 \citep{beck07} where diffusive propagation seems to dominate \citep{basu13}.
These contradictory results call for further studies of a larger sample of galaxies.

\subsection{Energy densities}
\label{sect:energy}

\begin{figure}[htbp]
\vspace{0.7cm}
\centerline{\includegraphics[width=0.45\textwidth]{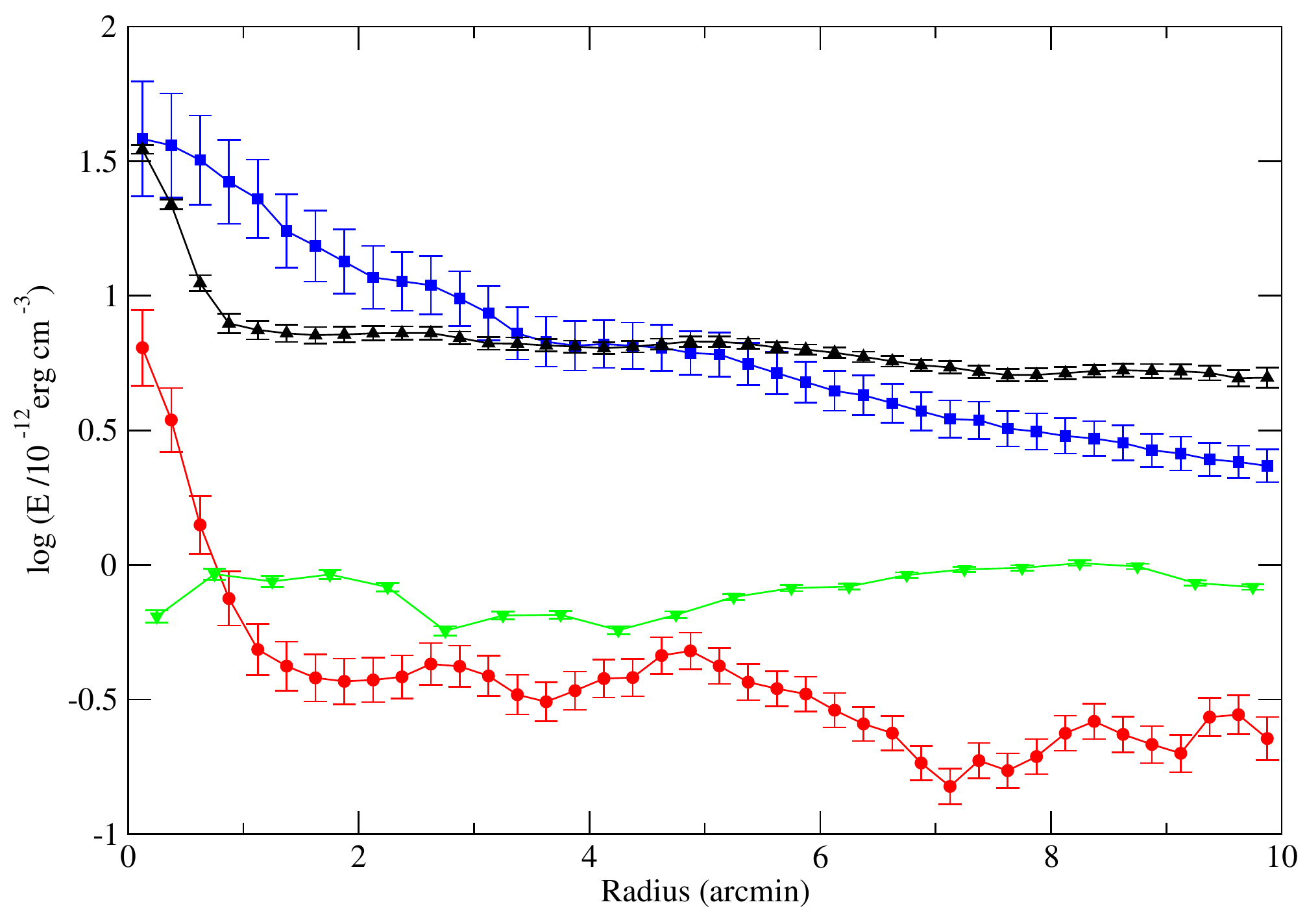}}
\caption{Mean energy densities as a function of radius in IC~342. Red circles: thermal energy of the ionized gas;
blue squares: turbulent kinetic energy of the neutral gas; black upwards triangles: magnetic energy of the total field;
green downwards triangles: magnetic energy of the ordered field. The error bars include only noise errors
in the maps from which the energy densities are derived. No systematic errors are included, e.g. those
due to a radial variation of thermal gas temperature, of the filling factor or of the turbulent gas
velocity, nor errors due to deviations from the equipartition assumption.}
\label{en}
\end{figure}

The energy densities of the various components of the ISM in IC~342 (Fig.~\ref{en})
are determined in rings in the plane of the galaxy that is inclined by $i=31\degr$.
The energy density of the total equipartition magnetic field ($B^2_\mathrm{tot}/8\pi$) in the disk,
identical to that of the total cosmic rays in case of the equipartition assumption, is derived
from the map of synchrotron intensity at \wave{6.2} at 25\arcsec\ resolution
(Fig.~\ref{th} right), assuming a
constant pathlength through the emitting medium of 1\,kpc/cos(i). The energy density of the ordered
magnetic field is derived from the average synchrotron intensity and the degree of synchrotron
polarization at \wave{6.2} in the same rings (Fig.~\ref{perc} right).
The thermal energy density (${3\over2} \langle n_\mathrm{e} \rangle k T$) of the
warm ionized gas ($T\simeq 10^4$\,K) is calculated from the map of thermal radio emission
(giving the emission measure $EM = L \langle n_\mathrm{e}^2 \rangle$, which is dominated by
$\HII$ regions (see Fig.~\ref{th} left), using a pathlength $L$ of 100\,pc/cos(i) and a constant
volume filling factor $f$ of 5\% \citep{ehle93}. The average number density
$\langle n_\mathrm{e} \rangle = \sqrt{f \langle n_\mathrm{e}^2 \rangle}$ of the $\HII$ regions
decreases from about $0.3\ccm$ at 1\arcmin\ radius to about $0.2\ccm$ at 5\arcmin\ radius.

The number density $\langle n \rangle$ of the total neutral gas (molecular + atomic) is taken
from \citet{crosthwaite01}. Here, $\langle n \rangle$ decreases from about
$30\ccm$ at 1\arcmin\ radius to about $7\ccm$ at 5\arcmin\ radius.
The molecular gas traced by the CO emission dominates the total neutral gas until about 3\arcmin\ radius.
To compute the kinetic energy density (${1\over2} \rho$~v$_\mathrm{tur}$) of the turbulent motion of the
total neutral gas, its turbulent velocity is assumed to be v$_\mathrm{tur}=10\kms$, taken from the
average velocity dispersion derived for the $\HI$ gas, which is almost constant in the inner disk of
IC~342 \citep{crosthwaite00}. The velocity dispersion derived for the molecular gas in the inner 3\arcmin\
is larger \citep{crosthwaite01}, but this is due to non-circular motions on larger scales than the turbulent scale.
Similar values for the velocity dispersions in $\HI$ and CO of about $12\kms$ have been derived by
\citet{caldu13} for a sample of 12 nearby spiral galaxies.

The energy densities of the turbulent motion and the total magnetic field are similar in the
inner disk (Fig.~\ref{en}), and both are more than an order of magnitude higher than the thermal
energy density. Turbulence in IC~342 is {\em \emph{supersonic,}}\ and its ISM is a
{\em \emph{low--$\beta$ plasma}}\ (where $\beta$ is the ratio of thermal to magnetic energy densities).

Beyond about 5\arcmin\ radius, the magnetic energy density decreases only slowly and is higher than that
of all other energy densities. This result is similar to those found in NGC~6946 \citep{beck07},
M~33 \citep{taba08} and the four galaxies studied by \citet{basu13}.
The total magnetic energy density is overestimated if the field strength fluctuates along the line of sight
(Sect.~\ref{sect:mf}), but can hardly remove the dominance of magnetic energy. On the other hand,
the dominance of magnetic energy is probably even greater in the outer parts of the galaxy
because of energy losses of cosmic-ray electrons (Sect.~\ref{sect:mf}) and because v$_\mathrm{tur}$ decreases
with radius in the outer disk, as indicated by the $\HI$ velocity dispersion \citep{crosthwaite00}.

The energy density of the ordered field in IC~342 is roughly constant to about 5\arcmin\ radius \footnote{In
the innermost ring, the value is a lower limit due to beam depolarization.} and  increases in the outer galaxy
(Fig.~\ref{en}), which is a challenge for classical mean-field dynamo models (see Sect.~\ref{sect:dynamo}).
Other than in IC~342, a slow decrease was found in NGC~6946 \citep{beck07}.
However, the values for NGC~6946 were erroneously determined by assuming equipartition between cosmic rays
and the ordered magnetic field, leading to overestimates at short radial distances and underestimates at
large radial distances. Applying the correct method to the data of NGC~6946 leads to similar behaviour
to Fig.~\ref{en}.

\subsection{Magnetic field extent}
\label{sect:eq}

In case of equipartition and constant ratio $K$ between the number densities of cosmic-ray protons
and electrons, the scalelength of the total magnetic field $l_\mathrm{B}$ is
$(3-\alpha_\mathrm{syn}) \simeq $4\,times larger than the synchrotron scalelength $l_\mathrm{syn}$
at high frequencies, where flattening by CRE diffusion is unimportant. With $l_\mathrm{syn}\simeq6\arcmin\
\simeq 6$\,kpc in the outer disk (Table~\ref{tab:length}), $l_\mathrm{B}\simeq24$\,kpc. This value is still a
lower limit because $K$ is expected to increase with radius. {\em \emph{The total magnetic field of IC~342 probably
extends far out into intergalactic space}.} A similar result was obtained for NGC~6946 \citep{beck07}.

The weakest total and ordered fields detected in the outer disk of IC~342 are about $6\muG$ and $2.5\muG$,
respectively (Sect.~\ref{sect:mf}). At the low frequencies observable with LOFAR, the synchrotron
intensity is higher and maps of magnetic field strengths can be obtained with high accuracy, such as for M~51
\citep{beck13,mulcahy14}. Detection of even weaker fields \footnote{For total magnetic
fields of $\le 3.3 \, (1+z)^2 \muG$ strength at redshift $z$ the energy loss of cosmic-ray electrons
is dominated by the inverse Compton effect in the CMB background.} will become possible with more sensitive
instruments like the JVLA and the planned Square Kilometre Array (SKA) \citep{beck15a}.

Weak regular magnetic fields can also be measured by their Faraday rotation of polarized emission from
background sources, but also requires deep integrations to obtain a sufficiently large number of
sources \citep{stepanov08}. The observations presented in this paper detected only one
source behind the inner disk of IC~342 and another three behind the outer disk (see Sect.~\ref{sect:pol}),
insufficient to separate the Faraday rotation contributions of IC~342 from that of the Milky Way.
The SKA will be able to increase this number significantly.

\subsection{Dynamo action in the outer disk}
\label{sect:dynamo}

In IC~342 (like in NGC~6946 and M~33) the magnetic energy density decreases more slowly than the
turbulent energy density and dominates beyond about 5\arcmin\ radius, while models of field
amplification by small-scale and mean-field dynamos predict a close similarity between
these energy components \citep{beck96,gent13}. This may point towards an additional source
of turbulence supporting mean-field dynamo action in the outer disk. For example, the magneto-rotational
instability (MRI) \citep{balbus98} is able to explain the velocity dispersion in the $\HI$ line of galaxies
in the outer disk ($r \gtrsim r_{25}$) \citep{tamburro09}.

The radially increasing energy density of the ordered field in Fig.~\ref{en} is even more surprising.
Present-day Faraday rotation data (Sect.~\ref{sect:rm}) are insufficient to decide whether the ordered field
in the outer disk is regular (i.e. of dynamo type) or anisotropic turbulent (due to enhanced shear or
compression). In the latter case, the field anisotropy would need to increase with radius. One source of anisotropy
is shear ($S$). However, for a flat rotation curve, $S$ decreases with increasing radius, so that other sources
of enhanced shear would be needed, for example, through the interaction with a companion galaxy
or with the intergalactic medium.

If the ordered field in the outer disk is regular, the mean-field dynamo should have a radially increasing
efficiency to overcome the general decrease in the total equipartition field. The mean-field
dynamo can also operate in the outer disk of galaxies \citep{mikhailov14}. The dynamo efficiency in the
thin-disk approximation can be estimated by a few parameters, rotational shear $S$ at radius $r$ given by
$S = r \, (\partial \Omega/\partial r)$, turbulent velocity v$_\mathrm{tur}$, and scale height $h$ of the ionized
gas \citep{arshakian09}. For a flat rotation curve and
constant v$_\mathrm{tur}$, the dynamo efficiency scales as $(h/r)^2$ and may increase with radius if $h$
increases faster than linearly with radius $r$. However, the radial variation of magnetic pitch angles
(Fig.~\ref{pitch}) gives no indication for a significant increase in $h$ until 10\arcmin\ radius.

The regular field generated by a mean-field
dynamo could be transported to the outer disk by the
joint action of a dynamo and turbulent diffusivity \citep{moss98c}. With a propagation speed of a few kpc
per Gyr \citep{mikhailov14}, this process is slow and needs continuous dynamo action without disturbing
merger events.

Finally, advanced dynamo simulations indicate that the mean-field dynamo (that dominates in the outer disk) can
generate regular fields with an energy density significantly larger than that of turbulent gas motions \citep{shapovalov11,brandenburg14}. The saturation level is given by the balance between the Lorentz force
and forces exerted by large-scale gas motions. A radially increasing strength of the regular field would
require a higher energy density of gas motions, such as an increasing rotation curve, as indicated from
$\HI$ data in the outer disk \citep{newton80a,crosthwaite00}.

Vice versa, \citet{ruiz10} propose that the extended regular magnetic field exerts forces to the gas in
the outer parts of the galaxy and may explain the increase in the rotation velocity in M~31 beyond
30\,kpc radius. The velocity increase in IC~342 beyond about 20\,kpc \citep{crosthwaite00,meidt09}
may also have a magnetic origin. A proper estimate of magnetic forces will become possible when the radial
profiles of the turbulent and regular magnetic fields are measured with high accuracy \citep{elstner14}.

Future sensitive Faraday rotation measure ($RM$) data, preferably measuring an $RM$ grid of polarized background
sources with the Square Kilometre Array (SKA) and its precursor telescopes ASKAP and MeerKAT could shed light
on the mystery of the strong magnetic fields in outer galactic disks \citep{beck15a}.

\subsection{The polarization spiral arms in IC~342}
\label{sect:features}

Five types of polarization spiral arms are observed in IC~342:

(1) A narrow arm (about 300\,pc wide, corrected for beam smoothing) seen at \wave{6.2} (Fig.~\ref{spitzer} left,
north of DEC=+68\degr\ 06\arcmin) is located at the inner edge of the dust arm, east of the central region,
and runs parallel to the adjacent material arm with an inward shift
of about 200\,pc. This indicates compression by density waves, as in the galaxy M~51 \citep{patrikeev06}.
The eastern arm is similarly narrow (200--300\,pc) in both dust and total radio emission (Fig.~\ref{wise} right),
indicating compression of magnetic field and gas by a density wave. Indeed, the CO line emission shows
distortions in the velocity field \citep{crosthwaite01}.
The magnetic pitch angle is aligned with the spiral structure (Fig.~\ref{cm3} right), as expected from
density-wave compression. The polarization degree of synchrotron emission of about 16\% at \wave{6.2}
is similar to that observed in the spiral arms of M~51 \citep{fletcher11}.

(2) A broad arm (300--500\,pc wide) with a ridge line that oscillates around the northern dust arm
(no.~4 in Fig.~\ref{cm6} right, seen also in Fig.~\ref{cm3} right and \ref{spitzer} left).
Polarized intensity, magnetic pitch angle, and Faraday rotation
measures are also oscillating. These field distortions are probably due to the Parker instability
(Sect.~\ref{sect:parker}). Why this phenomenon occurs only in this region of the galaxy remains unclear.

(3) A broad arm south of the north-western spiral arm seen at \wave{6.2} (no.~3 in Fig.~\ref{cm6} right) that is
located between the main dust arms and associated with weak dust emission (Fig.~\ref{spitzer} left).
This spiral arm has a constant pitch angle (Fig.~\ref{spiral} left) and a width of about 500\,pc.
Location, width, and pitch angle qualify it as a magnetic arm. However, the average polarization degree of
synchrotron emission of 18\%, the extent of about 5\,kpc
and the ordered field strength of about $6\muG$ are significantly lower than those for the magnetic arms
of NGC~6946 \citep{beck07}, so that this arm is called a \emph{\emph{rudimentary magnetic arm}}.

(4) Broad and long spiral arms in the outer south-eastern galaxy seen at \wave{20.1}, roughly coincident with
arms of dust and total neutral gas (nos.~1, 2 and 5 in Fig.~\ref{cm20b} right, also seen in Figs.~\ref{spitzer}
right and \ref{gas}). The typical degrees of polarization at 25\arcsec\ resolution are 10\%, 20\%, and 30\% for
the inner, middle, and outermost arms. The ordered field strength is about $8\muG$.
There are no signs of crossing between the polarization and gaseous arms, as claimed by \citet{krause89a}
based on older $\HI$ data. Aligned arms of gas and magnetic fields can be a signature of fast MHD density waves
\citep{lou99,lou02}. The almost constant pitch angles of the polarization arms (Table~\ref{tab:pitch})
further support that this is a large-scale wave phenomenon. In mean-field dynamo models, the magnetic
pitch angle is independent of the spiral structure in the gas, so that no alignment is expected from dynamo
action alone.

(5) Fuzzy features in the outer south-western galaxy seen at \wave{20.1}, roughly coincident with similar
features of total neutral gas (Figs.~\ref{cm20b} right and \ref{gas} left). Tidal interaction with the Local Group
\citep{buta99} may have distorted the spiral structure there.

\subsection{A helically twisted Parker loop}
\label{sect:parker}

\begin{figure*}[htbp]
\vspace{0.7cm}
\includegraphics[width=0.45\textwidth]{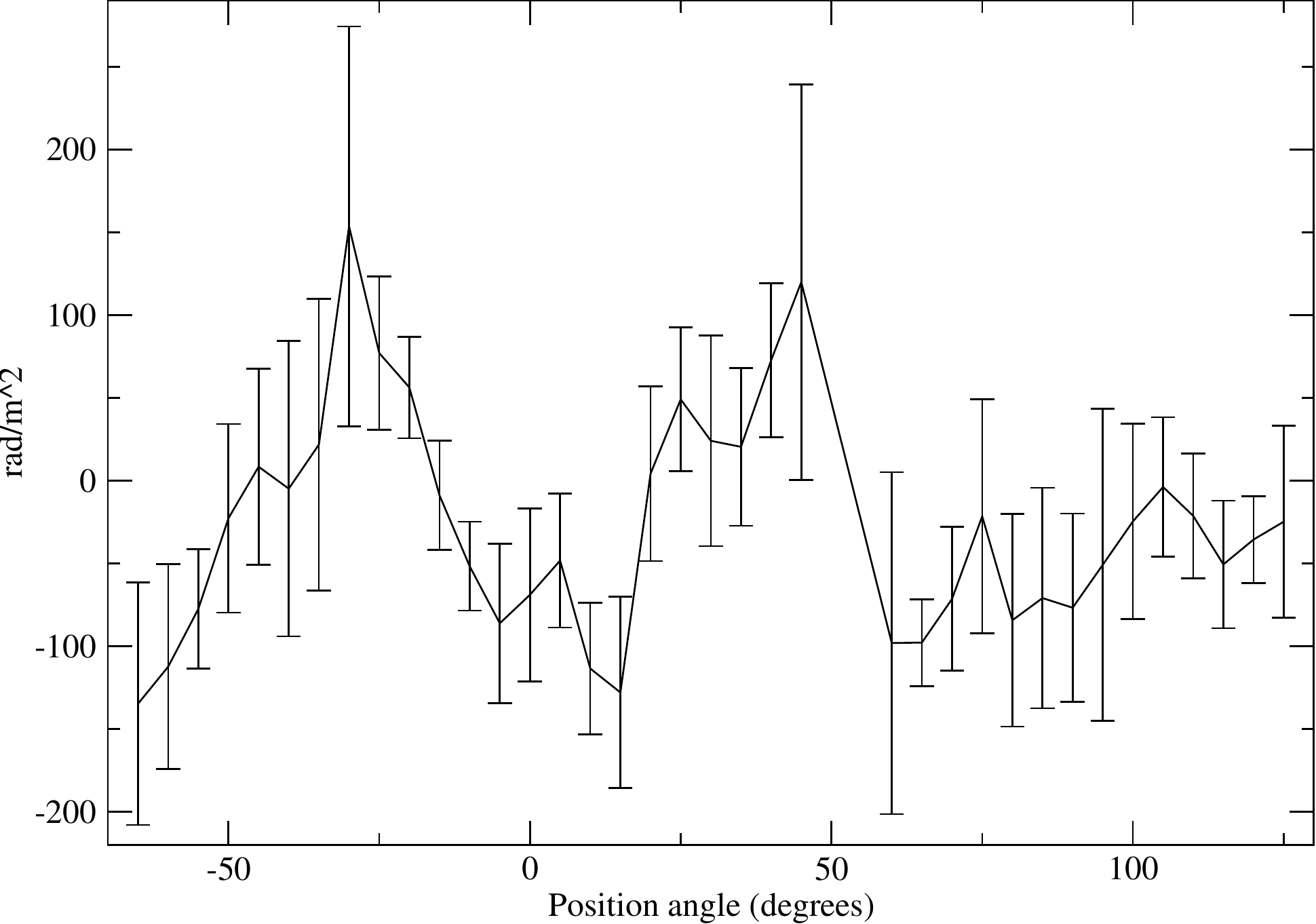}
\hfill
\includegraphics[width=0.45\textwidth]{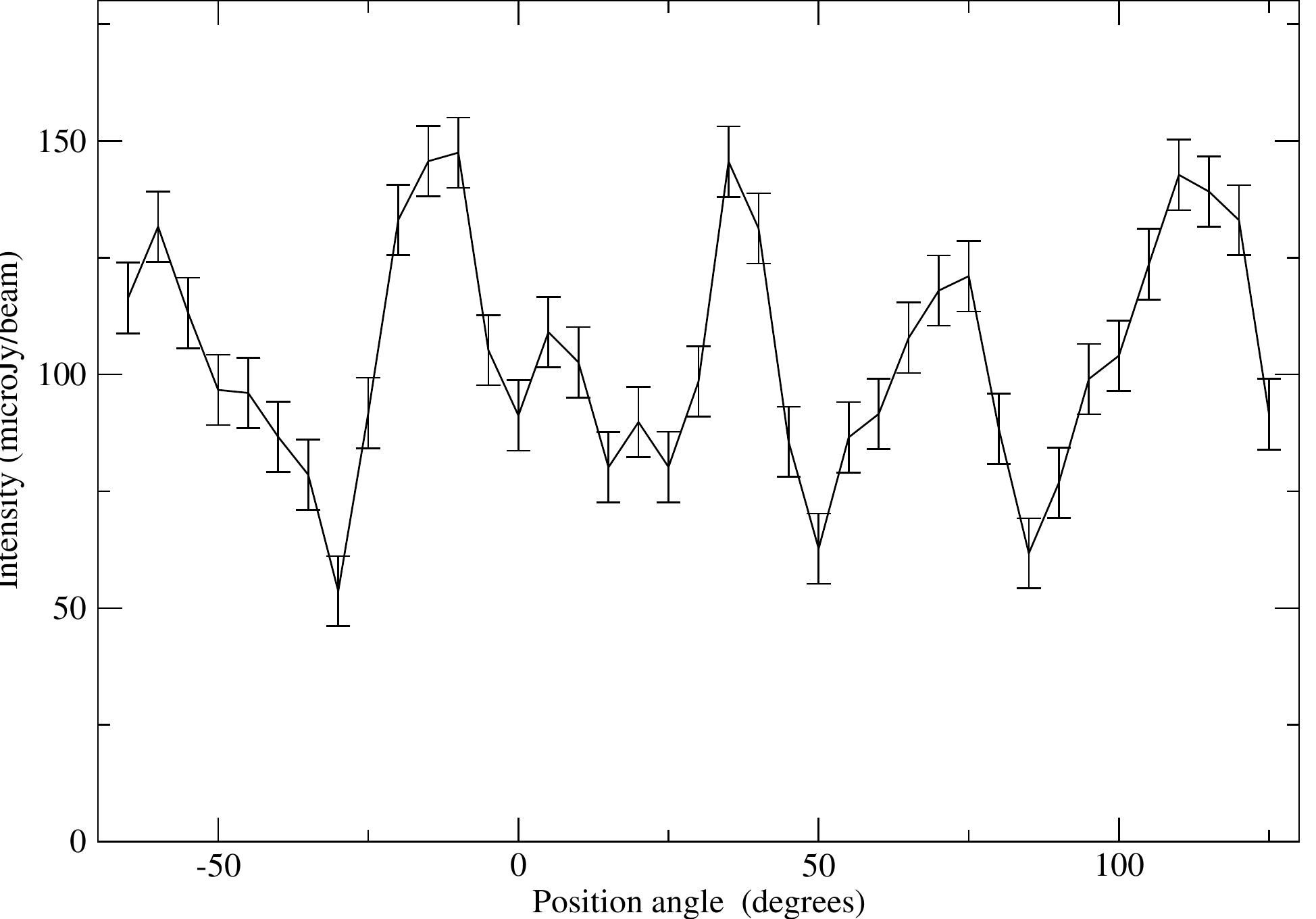}
\caption{{\it Left:\/} Average Faraday rotation measures $RM$ between \wave{3.5} and \wave{6.2} (VLA only)
at 25\arcsec\ resolution along a half ring between 3\arcmin\ -- 4\arcmin\ radius in the sky plane,
with its centre at RA, DEC (J2000) = $03^\mathrm{h}\ 46^\mathrm{m}\ 26\fs 9$, +68\degr\ 06\arcmin\ 07\arcsec\
(offset from the galaxy's centre), which best
delineates the inner spiral arm from north-west to south-east. The position angle along the ring starts at
-65\degr\ and runs counter-clockwise (0\degr\ is north and 90\degr\ is east). An interval in position angle of
50\degr\ corresponds to 3\arcmin\ (3\,kpc) along the ring.
{\it Right:\/} Average polarized intensity at \wave{6.2} (VLA only) at 25\arcsec\ resolution in the same half ring.
}
\label{circle}
\end{figure*}

Faraday rotation measures $RM(3/6)$ (Fig.~\ref{rm} left) are found to vary quasi-periodically along a
half-circle delineating the northern spiral arm (no.~4 in Fig.~\ref{cm6} right), with two pronounced maxima
at about $-30$\degr\ and +40\degr\ position angles, one pronounced minimum at about +10\degr\ and a broad
minimum around +80\degr\ (Fig.~\ref{circle} left). The average distance between
extrema of about 35\degr\ corresponds to about 2.1\,kpc at the radius of the half circle.
This feature can be interpreted as a magnetic loop (``Parker instability'') extending out
of the galaxy's disk and out of the sky plane, giving rise to a periodic pattern in $RM$.
A regular field bending out of the plane should lead to minima in polarized intensity (tracing the field
component in the sky plane) at locations where $|RM|$ (tracing the field component along the line of sight)
is at its maximum. Indeed, Fig.~\ref{circle} (right) shows minima of polarized intensity roughly at the locations of
extrema in $RM$, with a similar distance between the maxima or minima of about 37\degr\ or 2.2\,kpc.

This is the second indication of a Parker loop in the magnetic field of a nearby galaxy, after M~31 \citep{beck89}.
The numerical models by \citet{kim02} (who assume $\beta=1$) predict a wavelength of the most unstable symmetric mode
between $4\,\pi\,h$ and $17\,h$, where $h$ is the scale height of the $\HI$ gas. The peak-to-peak wavelength
of about 4\,kpc measured in IC~342 corresponds to $h\approx 230-320$\,pc, which is larger than typical scale
heights of $\HI$ gas disks of spiral galaxies \citep{bagetakos11}. The reason for this discrepancy may be that the assumption of $\beta$ is incorrect, because observations indicate $\beta<1$ (Sect.~\ref{sect:energy}).

The magnetic field in the northern arm diverts not only in the vertical direction, but also in the disk plane.
The ridge line of the polarization spiral arm (Fig.~\ref{spitzer} left), as well as the magnetic pitch angle at
\wave{3.5} (Fig.~\ref{cm3} right), oscillates around the northern dust arm with similar periodicity. This gives
indication for a large-scale, helically twisted flux tube, as predicted by models of the Parker instability
\citep{shibata91,hanasz02}.

Further radio polarization observations with higher resolution may uncover more helical loops in spiral galaxies.

\subsection{Origin of magnetic arms}
\label{sect:ma}

In most spiral galaxies observed so far, the highest polarized intensities (i.e. strongest ordered fields) are
detected {\em \emph{between}}\ the optical arms, filling a large fraction of the interarm space, sometimes concentrated
in magnetic arms, as in NGC~6946 \citep{beck07}. The southern-sky spiral galaxy
NGC~2997 hosts compressed magnetic fields at the inner edges of material arms, as well as one well-developed
magnetic arm \citep{han99}. The stronger density waves in NGC~2997 lead to higher
degrees of polarization at \wave{6.2} of typically 25\% at the inner edge of the northern arm, similar to those
in M~51 \citep{fletcher11}. The degree of polarization of the magnetic arm of 40\% and its length of at least
10\,kpc are similar to that in NGC~6946; both values are much higher than in IC~342.
Like IC~342, M~51 has a short, rudimentary magnetic arm with a low degree of polarization \citep{fletcher11}.
NGC~1566 shows signatures of a magnetic arm in the south-east \citep{ehle96},
but the angular resolution of these radio observations was too coarse to clearly detect magnetic arms.
In large barred galaxies, ordered magnetic fields also fill most of the interarm space, but do not form well-defined
magnetic arms, as seen in NGC~1672 \citep{beck02}, NGC~2442 \citep{harnett04}, NGC~1097, and NGC~1365 \citep{beck05c}.

Several mechanisms have been proposed to explain the higher degree of field order in interarm regions:\\
(1) Magnetic field ropes as a result of a magnetic buoyancy instability in a turbulent high--$\beta$ plasma
\citep{kleeorin90} however, the ISM of galaxies is a low--$\beta$ plasma,
meaning that the thermal pressure is lower than the magnetic pressure (Sect.~\ref{sect:energy});\\
(2) slow MHD density waves in the rigidly rotating inner region of a galaxy \citep{lou99,poedts02}  however,
slow MHD density waves in 3-D can be subject to the Parker and shearing instabilities \citep{foglizzo95};\\
(3) more efficient action of the mean-field dynamo between the optical
arms due to lower turbulent velocity in interarm regions \citep{moss98a,shukurov98};
however, the observed turbulent velocity is {\em \emph{not}}\ lower in the interarm regions \citep{crosthwaite01};\\
(4) more efficient mean-field dynamo action between the optical arms due to a smaller correlation length in
interarm regions \citep{rohde99}; however, no indication of such a variation in correlation length has
been found from observations so far;\\
(5) introduction of a relaxation time of the magnetic response in the dynamo equation, leading to a phase
shift between the material and magnetic spiral arms \citep{chamandy13a,chamandy13b} however, the resulting
magnetic arms are restricted to a relatively small region around the corotation radius and have a much smaller
pitch angle than do gaseous arms;\\
(6) magnetic arms generated by multiple interfering spiral patterns \citep{chamandy14} they are more extended than
those mentioned under (5), but these still have too small pitch angles;\\
(7) drift of magnetic fields with respect to the gaseous arms in a non-axisymmetric gas flow caused by a spiral
perturbation \citep{otmian02} or by a bar \citep{kulpa11};\\
(8) suppression of the mean-field dynamo in the material arms by continuous injection and amplification of
turbulent fields by supernova shock fronts \citep{moss13};\\
(9) suppression of the mean-field dynamo in the material arms by star-formation-driven outflows
\citep{chamandy15};\\
(10) magnetic arms as a transient phenomenon during the evolution of galactic magnetic fields,
possibly related to the short lifetimes of spiral patterns seen in numerical simulations
\citep[e.g.][]{sellwood11,wada11,dobbs14}.

Models (7) -- (10) are the most promising ones. However, all these models are simplified and consider either
gravitational perturbations or dynamo action.
Self-consistent MHD models of galaxies, including the gravitational potential
with spiral perturbations and mean-field dynamo action, are still missing.\\

The origin of the rudimentary magnetic arm in IC~342 deserves a detailed discussion.
The absence of long magnetic arms in IC~342 is surprising in view of the apparent similarities between
IC~342 and NGC~6946, like the rotation curve \citep{crosthwaite00,boomsma08} and the star-formation surface
density $\Sigma_\mathrm{SFR}$ \citep{calzetti10}.
One difference is the ``lopsided'' spiral structure of NGC~6946, with a spiral arm with massive
star formation in the north-east and an asymmetric rotation curve in $\HI$ \citep{crosthwaite00}.
Although lopsidedness enhances angular momentum transport \citep{jog09}, it hardly modifies the mass distribution
and hence cannot stabilize the spiral pattern.

Increasing the star-formation surface density $\Sigma_\mathrm{SFR}$ in mean-field dynamo models decreases
the general field order, while magnetic arms become more pronounced \citep{moss13}.
However, the star-formation rate per surface area ($\Sigma_\mathrm{SFR}$) is only about 10\% higher in
NGC~6946 than in IC~342 \citep{calzetti10} and hence cannot explain the difference in the magnetic pattern
between these two galaxies.

The timescale of mean-field dynamo amplification in a galaxy with a flat rotation curve can be approximated as
$t_\mathrm{dyn} \simeq 3 \, h \, r / $(v$_\mathrm{rot} \, l_\mathrm{tur})$ \citep{giess14}.
The maximum rotation velocity v$_\mathrm{rot}$ is almost the same in both galaxies. Assuming that
$h$ and $l_\mathrm{tur}$ are also similar, the dynamo timescale of IC~342 is about 1.7\,times larger
owing to its larger disk radius $r$, measured by the optical radius $r_{25}$. The timescale of field ordering
also depends linearly on radius $r$ \citep{arshakian09}, so that the dynamo in IC~342 needs more time
to develop a stable magnetic pattern.

The evolution of a spiral magnetic pattern in IC~342 may be hampered further by the possible tidal interaction
with the Local Group \citep{buta99} and/or by the complex spiral pattern in the gas of IC~342, described by a
superposition of a two- and four-armed spiral pattern with different pattern speeds \citep{meidt09}, which may
lead to an unstable spiral pattern \citep{crosthwaite00}. The mean-field dynamo needs a least a few rotation
periods to build up a regular field \citep[e.g.][]{moss12}. A short lifetime of a stable pattern may not allow
the formation of magnetic arms.  The existence of two regions of magnetic fields with opposite
directions in disk and central region (Sect.~\ref{sect:central}) and the relatively weak ordered field
(Sect.~\ref{sect:flux}) are further hints that the large-scale field structure is not (yet) fully developed
in IC~342.

The spiral pattern in the gas of M~51 is also distorted and short-lived, so that only a weak large-scale
regular field and a rudimentary magnetic arm could develop. On the other hand, NGC~6946 has a two-armed
spiral pattern with a well-defined pattern speed \citep{fathi07}, the spiral field extends smoothly
into the central region \citep{beck07} and the large-scale regular field is strong \citep{ehle93}

In summary, {\em \emph{the existence of magnetic arms may indicate a stable spiral arm pattern over
several galactic rotation periods,}} which may not be given in IC~342.

Models of spiral perturbations in stars and gas \citep[e.g.][]{wada11} have neglected the effect of magnetic
fields so far. The only dynamo model for the amplification and ordering of magnetic fields including the
spiral perturbations of gas density and gas velocity is the kinematical model by \citet{otmian02}, in which
the back reaction of the field onto the gas flow was not included. The MHD model of \citet{pakmor13}
includes self-gravity and spiral perturbations, but no mean-field dynamo action. There is urgent need for
a synergy between these approaches to achieve a comprehensive description of the evolving magnetized
ISM in galaxies.

Further polarization observations with higher resolution of IC~342, NGC~2997, NGC~6946, and other galaxies with
magnetic arms in different evolutionary stages are needed for a better understanding of this phenomenon.


\begin{figure*}[htbp]
\begin{center}
\includegraphics[width=0.85\textwidth]{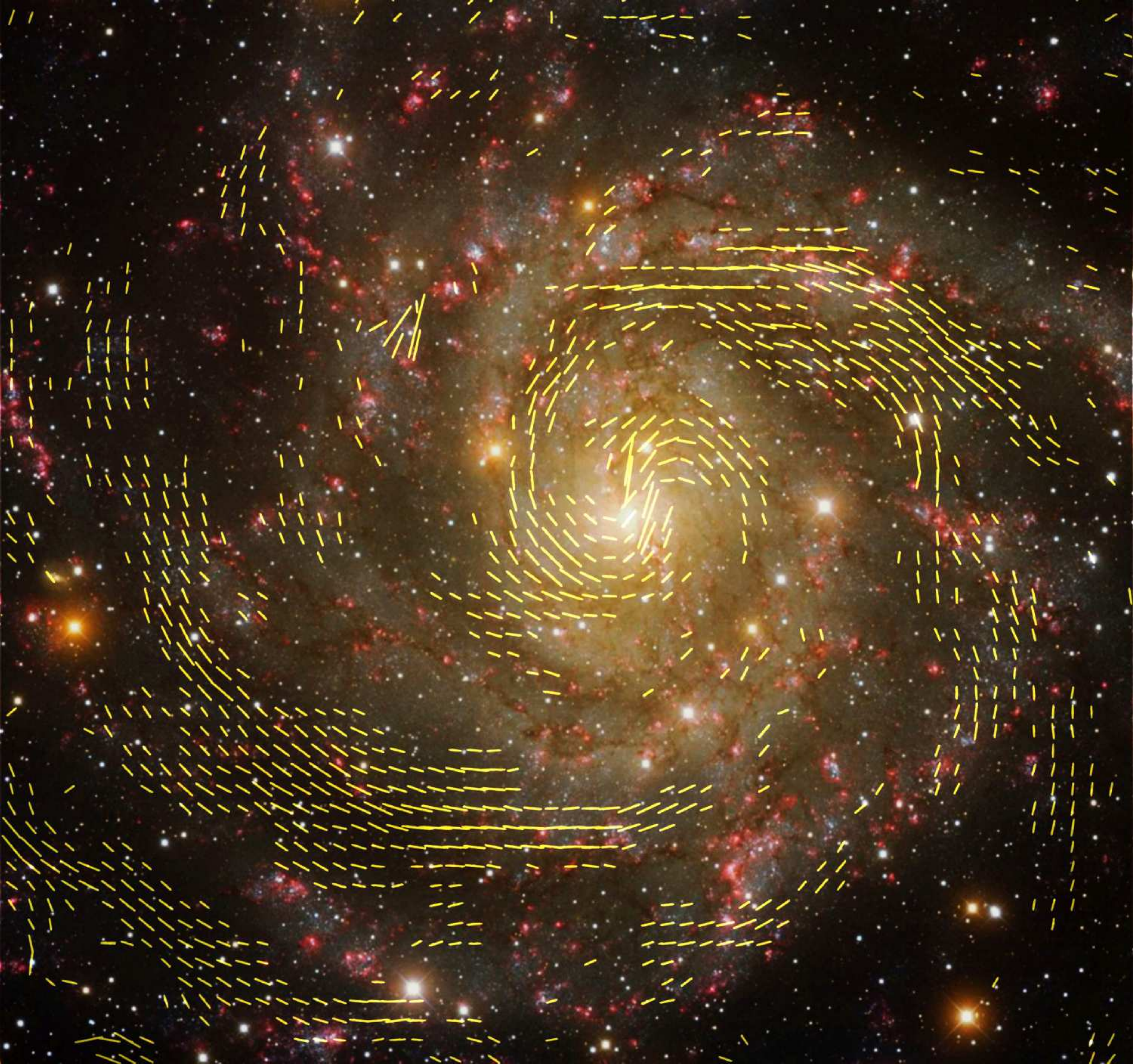}
\caption{$B$ vectors ($E$+90\degr) of IC~342 at \wave{6.2} (VLA+Effelsberg) at 25\arcsec\ resolution,
overlaid on a colour image from the Kitt Peak Observatory (credit: T.A. Rector, University of Alaska Anchorage,
and H. Schweiker, WIYN and NOAO/AURA/NSF).}
\label{kitt}
\end{center}
\end{figure*}

\section{Summary and conclusions}

The total and polarized radio continuum emission from the nearby spiral galaxy IC~342 was observed
in four wavelength bands with the Effelsberg and VLA telescopes. The main results are the following.

(1) The total intensity maps of IC~342 show many spiral filaments embedded in a diffuse disk. The VLA map at
\wave{6.2} alone misses most of the diffuse emission and contains only about 38\% of the total flux density,
while about 75\% of the polarized emission at \wave{6.2} was detected with the VLA alone.
The diffuse disk of polarized emission seen in the Effelsberg maps at \wave{6.2} and \wave{11.2} is
resolved into spiral arms of ordered fields with the VLA, while in total emission, the diffuse disk of
turbulent magnetic fields is seen at low and high resolution.

(2) The total radio emission closely follows
the dust emission in the infrared at 8\,$\mu$m (SPITZER) and at 22\,$\mu$m (WISE). This relation shows
that the amplification of total (mostly turbulent) magnetic fields occurs in star-forming regions, probably by
the turbulent dynamo \citep{abeck12,schleicher13,pakmor14}.

(3) The radial distributions of the total synchrotron emission at \wave{6.2} and \wave{20.1} can be described by
different exponential functions in the inner and outer disks with a break radius at about 6\arcmin, similar to
the break radius in the distribution of the thermal emission and of the total neutral gas between the inner
(strongly star-forming) and outer disks.
The radial scalelength in the outer disk at \wave{20.1} is about 1.7\,times larger than at \wave{6.2},
indicating a larger propagation length of the lower-energy cosmic-ray electrons (CREs)
radiating at the longer wavelength, probably via the streaming instability.
The average radial scalelengths of the radio disk of IC~342 of $4.9\pm0.2$\,kpc at \wave{6.2} and $7.5\pm0.3$
at \wave{20.1} are the largest ones found in any galaxy so far, which may be related to fast CRE propagation
by the streaming instability.

(4) The equipartition strength of the total field in the main spiral arms is typically $15\muG$, that of the
ordered field $5\muG$. The strongest ordered fields of about $8\muG$ are found in the outer southern and
south-eastern arms. The average strengths of the total and ordered fields within a 7\arcmin\ radius of the disk
are $13\muG$ and $4\muG$, respectively. The average ordered field of $4.3\muG$ is about 30\% weaker than that in NGC~6946, indicating less efficient mean-field dynamo action in IC~342.

(5) The average surface brightness of the total radio synchrotron emission of IC~342 is less than half that of the
spiral galaxy NGC~6946. Consequently, the average total equipartition field in IC~342 is about 20\% weaker than in
NGC~6946, which is smaller than expected from the star-formation rate per surface area. Other parameters seem to
influence the total field strength, such as the efficiency of magnetic field amplification or the propagation
characteristics of cosmic-ray electrons.

(6) The energy densities of the turbulent gas motions and the total magnetic field are similar in the
inner disk, and both are more than an order of magnitude higher than the thermal energy density.
Turbulence in IC~342 is supersonic, and its ISM is a low--$\beta$ plasma.
Beyond about 5\arcmin\ radius the magnetic energy density becomes dominant, indicating that the
saturation level of the field is determined by forces exerted by large-scale gas motions.

(7) The average magnetic pitch angle of the ordered field varies from about $-23\degr$ at 1.5\arcmin\ radius
to about $-10\degr$ at 13\arcmin\ radius. In terms of the mean-field dynamo, such a variation
indicates an increase in the scale height of the ionized gas by about a factor of two between 12\,kpc and 14\,kpc
radius, similar to the flaring of $\HI$ disks.

(8) Faraday rotation measures (RM) of the polarization angles seen in the Effelsberg maps confirm that
the large-scale regular field has an underlying axisymmetric spiral (ASS) structure, as previously found by
\citet{krause87}, \citet{graeve88}, and \citet{krause89a}.
The strength of the ASS field of $\simeq0.5\muG$ is about one order of magnitude weaker than the
field observed on scales of a few 100\,pc in the $RM$ map derived from the VLA data.

(9) The ridge line of polarized emission in the north (no.~4 in Fig.~\ref{cm6} right), the polarization
angles, and the Faraday rotation measures show that the magnetic field oscillates around the northern arm
on a scale of about 2\,kpc, indicative of a helically twisted flux tube, which was probably generated by the Parker
instability.

(10) In the eastern inner spiral arm (near to the central region),
the magnetic field is compressed by a density wave, forming a narrow polarization arm
of about 300\,pc width that is displaced inwards with respect to the dust arms by about 200\,pc.
The polarization degree of synchrotron emission of about 16\% indicates mild compression.
The magnetic field is well aligned along the ridge line.

(11) A rudimentary magnetic arm with constant pitch angle is located in an interarm region in the
north-west (no.~3 in Fig.~\ref{cm6} right).
The much longer magnetic arms in NGC~2997 and NGC~6946 are possibly due to more efficient mean-field dynamo
action in a more stable density-wave environment, while the dynamo in IC~342 is slow and weak, probably
disturbed by the bar, tidal interaction, or a transient spiral pattern. Magnetic arms may be a signature of
a mean-field dynamo supported by a spiral arm pattern that is stable over several rotation periods.

(12) Several broad polarization arms in the outer galaxy (nos.~1, 2, and 5 in Fig.~\ref{cm20b} right)
are of spiral shape and mostly coincide with arms
in the total neutral gas. The magnetic pitch angles are as small as the pitch angle of the structure itself.
Fast MHD density waves can generate such coincident spiral arms in the gas and magnetic field.

(13) The magnetic field in the central region resembles that of large barred galaxies like NGC\,1097. It has a
regular spiral pattern with a large pitch angle and is directed outwards, opposite to the direction of the
disk field. Independent mean-field dynamos may operate in the central region and in the disk.

(14) Polarized emission at \wave{20.1} is strongly affected by Faraday depolarization in the northern part
of the galaxy, where the galaxy's rotation is receding. Similar asymmetries
were found in many other mildly inclined galaxies. Helical fields extending from disk to halo as
predicted by mean-field dynamo models can account for this.

\begin{acknowledgements}

This work is based on observations with the VLA, operated by the NRAO, and the 100-m telescope of the MPIfR (Max-Planck-Institut f\"ur Radioastronomie) at Effelsberg. --

This work is based in part on archival data obtained
with the Spitzer Space Telescope, which is operated by the Jet Propulsion Laboratory, California Institute
of Technology under a contract with NASA.
This publication makes use of data products from the Wide-field Infrared Survey Explorer (WISE), which is a
joint project of the University of California, Los Angeles, and the Jet Propulsion Laboratory/California
Institute of Technology, funded by the National Aeronautics and Space Administration. The author thanks the operators at the Effelsberg telescope for their support, the VLA team for performing
the observations and help during data reduction, Bill Sherwood for processing the Effelsberg data at \wave{2.8}
and Uli Klein for help with generating Fig.~\ref{kitt}. Aritra Basu, Michal Hanasz, Marita Krause, Woong-Tae Kim,
and Fatemeh Tabatabaei are acknowledged for useful discussions and comments.
Special thanks go to Marita Krause and Elly M. Berkhuijsen for careful reading of the manuscript,
and to the anonymous referee for suggestions that helped to improve the paper.
Financial support from DFG Research Unit FOR1254 is acknowledged.
\end{acknowledgements}

\bibliographystyle{aa} 
\bibliography{ic342}

\end{document}